\documentclass[amstex,epsf,amssymb,usenatbib]{mnras}
\usepackage{epsf,graphics,subfig}
\usepackage{amsmath,amssymb,natbib}
\usepackage{lmodern}
\usepackage{cite}

\usepackage{rotating,times,graphicx,latexsym,color,longtable,lscape} 

\def\aj{AJ}                   
\def\araa{ARA\&A}             
\def\apj{ApJ}                 
\def\apjl{ApJ}                
\def\apjs{ApJS}               
\def\aap{A\&A}                
\def\aapr{A\&A~Rev.}          
\def\aaps{A\&AS}              
\def\mnras{MNRAS}             
\def\nat{Nature}              

\newcommand{\degsy}{$^\circ$}

\title[Classification of distant radio galaxies]{The morphological classification of distant radio galaxies explored with three-dimensional simulations}
    
\author[M.D. Smith. \& J. Donohoe]
{ Michael D. Smith $^{1}$\thanks{E-mail: m.d.smith@kent.ac.uk} \&
{Justin Donohoe $^{1}$\thanks{E-mail: donohoejustin@googlemail.com }  }\\
$^{1}$Centre for Astrophysics \& Planetary Science, The University of Kent, Canterbury, Kent CT2 7NH, U.K. }                                                                                                                                                             
\date{Accepted .....
      Received ..... ;
      in original form .....}

\pagerange{\pageref{firstpage}--\pageref{lastpage}}
\pubyear{2019}

\begin{document}
                                                                                                                                                             
\maketitle
                                                                                                                                                             
\label{firstpage}
                                                                                                                                                             
\begin{abstract}
We explore the observational implications of a large systematic study of high-resolution three dimensional simulations of radio galaxies driven by supersonic jets. For this  fiducial study, we employ non-relativistic hydrodynamic adiabatic flows from nozzles into a constant pressure-matched environment. Synchrotron emissivity is approximated via the thermal pressure of injected material.  We find that the morphological classification of a simulated radio galaxy depends significantly on several factors with increasing distance (i.e. decreasing observed resolution) and decreasing orientation often causing re-classification from FR\,II (limb-brightened) to FR\,I (limb-darkened) type. We introduce the Lobe or Limb Brightening Index (LBI) to measure the radio lobe type  more precisely.  The jet density also has an influence as expected with lower density leading to broader and bridged lobe morphologies as well as brighter radio jets.  Hence, relating observed source type to the intrinsic jet dynamics is not straightforward.
Precession of the jet direction may also be responsible for wide relaxed sources with lower LBI and FR class as well as for X-shaped and double-double structures. Helical structures are not generated because the precession is usually too slow.
 We conclude that distant radio galaxies could appear systematically more limb-darkened  due to merger-related re-direction and precession as well as due to the resolution limitation.  
\noindent 
\end{abstract} 
  
\begin{keywords}
 hydrodynamics --   galaxies: active -- galaxies: jets --  radio continuum: galaxies
 \end{keywords}                                                                                                                                           
\section{Introduction} 
\label{intro}

Radio galaxies have long been known to be powered by twin jets which inflate the radio lobes and create giant cavities in the ambient medium \citep{1974MNRAS.169..395B,1974MNRAS.166..513S}.  The jets have taken on increased importance as we begin to understand their feedback role on galaxy formation through the conveyance of momentum and energy from Active  Galactic Nuclei into the cluster gas \citep{2012ARA&A..50..455F,2019A&A...622A..12H}. However, the origin, contents and dynamics of the jets  remain issues of debate and approach. The resulting uncertainty in interpretation of their structure, classification and evolution prohibits their employment as diagnostic tools for both the nature of the driving source and their overall cosmological evolution \citep{2012ajb..book.....S}. Indeed, if seen as analogous to smoke out of a fired gun, we may not find them very useful as standard candles or otherwise.

One approach is to perform simulations and then derive synthetic images to compare to observations. However, to be able to relate intrinsic dynamical properties to the diverse shapes of radio galaxies requires a large systematic exploration rather than one-off comparisons.
 In this programme, we have sought to achieve this by first generating systematically a large set of simulations using three dimensional hydrodynamics \citep[see Paper 1:][]{2016MNRAS.458..558D}.  We continue now by deriving the observational properties, covering a range in Mach number, density contrast and precession properties. 

In particular, radio galaxies generated by jets within a wide precession cone would extend the reach  of feedback into the intracluster medium. 
This may help spread support of the gas across an entire cooling-flow \citep{2019MNRAS.485.5590R} and also inhibit galaxy growth, two major issues recently summarised by \citet{2019A&A...622A..12H}. Precession may prove to be a prominent factor in distant radio galaxies as they form in proto-clusters. Therefore, we concentrate on the effects of precession  and the influence this may have on their apparent morphological classification.

Synthetic radio images were first derived by  \citet{1985MNRAS.214...67S} from a single 2D simulation. The study exposed a strongly time-varying structure for the hotspot. The emissivity was taken to be a function of the thermal pressure  as also assumed by \citet{2013MNRAS.430..174H}, employing constant energy partition conditions.  Others have taken  similar approaches to generate synthetic images from hydrodynamical simulations
\citet{1990MNRAS.242..623M,2007ApJS..173...37S,2016MNRAS.458..802N}. 
Radio images  have also been presented more rigorously by \citet{2008AJ....136.2473T}  and \citet{2014MNRAS.443.1482H} where three dimensional simulations with a magnetic field were analysed.
These simulations explored the dependence on the ambient density profile and demonstrated that the total radio flux will essentially  level off \citep{2016MNRAS.461.2025E}.  This is due to the emission being dominated by high pressure warm regions  of constant volume through which jet material is advected at a constant rate. In addition, the difficulty in working from the total luminosity of a radio galaxy to extract the injected jet power 
was emphasised with orders of magnitude variation. Given these findings, we search here for systematic relationships. 

The well-known dichotomy between limb-brightened (FR\,II) and limb-darkened (FR\,I) sources \citep{1974MNRAS.167P..31F} has remained.  Responsibility has been placed on the environment through the influence on the jet Mach number \citep{1985PASAu...6..130B} and the ability to disrupt at a galactic halo transition \citep{1982ApJ...259..522S}. Others have accused (1) the accretion rate since it controls the jet power and penetration \citep{2004A&A...426L..29M}, (2) the black hole spin which triggers a magnetic switch \citep{1999ApJ...522..753M} and (3) hydrodynamic or magnetohydrodynamic instabilities that disturb the jet \citep[e.g.][]{2016MNRAS.461L..46T}. 
In theory, the jet power could well be the decisive factor but three dimensional simulations are necessary to permit turbulent structures to cascade correctly \citep{2016A&A...596A..12M}.  However, observationally,  a radio source produced by a low power jet can be either edge brightened or edge darkened \citep{2017arXiv170303427C}, leaving us in a quandary.

The degree of limb-darkening  or limb-brightening can be parameterised by taking the ratio of the separation between brightness maxima (hotspots) to the total source extent or by calculating one-dimensional moments along the major axis \citep{1976MNRAS.175..461B}.  
 An associated issue surrounds the precise hotspot location in the FR\,IIs. The brightest spot appears very close to the edge in more sources than would be expected given that projection effects should relocate it when the jet is not in the plane of the sky \citep{1990PASAu...8..254J}. In any case, there is considerable freedom to choose a suitable strategy for measuring both the hotspot and lobe length for sources with diverse structure and resolution \citep{2012ApJ...756..116A}. 

A third controversy  surrounds the energy distribution and pressure balance between the lobes and the ambient medium.
 \citet{2000MNRAS.319..562H}  analysed 63 FR\,II radio galaxies embedded in the X-ray radiating gas of galaxy clusters and concluded that measured pressures internal to the lobes are a factor of a few lower than in the surrounding gas. 
However, \citet{2017MNRAS.467.1586I} have now found that
the pressures of FR\,II radio lobes are typically electron dominated by a small factor and are over-pressured relative to the ambient medium in their outer  parts.  This implies that there is no need for an  energetically significant proton population in the lobes of FR\,II radio galaxies (unlike for FR\,Is). 

A fourth issue concerns the relevancy of self-similar solutions.  In general, the re-expansion of gas flowing from the hotspot into the lobe brings the pressure back down to that of the ambient medium, probably to the extent that self-similar solutions are not applicable \citep{2013MNRAS.430..174H}. However, for jets with extremely low densities, the picture changes because the lobe forms a broad protective cocoon around the jet. The higher pressure then acts to squeeze the jet as well as generate a somewhat over-pressured lobe \citep{2016MNRAS.458..558D}. 

It should also be remarked that a further class of compact radio source has been identified. The FR\.0s have a deficit of extended radio emission \citep{2015A&A...576A..38B} and have been interpreted as driven by low-speed jets from black holes of low spin.

Asymmetries across the lobes are not surprising given the host-galaxy motion  through an intracluster medium although
relativistic motions may also contribute. The manifestations of time-varying jets as  X-shaped and double-doubles can be also associated with merger-related accretion or binary black-hole  interactions \citep[e.g.][]{1978Natur.276..588E,2016A&A...595A..46M}.

The relationship between radio power and jet power has emerged as an important quantity \citep{2004ApJ...607..800B}. This is critical to the understanding of momentum and energy feedback from the supermassive black hole to the galaxy, to determine if galaxy construction can be regulated. Knowledge of the jet power, also termed beam power, would help constrain the driving black hole properties.   By  extrapolating to conditions at high redshift, radio galaxies may well provide diagnostic tools for conditions during the development of cluster scale structure. 
The key parameter that we can test is Q, the ratio of jet power to total radio luminosity. In fact. only a small fraction of jet power is converted into radio power with Q as high as 100 at low radio powers \citep{2010ApJ...720.1066C,2012MNRAS.423.2498D}. Furthermore,
 \citet{2016MNRAS.456.1172G} find that in previously used samples of high-power sources, no evidence for an intrinsic correlation is present when the effect of distance is accounted for.

The rather surprising conclusion is that radio luminosity is not a good measure of jet power or the manner in which jet power is deduced is very convoluted and subject to large error propagation \citep[e.g.][]{2012ApJ...756..116A}.
In order to support the intergalactic medium, heating due to adiabatic expansion of the lobes is too small to offset radiative cooling by a factor of at least six. In contrast, the jet power estimates are an order of magnitude larger than would be required to counteract cooling. If this is correct, then heating must be primarily due to another mechanism, possibly through the transmission of precession-triggered 
sound waves. In Paper\,1, we indeed found that the injected jet energy is about ten times higher than contained in the  lobes with, instead, 
the energy efficiently  transferred into thermal form in the ambient medium.

 In this paper, we analyse the observable evolution of a radio galaxy as a function of density contrast, Mach number and precession angle. We are motivated by the issues raised above: the relationship between emission properties and the underlying physics and dynamics. We shall show, however, that source morphology depends on both orientation and resolution.  Distant radio galaxies may thus be classified differently to nearby sources.

\begin{figure}
 \hspace*{-0.9cm} \includegraphics[width=0.53\textwidth]{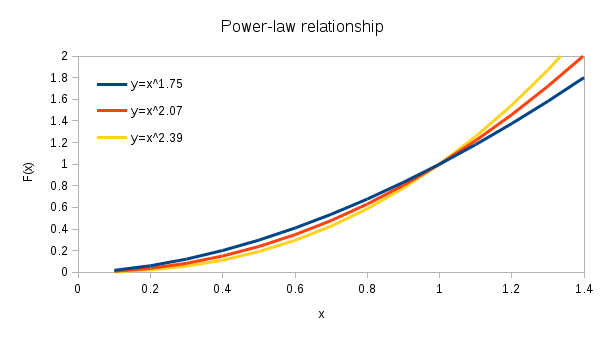} \vspace{-1cm}
 \caption{The three emissivity-pressure power-law relationships utilised in this work for the synchrotron radiation at a specific frequency. Lower indices  allow lower surface brightness regions to be highlighted whereas the opposite is true when a higher index is employed. The three values are chosen to correspond to (negative) spectral indices $\alpha$ of 0.7, 1.0 and 1.3.}
\label{3_power_law_relasionship}
\end{figure}

\begin{figure*}
   \begin{tabular}{ccc}
\includegraphics[width=0.36\textwidth]{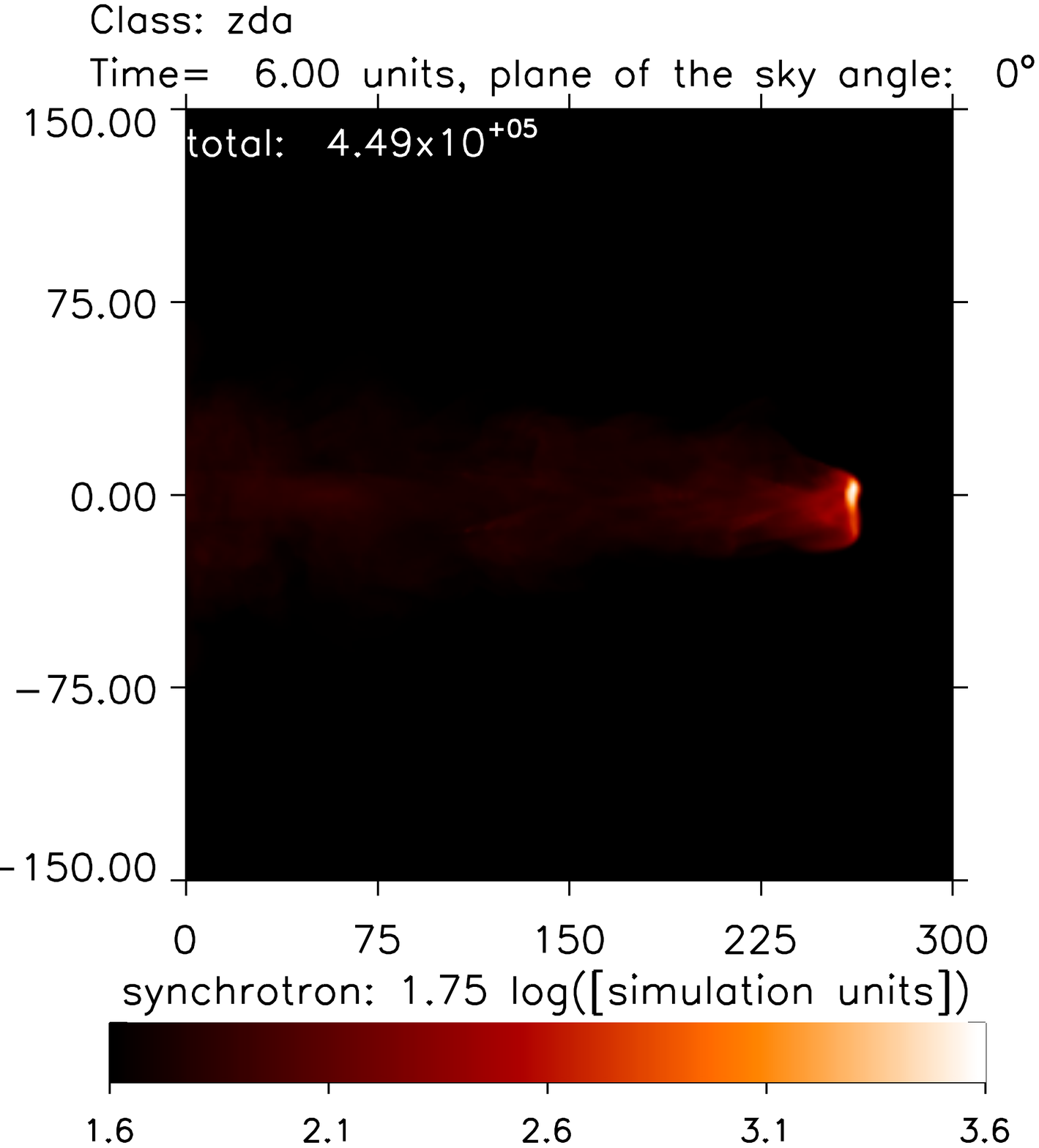} &  \hspace{-1.2cm}   
\includegraphics[width=0.36\textwidth]{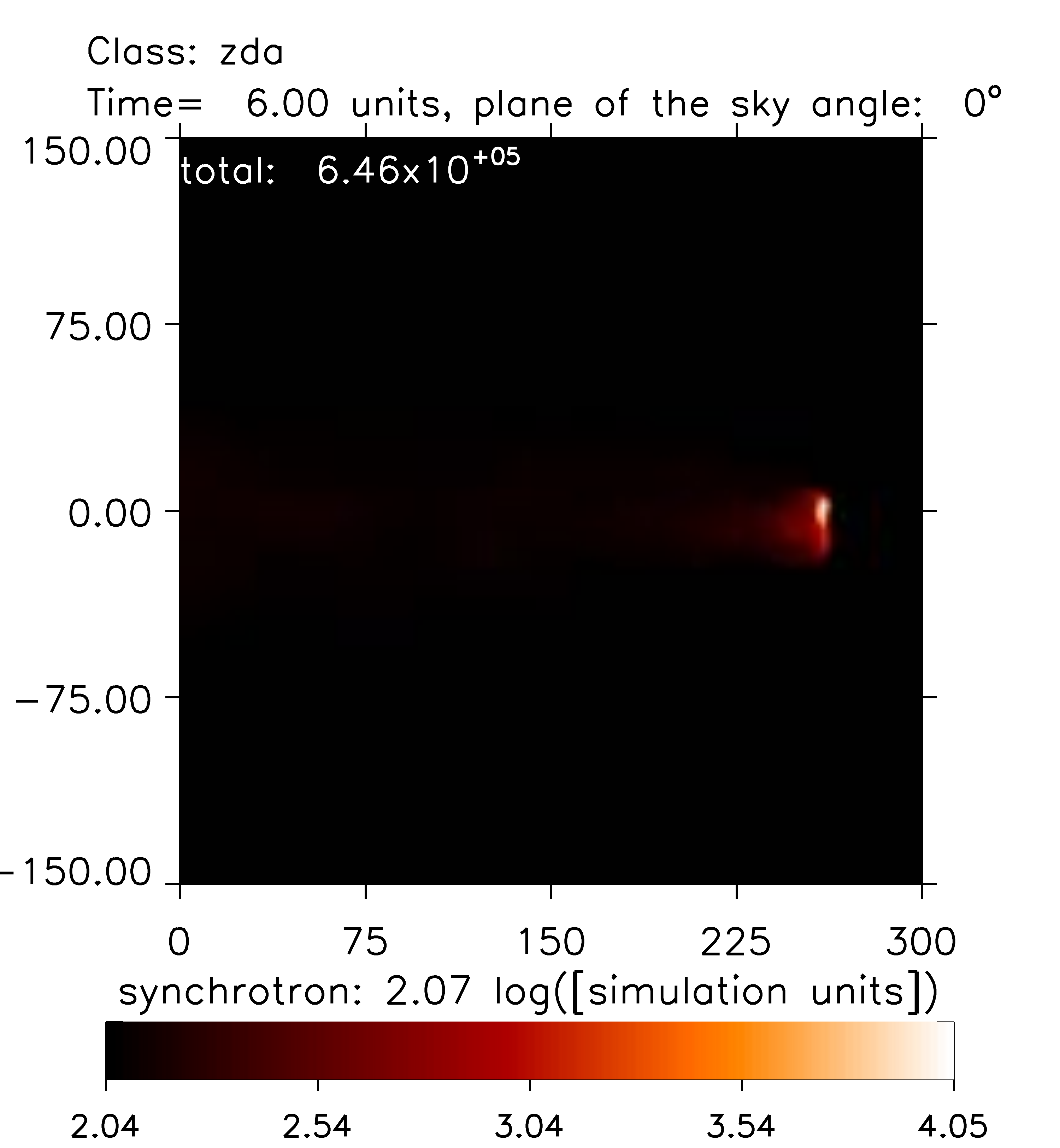} &   \hspace{-1.2cm} 
\includegraphics[width=0.36\textwidth]{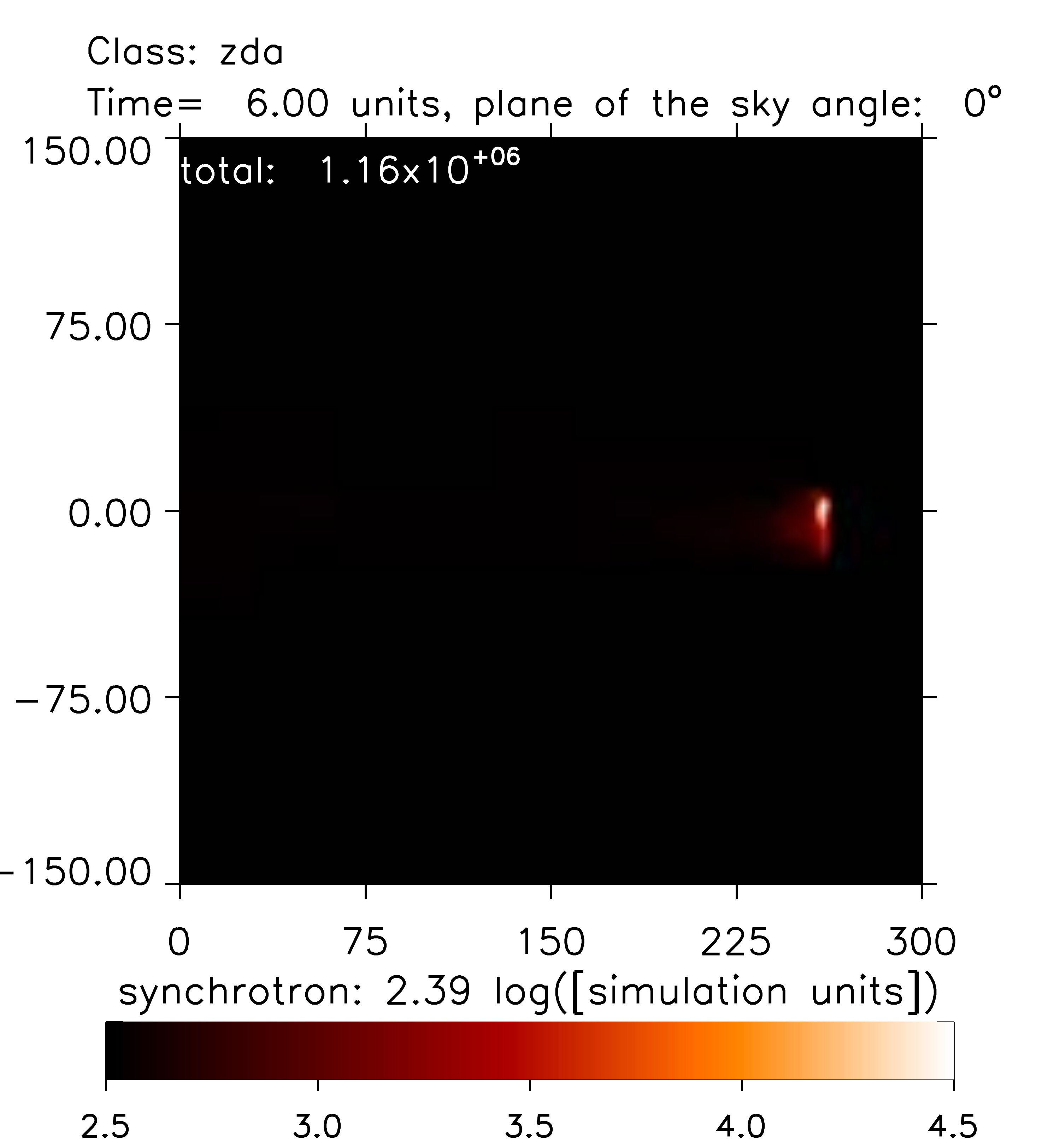}\\
(a) $\alpha$ = 0.7   & (b) $\alpha$ = 1.0  & (c) $\alpha$ = 1.3 \\ [6pt]
\includegraphics[width=0.36\textwidth]{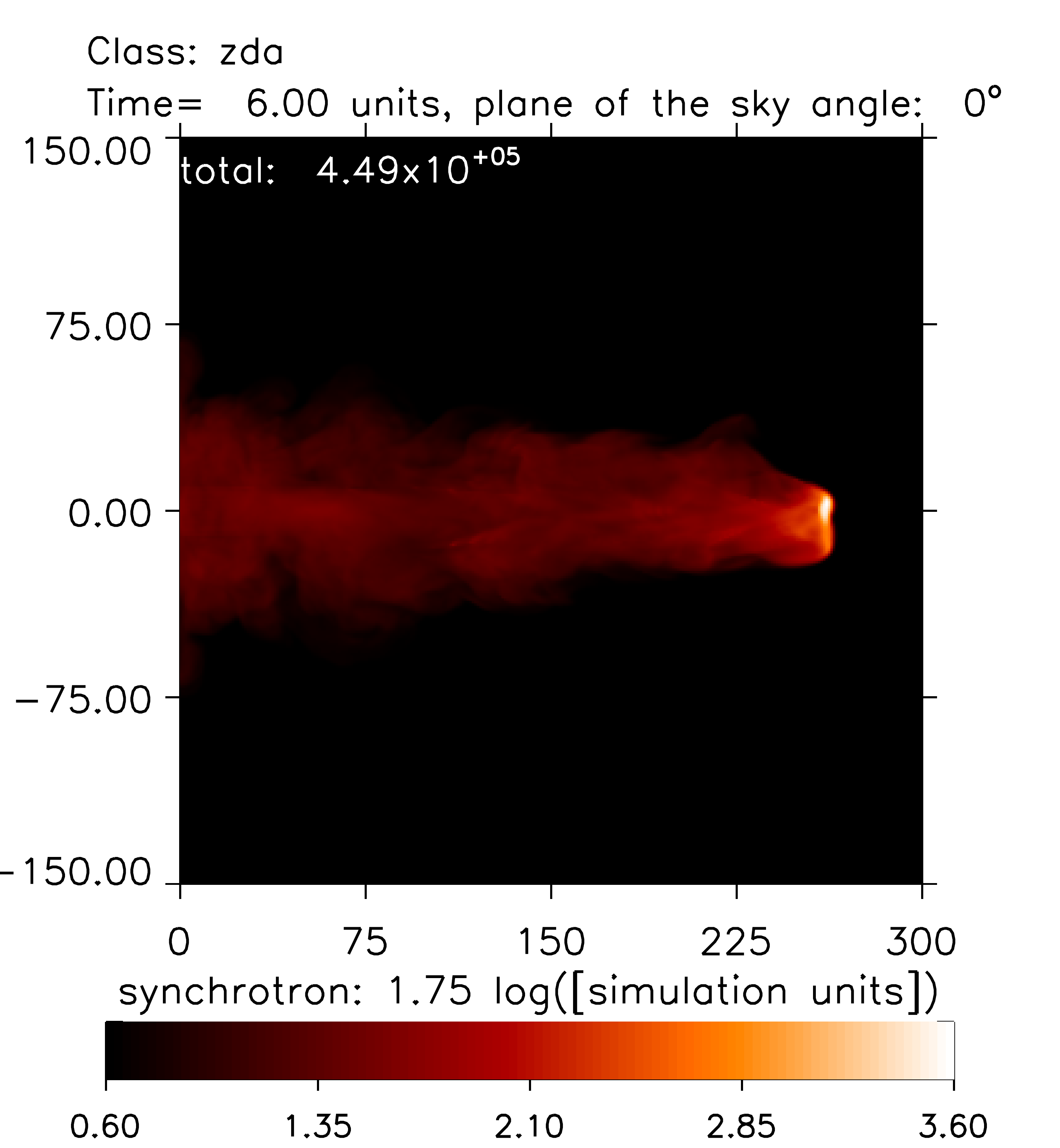} &  \hspace{-1.2cm}  
\includegraphics[width=0.36\textwidth]{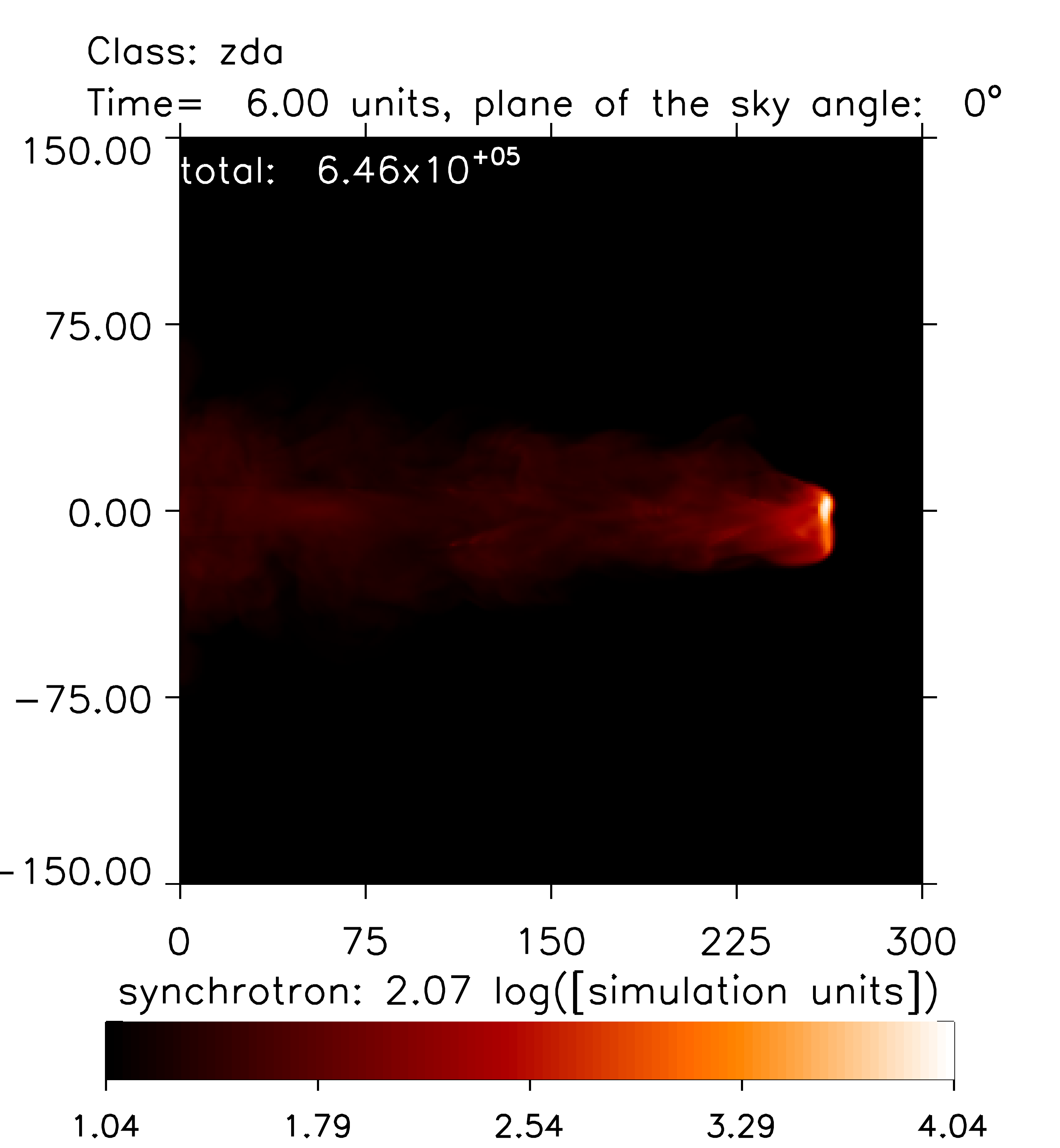} &   \hspace{-1.2cm} 
\includegraphics[width=0.36\textwidth]{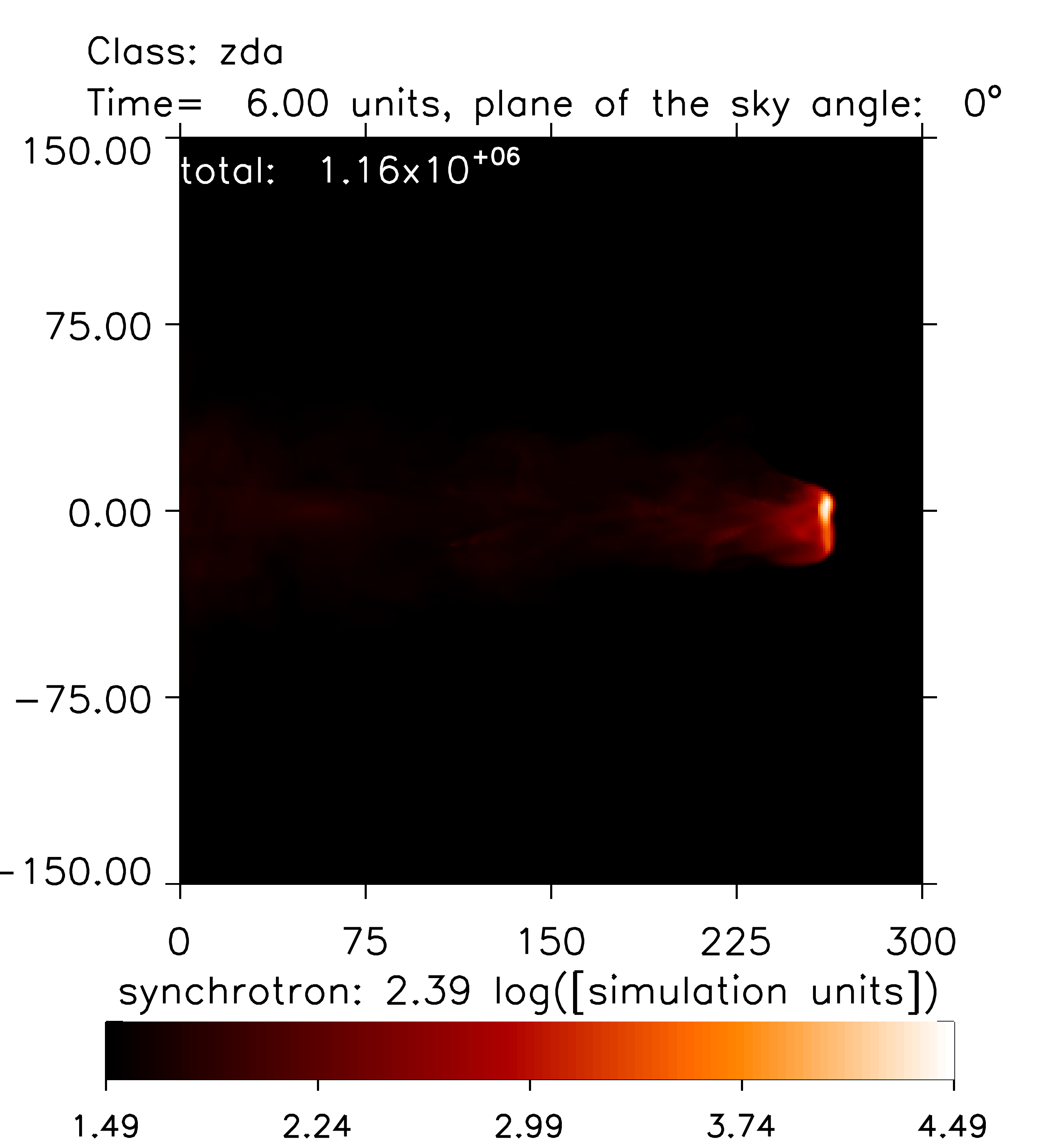} \\ 
(d)  $\alpha$ = 0.7   & (e) $\alpha$ = 1.0  & (f) $\alpha$ = 1.3  
\end{tabular}
\caption{ {\bf Straight jet with 1$^\circ$ precession.} Dependence on radio power-law index and dynamic range. Simulated synchrotron emission from a (perturbed) straight jet simulation with jet-ambient density ratio of 0.1 and  Mach number of  6  (sim. code zda). Both rows show the dependence on the radio spectral index $\alpha$ and the pressure power law index utilised which corresponds to  $\eta$ = 2.4 (left), 3.0 (middle)  and 3.6 (right). The upper row holds the signal range  limits to 100 and the lower row holds it to 1,000. The jet axis is in the plane of the sky.}
\label{zda_power_law}
\end{figure*}
\begin{figure*}
   \begin{tabular}{ccc}
      \includegraphics[width=0.36\textwidth]{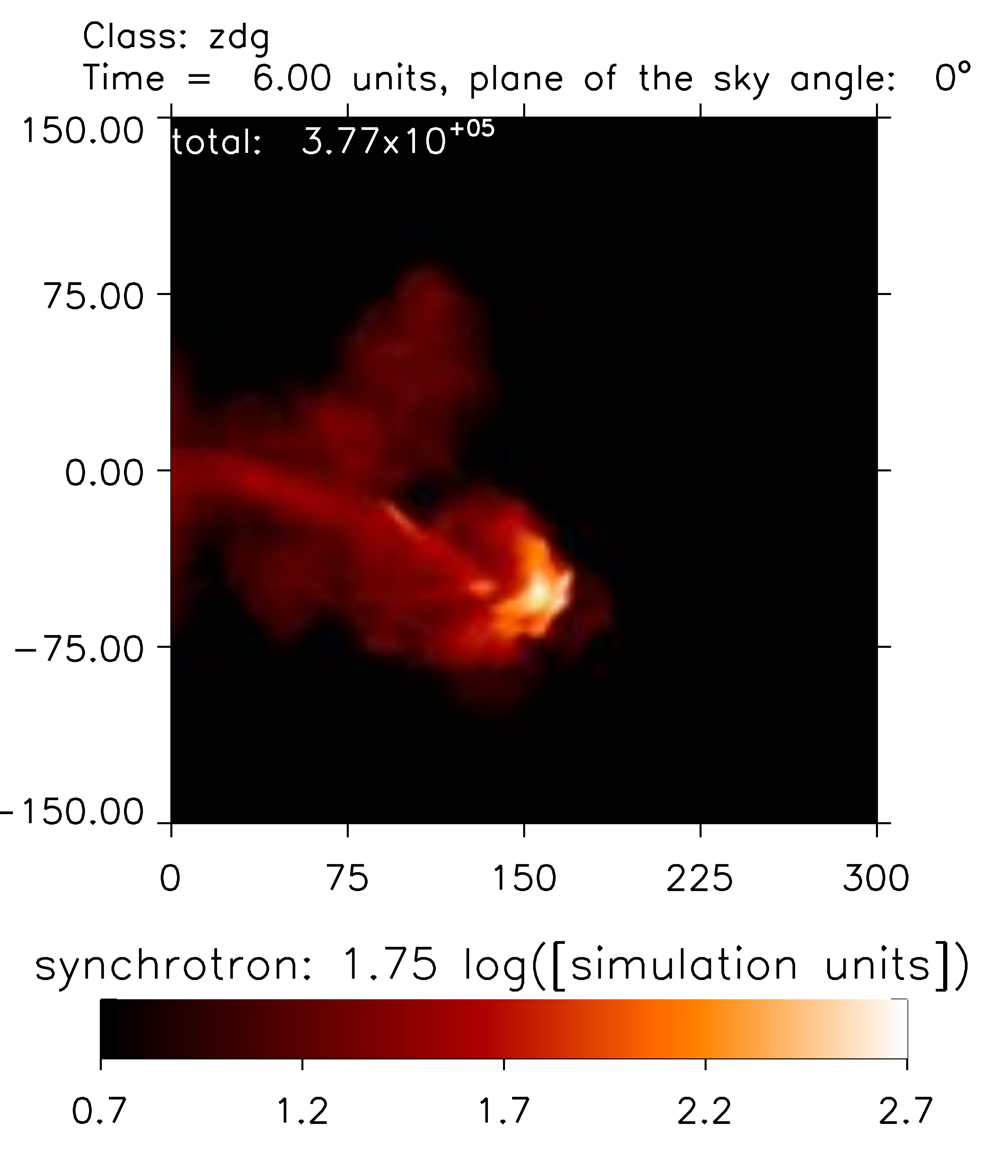} & \hspace{-1.2cm}  
                \includegraphics[width=0.36\textwidth]{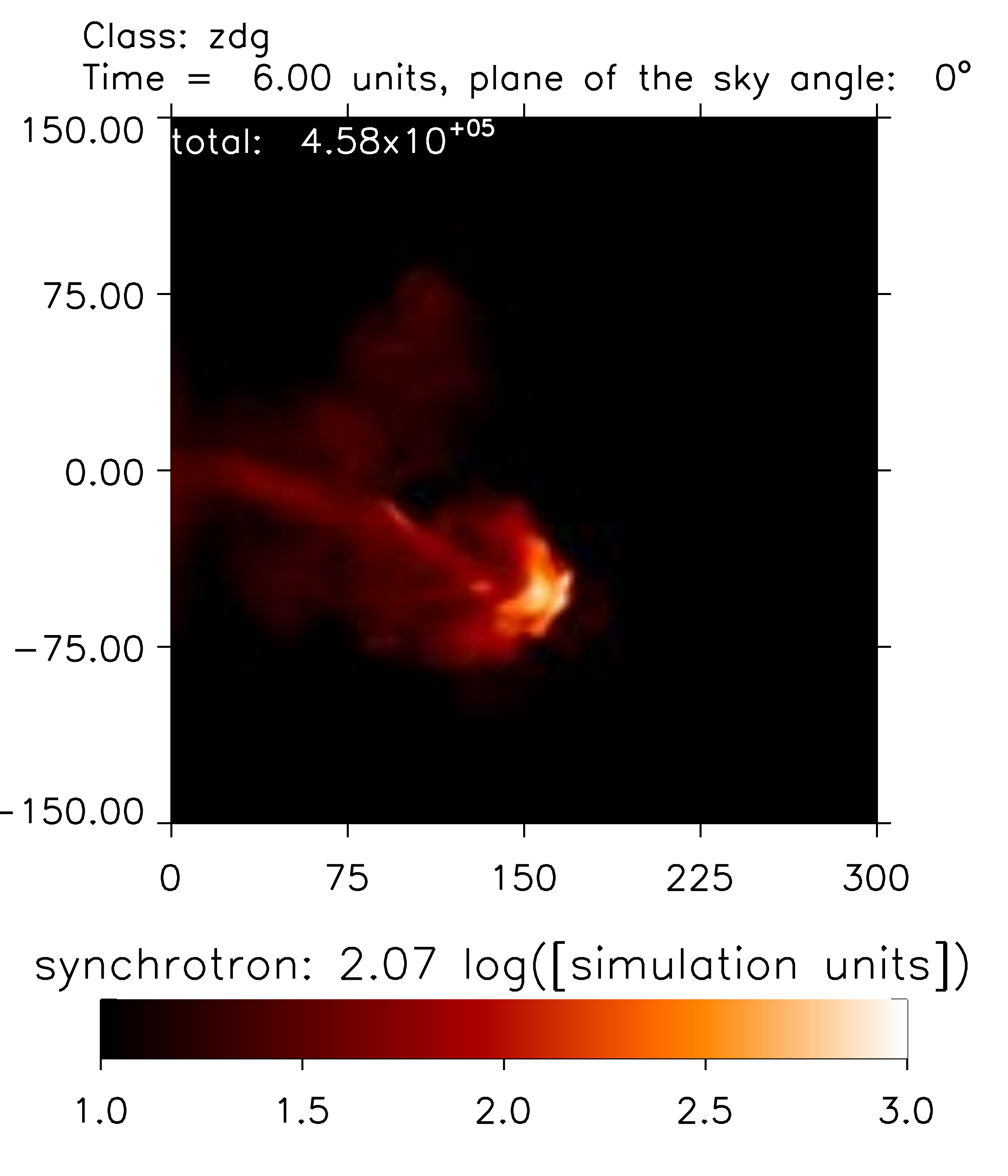} &  \hspace{-1.2cm}  
                 \includegraphics[width=0.36\textwidth]{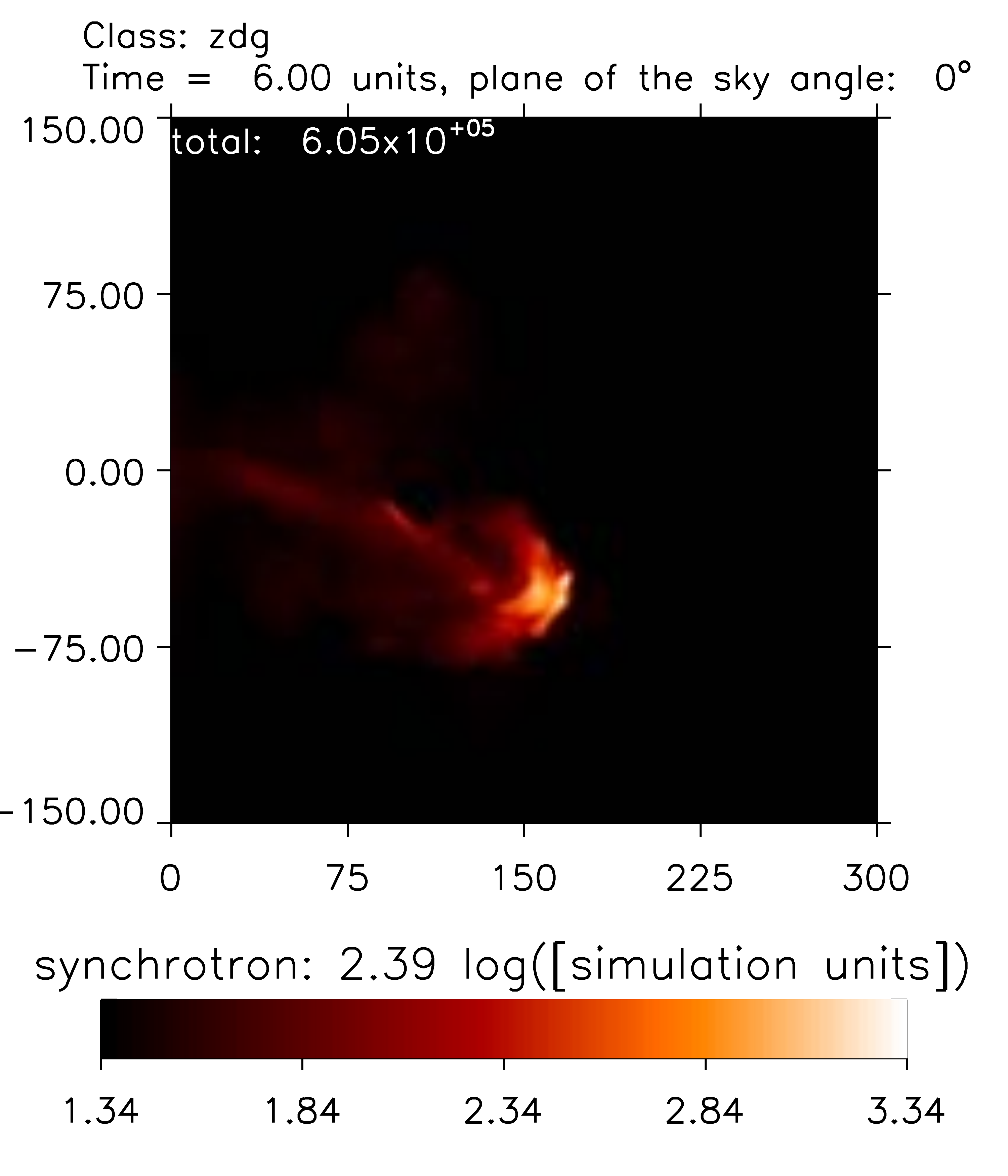}\\ 
 (a) $\alpha$ = 0.7   & (b) $\alpha$ = 1.0  & (c) $\alpha$ = 1.3   \\ [6pt] 
      \includegraphics[width=0.36\textwidth]{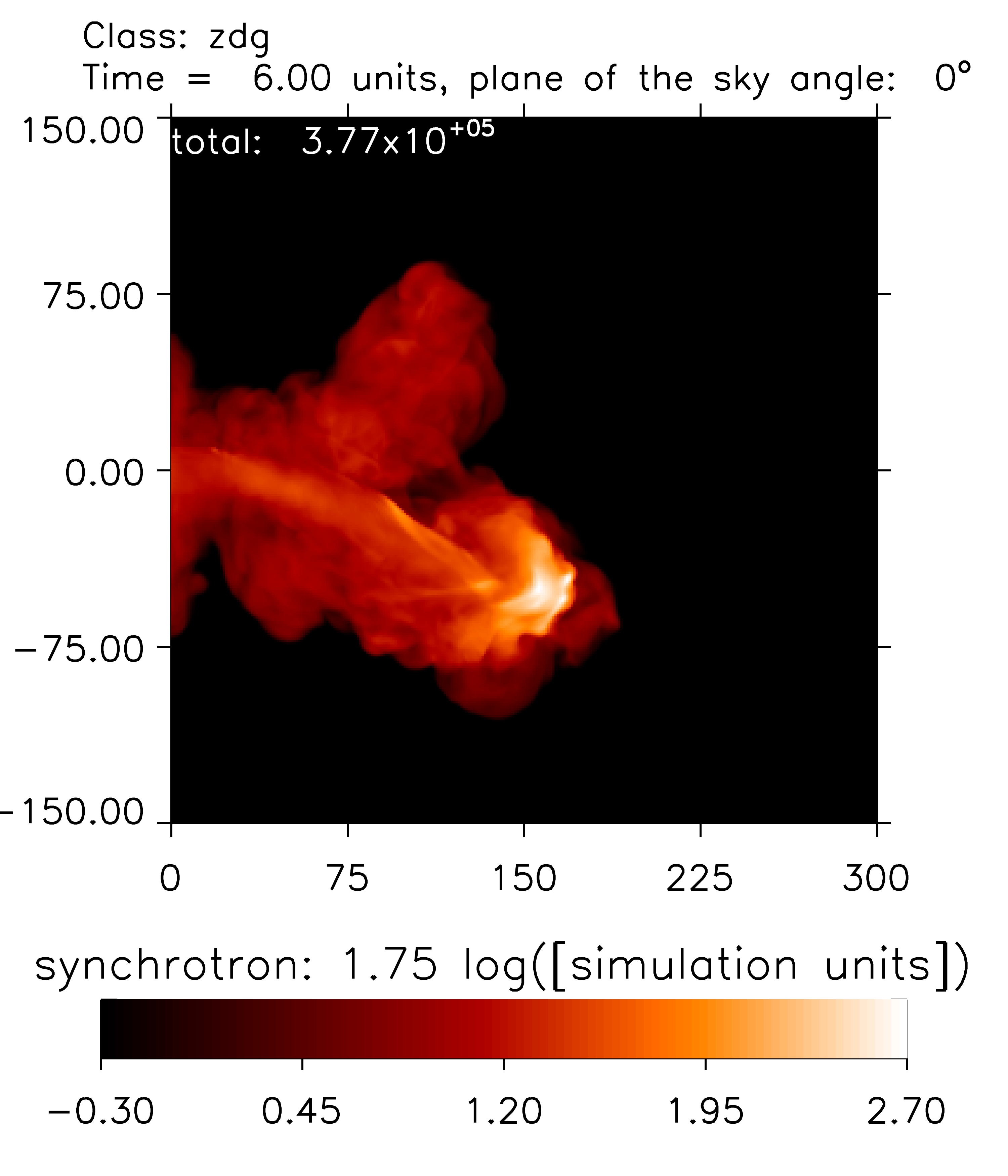} & \hspace{-1.2cm}       
      \includegraphics[width=0.36\textwidth]{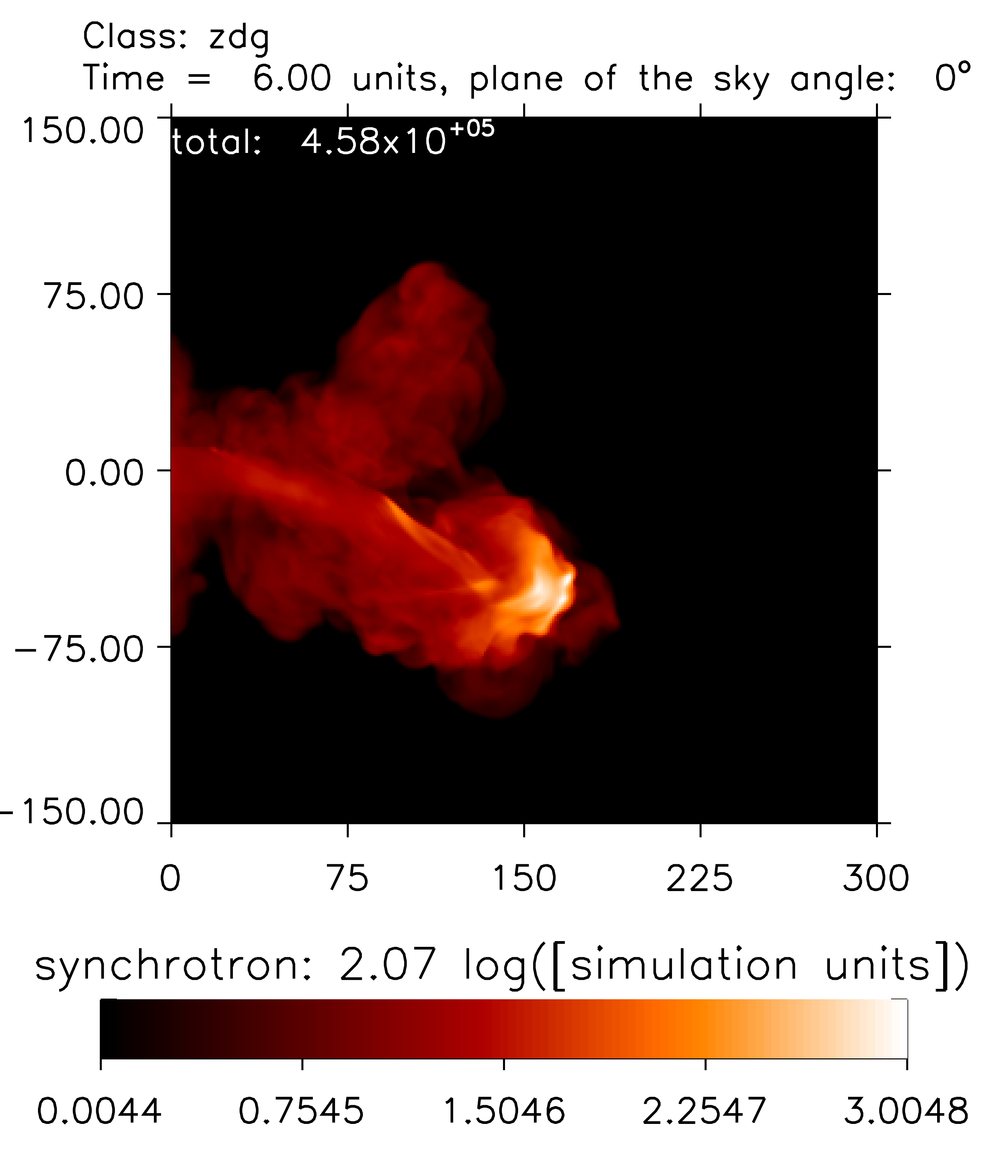} &  \hspace{-1.2cm} 
            \includegraphics[width=0.36\textwidth]{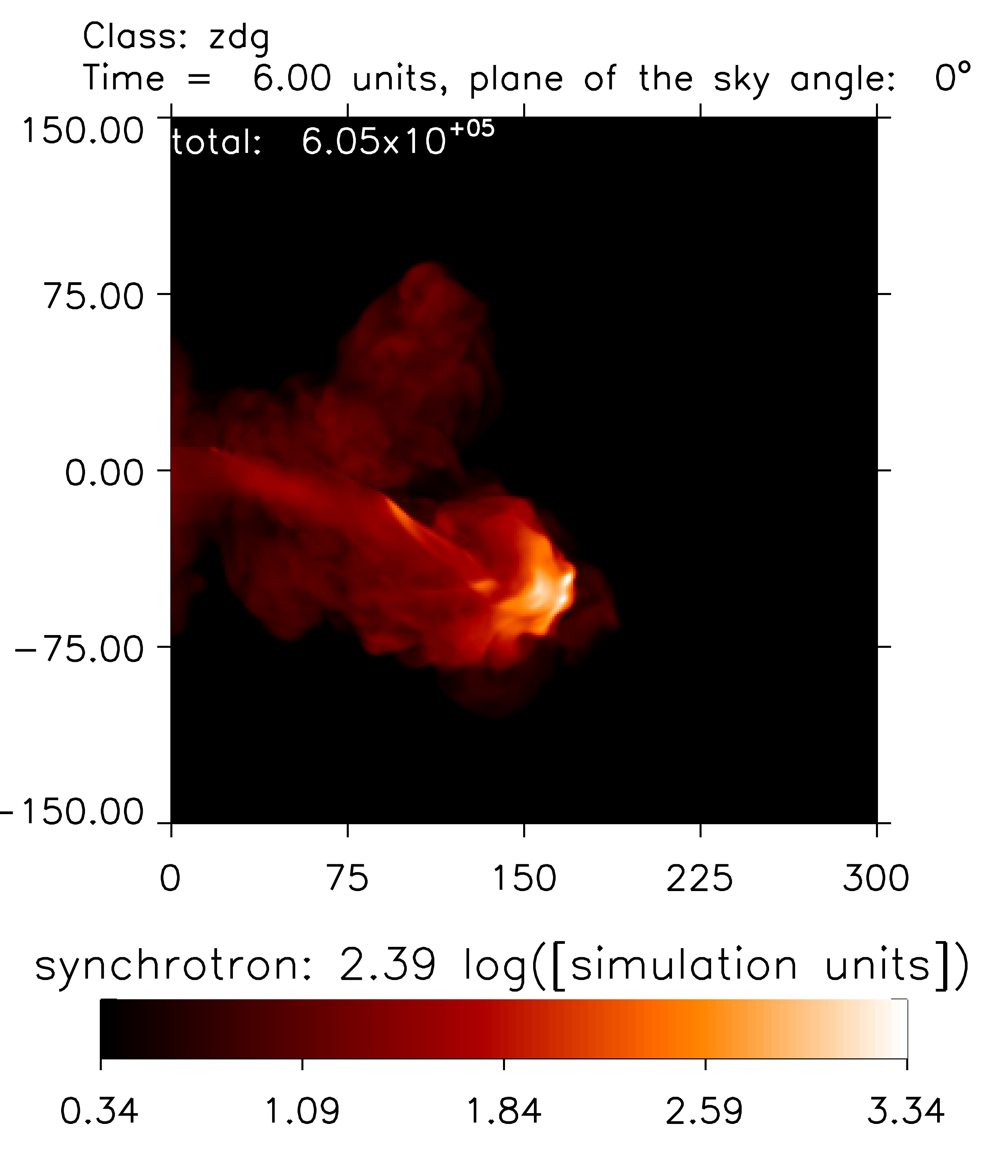} \\ 
      (d)  $\alpha$ = 0.7   & (e) $\alpha$ = 1.0  & (f) $\alpha$ = 1.3     
   \end{tabular}
\caption{ {\bf Wide angle precession.} Simulated synchrotron emission resulting from a radio galaxy driven by a jet with  a 20\degsy precession angle. Other parameters are identical to those of Fig.\,\ref{zda_power_law} with the radio spectral index $\alpha$ low (left panels) and steep (right panels), and the dynamic range 100 (top panels) and 1,000 (lower panels). }
\label{zdg_power_law20}
\end{figure*}

\section{Method} 
\label{Method}

\subsection{Parameters for the ambient medium}

We take a uniform  ambient medium of temperature $T = 3 \times 10^7$~K, corresponding to 2.575 keV, with a hydrogen nuclei density of $n = 4 \times 10^{-4}$~cm$^{-3}$ (Giant Radio Galaxy) or $6 \times 10^{-3}$~cm$^{-3}$ (Dense Cluster Environment). Whereas the Giant source covers 750~kpc, for the Dense Cluster Environment we take a 5~kpc radius jet propagating across a grid of size 150~kpc. 
These parameters provide an ambient thermal pressures of $4 - 60 \times 10^{-12}$~dyne~cm$^{-2}$ , covering the range expected including that corresponding to dense Abell clusters \citep{1997ApJ...474..580G}. The simulations were performed on grids of 150$^3$ and 150$^3$; further details can be found here in \ref{simulation_name}  and Paper\,1.

\subsection{Parameters for the jet}

 We are limited here by the input physics: non-relativistic hydrodynamics.  We do this in order to explore ranges in the dynamics in full three-dimensions with a reasonable resolution. Relativistic motions would also require the synthetic images to be calculated during the simulation to avoid an excess of data storage.
 
 An analysis of the energy balance on the kiloparsec scale reveals that sub-relativistic jet speeds of the order of 10,000 km\,s$^{-1}$ may be consistent with FR\,I jets \citep{1999MNRAS.306..513L}. But mildly relativistic  speeds are appropriate for FR\,II jets on the same kiloparsec scale \citep{2011ApJ...740...98M} with the possible existence of a two-velocity structure with a fast spine and a lower speed sheath  \citep{2001ApJ...552..508G}. Estimates derived from the jet-counterjet asymmetry indicate a range 0.2c Ð 0.7c   \citep{1999MNRAS.306..513L,1997MNRAS.286..425W,2006MNRAS.368..609J,2018ApJ...855...71S}. However, there is still no direct evidence on the kiloparsec scale. 
   
Here, we opt to fix the jet Mach number with the justification that this non-dimensionless parameter, along with density and pressure ratios,  is responsible for the major differences between adiabatic jet flows \citep{1983ASSL..103..227N}. The  injection speed is then determined by the density ratio while the jet injection pressure   is equated to the ambient value. This means that the jet speed rises as the jet-ambient density ratio falls, with speeds in units of the ambient sound speed.  However, the hotspot advance speed should be roughly constant since the momentum flow rate is independent of the density. In practice, this is not quite true with certain configurations more streamlined than others and able to penetrate faster (see Paper 1 for the details). Table~\ref{simulation_name}  lists the simulation parameters and conditions that we have investigated in detail.
  
\subsection{Synchrotron emission}
\label{synchrotron_emission}

To generate the images shown in all the figures below, we employ the usual convention for the power-law approximation across the radio.
 We  define the spectral index $\alpha$ through $F_{\nu} \propto  \nu^{-\alpha}$.

The synchrotron emissivity is taken to be a function of the thermal pressure, $p$,  as assumed by \citet{1985MNRAS.214...67S} and  \citet{2013MNRAS.430..174H}, employing constant energy partition conditions to justify taking an emissivity proportional to $p^{1.8}$. 
\citet{2016MNRAS.458..802N} employ  $p^{(3+\alpha)/2}$ where $\alpha$ is the synchrotron power-law spectral index as suggested by \citet{2007ApJS..173...37S}. These provide convenient approximations \citep{1990MNRAS.242..623M}. 

 We will assume optically thin  emission in the radio from the synchrotron process. It is generated by relativistic electrons accelerated in internal and termination shocks and passively advected without significant radiative energy loss. In the present study, we also assume that the population of relativistic electrons possesses a power-law energy distribution. The number density, $N$, is given by
\begin{equation}\label{eq_sync1}
   \frac{dN}{dE} = K \cdot E^{-\eta}  = \frac{N}{(\eta-1) \cdot E_{min}} \cdot \left[\frac{E}{E_{min}}\right]^{-\eta},
\end{equation}
between energies  $E_{min}$ and  $E_{max}$ with  $E_{min} << E_{max}$ and the index $\eta$ exceeding unity.

We now make the simplification that each electron, instead of  generating radiation over a wide frequency range, radiates at a frequency around $\nu_c = f E^2 B$ where $f$ is a constant. In addition, each electron  radiates a power of $P(E,B) = a E^2 B^2$ , where $a$ is a constant. 

We can calculate the emissivity $F$ and the flux density  $F_\nu$ at any specified frequency through
\begin{equation}\label{eq_sync2}
     F = \frac{a \cdot f^{(\eta-1)/2 }}{2 \cdot (\eta-1)} \cdot N  \cdot B^{\frac{1}{2}\cdot (\eta+1)} \int E_{min}^{\eta-1} \nu^{\frac{1}{2} \cdot (1-\eta)} d\nu ,
\end{equation}
where the critical integration limits are given by $B$ and chosen $E_{min}$ and $E_{max}$. Note that the (traditionally negative) spectral index is given by $\alpha = \frac{1}{2}(\eta-1)$, implying that most flux is produced at the highest energies only if $\alpha$ is less than unity. For the vast range of radio galaxies, $\alpha$ is spread throughout the range  0.6 to  1.5, which corresponds to $2.2 < \eta < 4.0$  \citep{2008A&ARv..15...67M,2010MNRAS.406..197B}.   Steeper spectrum sources appear to be related to denser cluster environments, and are highly relevant to find and model  distant powerful radio galaxies\citep{2008A&ARv..15...67M}.
    
Various assumptions and approximations can now be applied. In hydrodynamic simulations, we make a best guess as to how the presumed advected trace  magnetic field should depend on the density and pressure. Flux freezing with volumetric changes in  three dimensions yields $B \propto \rho^{2/3}$. On assuming $E_{min}$ and $E_{max}$ are constants over the entire radio lobe (often made to estimate a radio luminosity from an observed flux density), then $F_\nu \propto N  B^{\frac{1}{2} \cdot (\eta+1)}$. This yields $F_\nu \propto \rho^{\frac{1}{3} \cdot (\eta+4)}$.  
  
However, particle acceleration and magnetic field amplification at shock fronts, in addition to contributions from turbulent diffusion and across shear layers, are significant. This  is partly taken into account in simulations by converting the dependence from density to the thermal pressure \citep{1985MNRAS.214...67S, 2014MNRAS.443.1482H}. Assuming a dominant thermal pressure (as the case in these simulations),  yields $F_\nu \propto p^{\frac{1}{5}(\eta+4)}$.  
  
An alternative is  to make the adiabatic assumption for the relativistic electrons  such that $E_{min} \propto p^{1/3}$ and then to take the magnetic pressure proportional to the relativistic pressure, as would be consistent with equipartition arguments. This yields $F_\nu \propto N \cdot E_{min}^{\eta-1} \cdot B^{\frac{1}{2} \cdot (\eta+1)} \propto p^{\frac{3}{5}}  p^{\frac{1}{3} \cdot (\eta-1)} p^{\frac{1}{5} \cdot (\eta+1)}  $. Therefore   $F_\nu \propto  p^{(\frac{7}{15} + \frac{8}{15} \cdot \eta)}$. 

The shape of the relativistic electron distribution function may well depend on the redshift, size and luminosity of the radio galaxy for reasons related to those discussed above. On large scales, a truly flat spectrum with $\eta <$ 2 is not found. Hence, we test below three values for $\eta$: 2.4, 3.0, and 3.6. These values correspond to a gentle slope, a moderate slope and a steep slope. The relative values are shown in Fig.\,\ref{3_power_law_relasionship} and the important effect on the appearance of the diffuse lobes is demonstrated in Fig.\,\ref{zda_power_law}.

\begin{figure*}
 \begin{tabular}{ccc}  
 \includegraphics[width=0.33\textwidth]{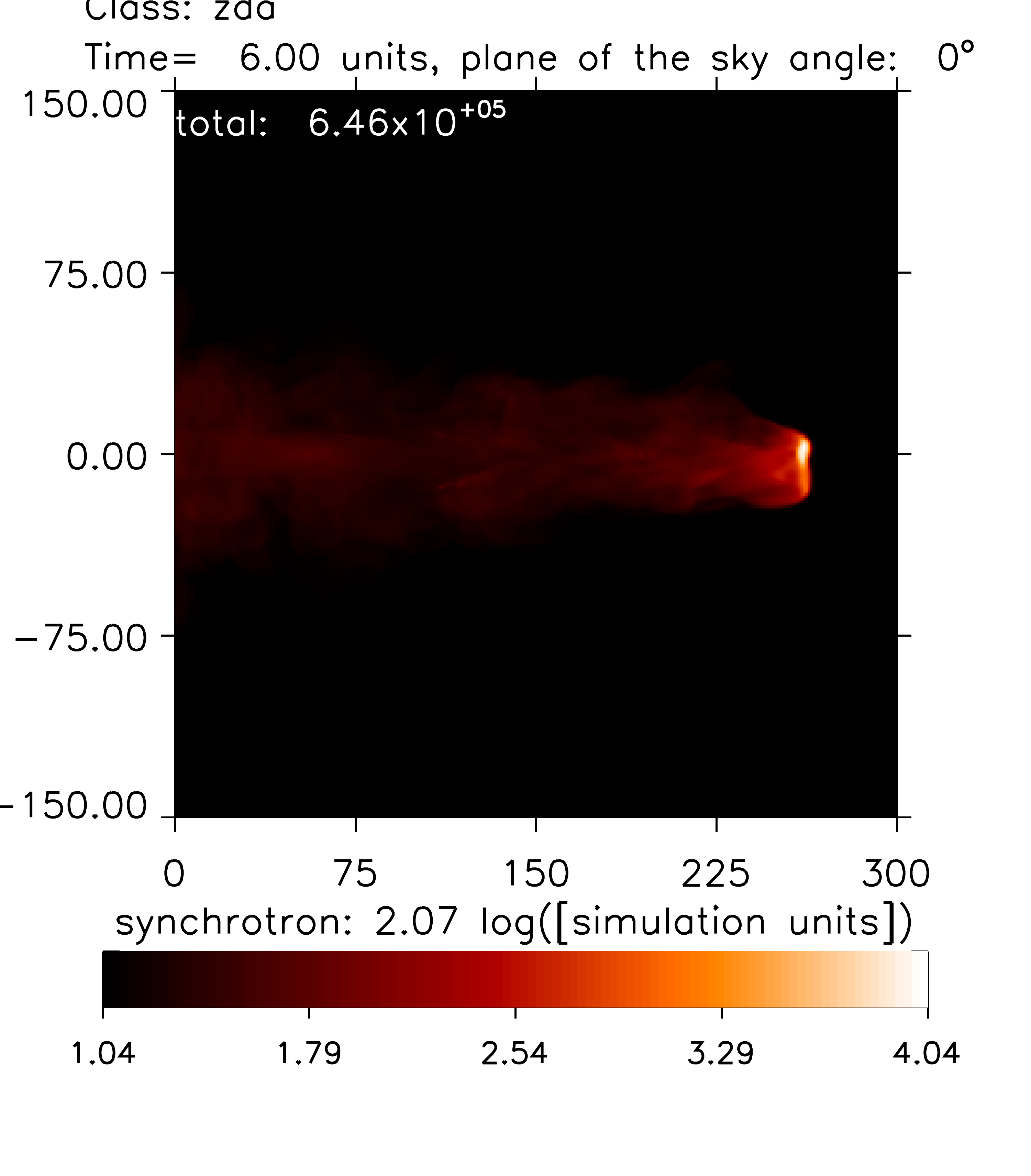}  & 
           \includegraphics[width=0.33\textwidth]{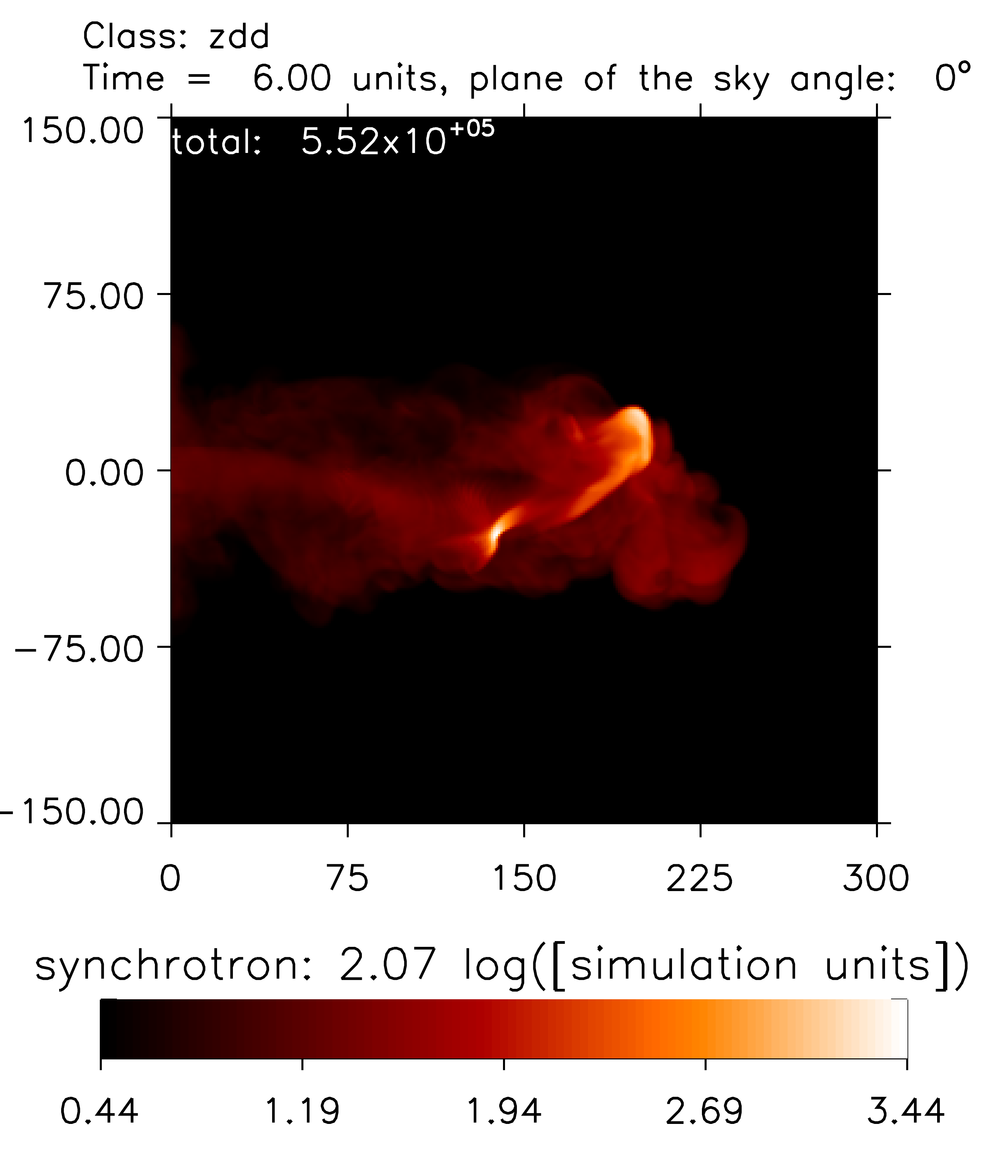} & 
      \includegraphics[width=0.33\textwidth]{1000mag-zdg-synch-207-observerAngle-000-static-060}\\ 
      (a) 1\degsy & (b) 10\degsy & (c)20\degsy\\
   \end{tabular}
   \caption{ {\bf The precession angle:} integrated line of sight synchrotron emission. Each simulation has the same condition apart from  the precession angle, $\theta$, employed; 1\degsy, 10\degsy and 20\degsy from left to right. The row locks the brightness range as 1,000. The radio spectral index is set to 1.0.}
   \label{zda_zdd_zdg-subt_amb-radio-00_axis_1}
\end{figure*}

\begin{figure*}
   \begin{tabular}{cccc}
    \hspace*{-0.6cm}   \includegraphics[width=0.29\textwidth]{100mag-zda-synch-207-observerAngle-000-static-060} & \hspace*{-1.2cm}
      \includegraphics[width=0.29\textwidth]{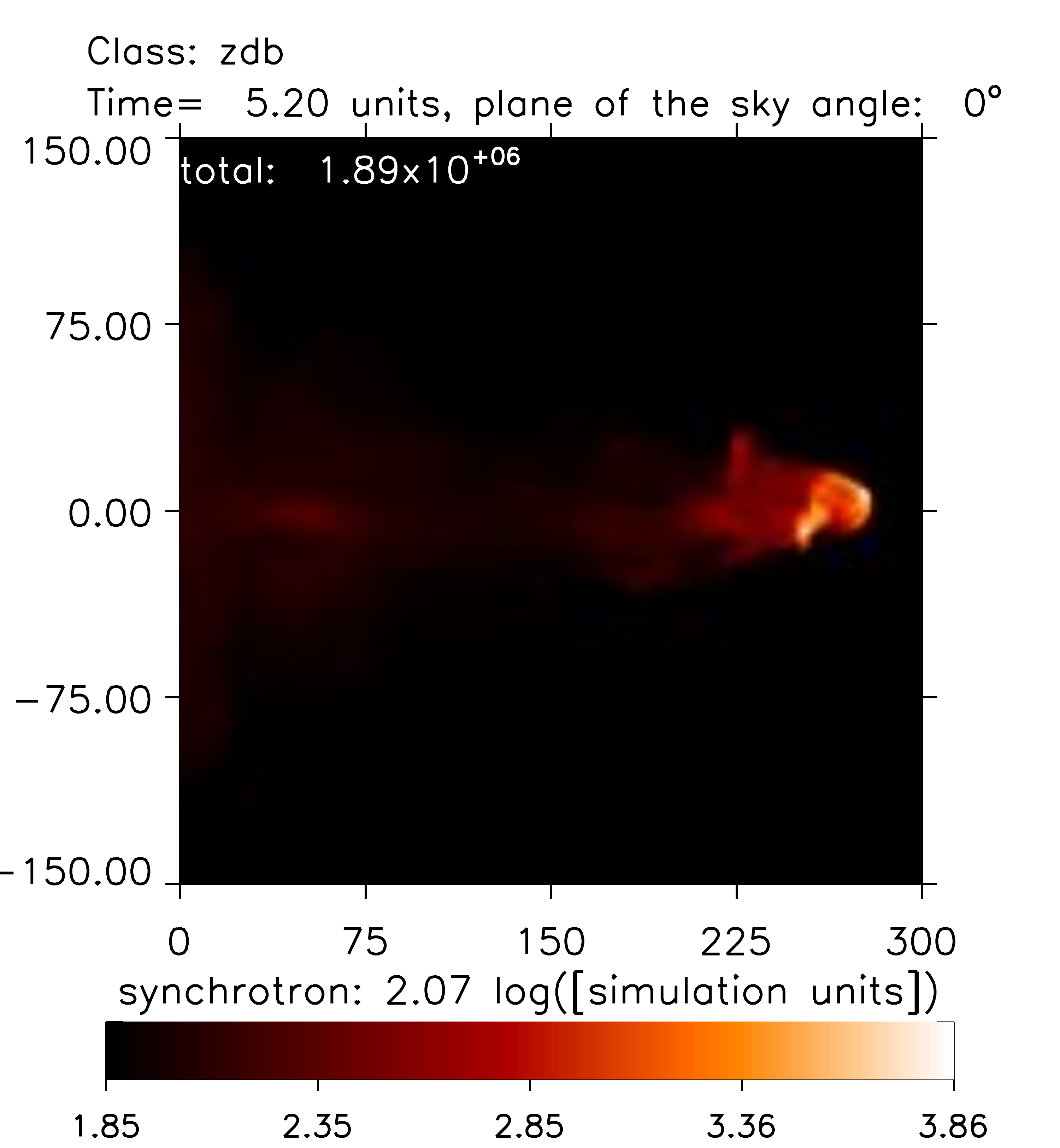} & \hspace*{-1.2cm}  
\includegraphics[width=0.29\textwidth]{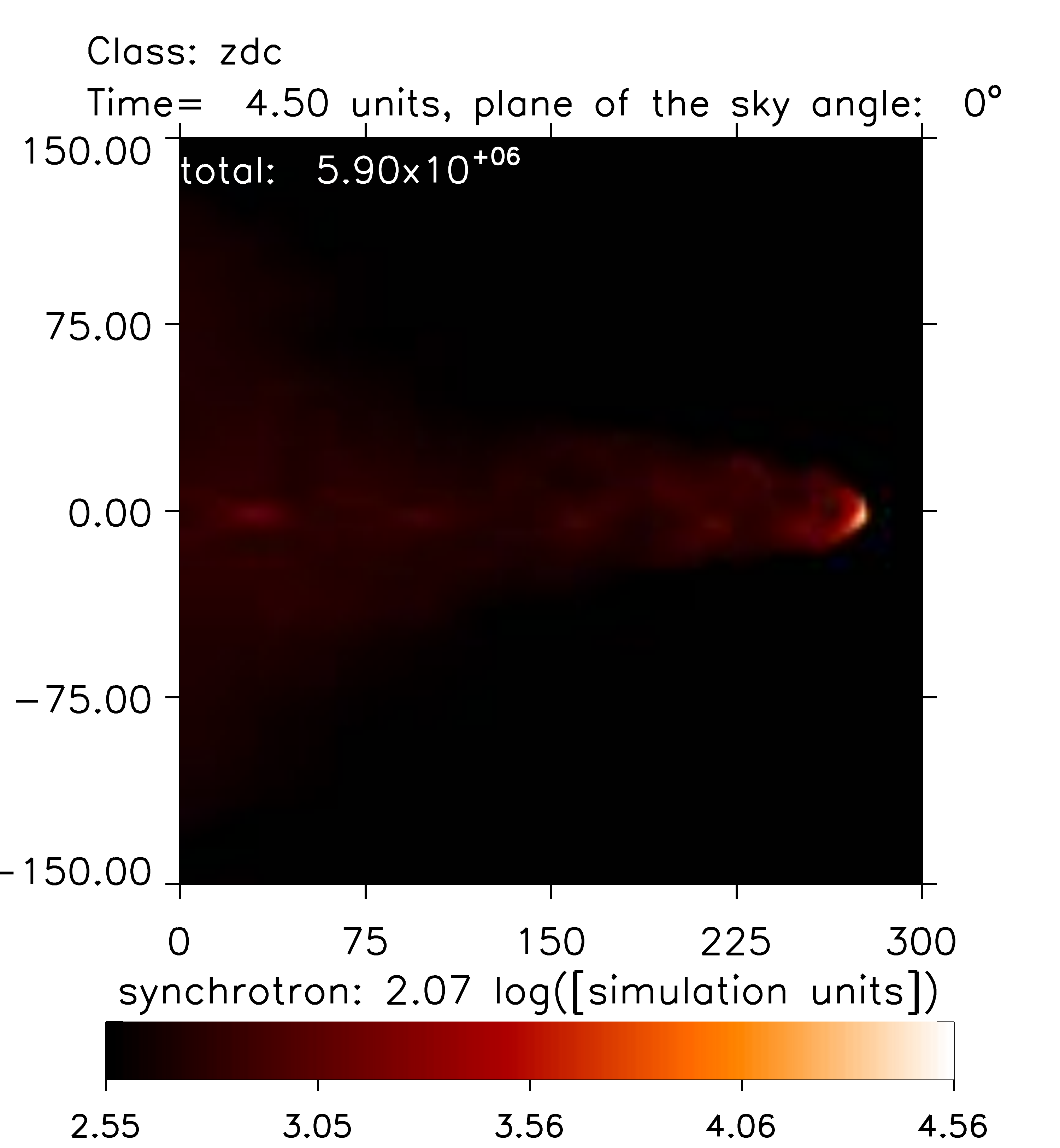} & \hspace*{-1.2cm}
\includegraphics[width=0.29\textwidth]{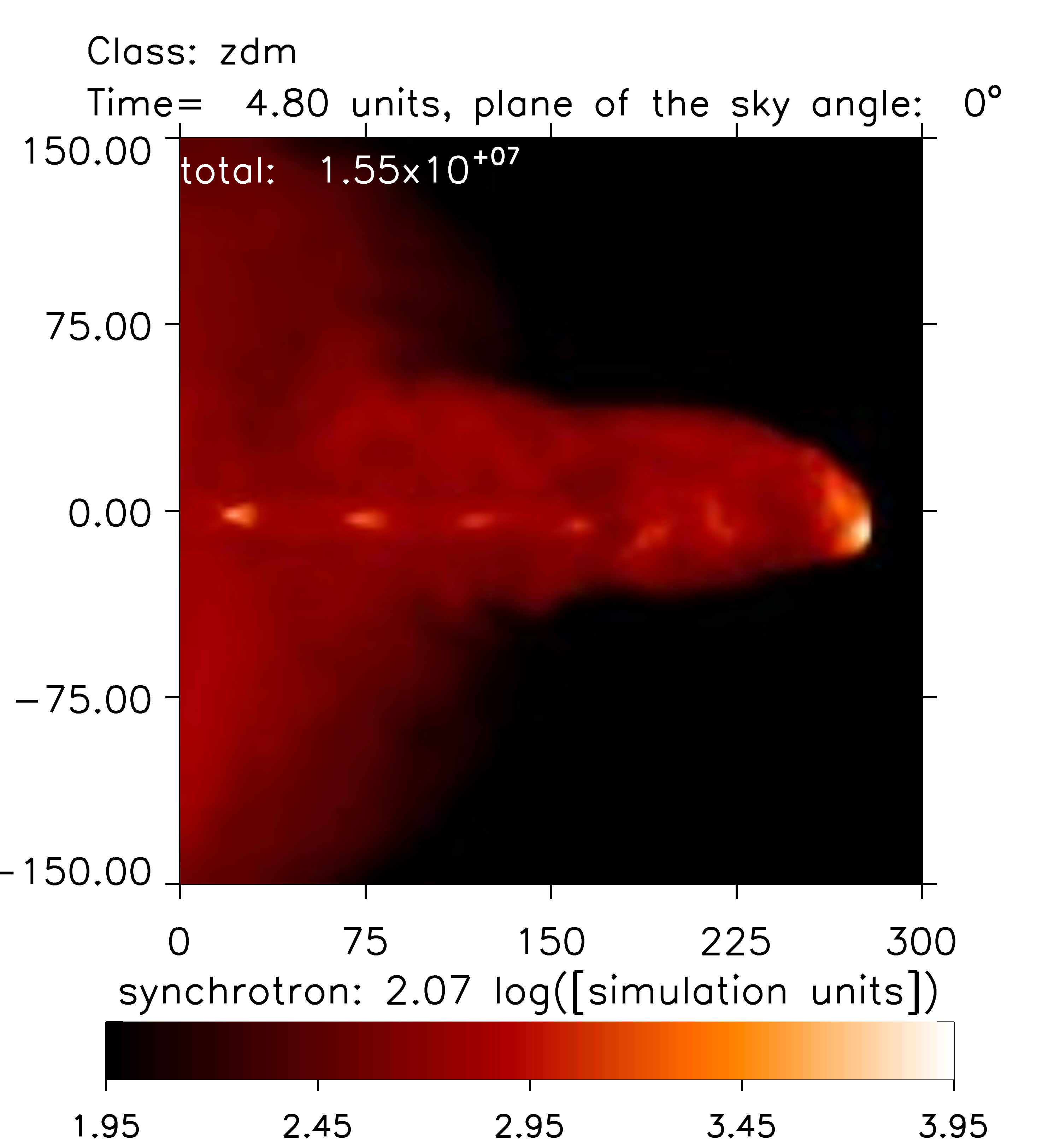} \\ 
(a) $\rho_j/\rho_a=0.1$ & (b) $\rho_j/\rho_a=0.01$ & (c) $\rho_j/\rho_a=0.001$ & (d) $\rho_j/\rho_a=0.0001$ \\ [6pt]
  \hspace*{-0.6cm}   \includegraphics[width=0.29\textwidth]{1000mag-zda-synch-207-observerAngle-000-static-060old} &  \hspace*{-1.2cm} 
 \includegraphics[width=0.29\textwidth]{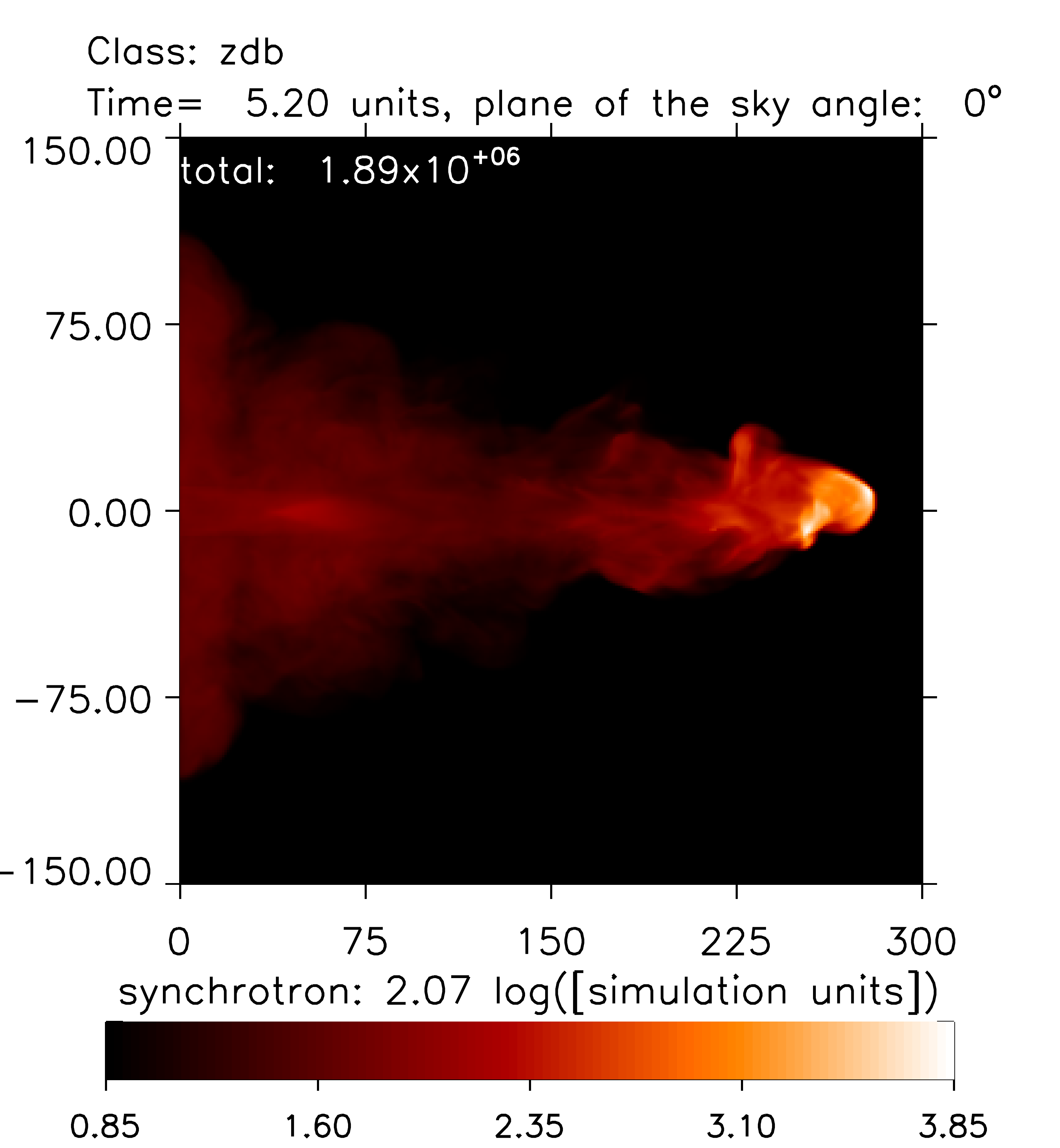} &   \hspace*{-1.2cm} 
\includegraphics[width=0.29\textwidth]{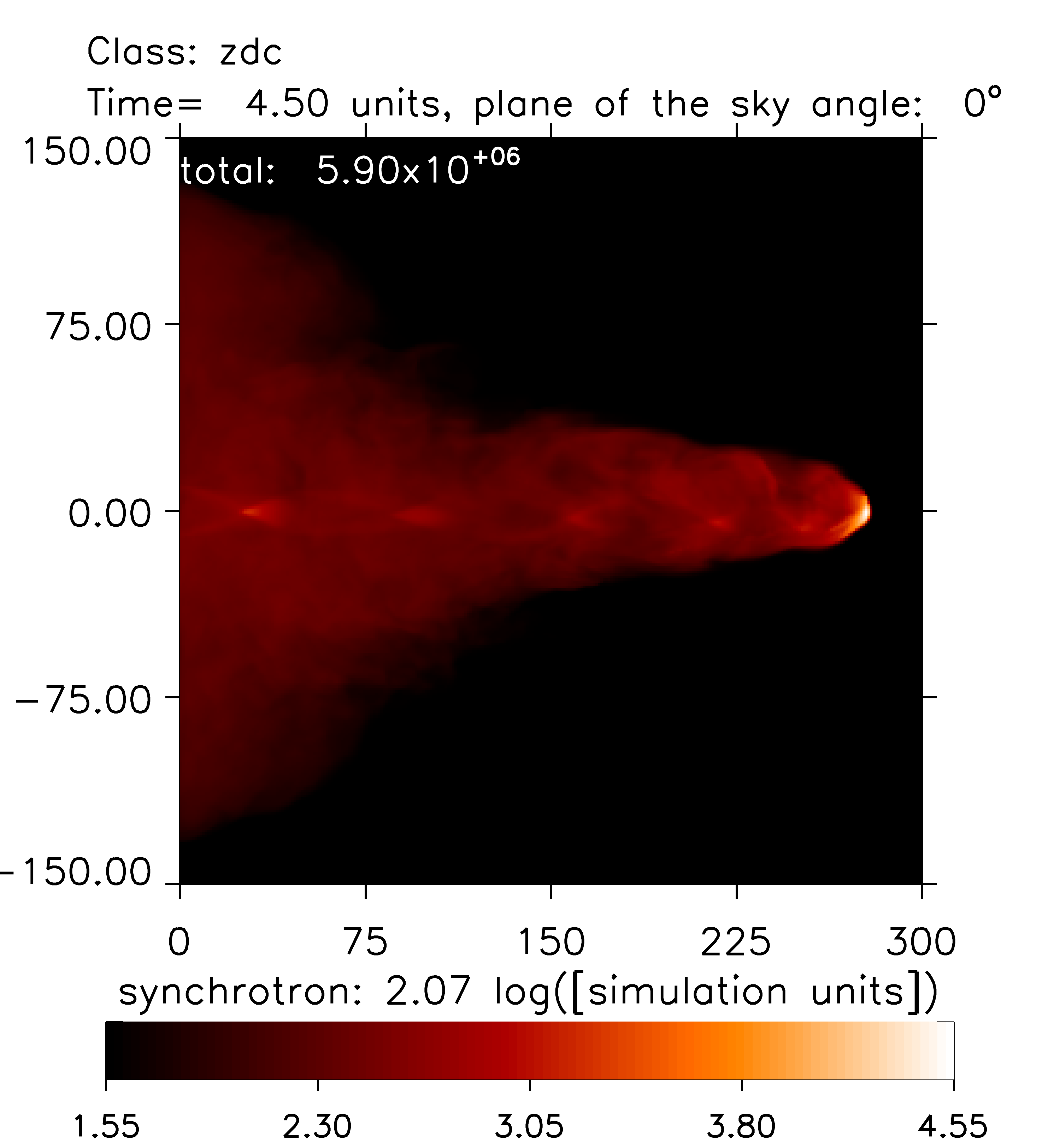} &  \hspace*{-1.2cm}  
\includegraphics[width=0.29\textwidth]{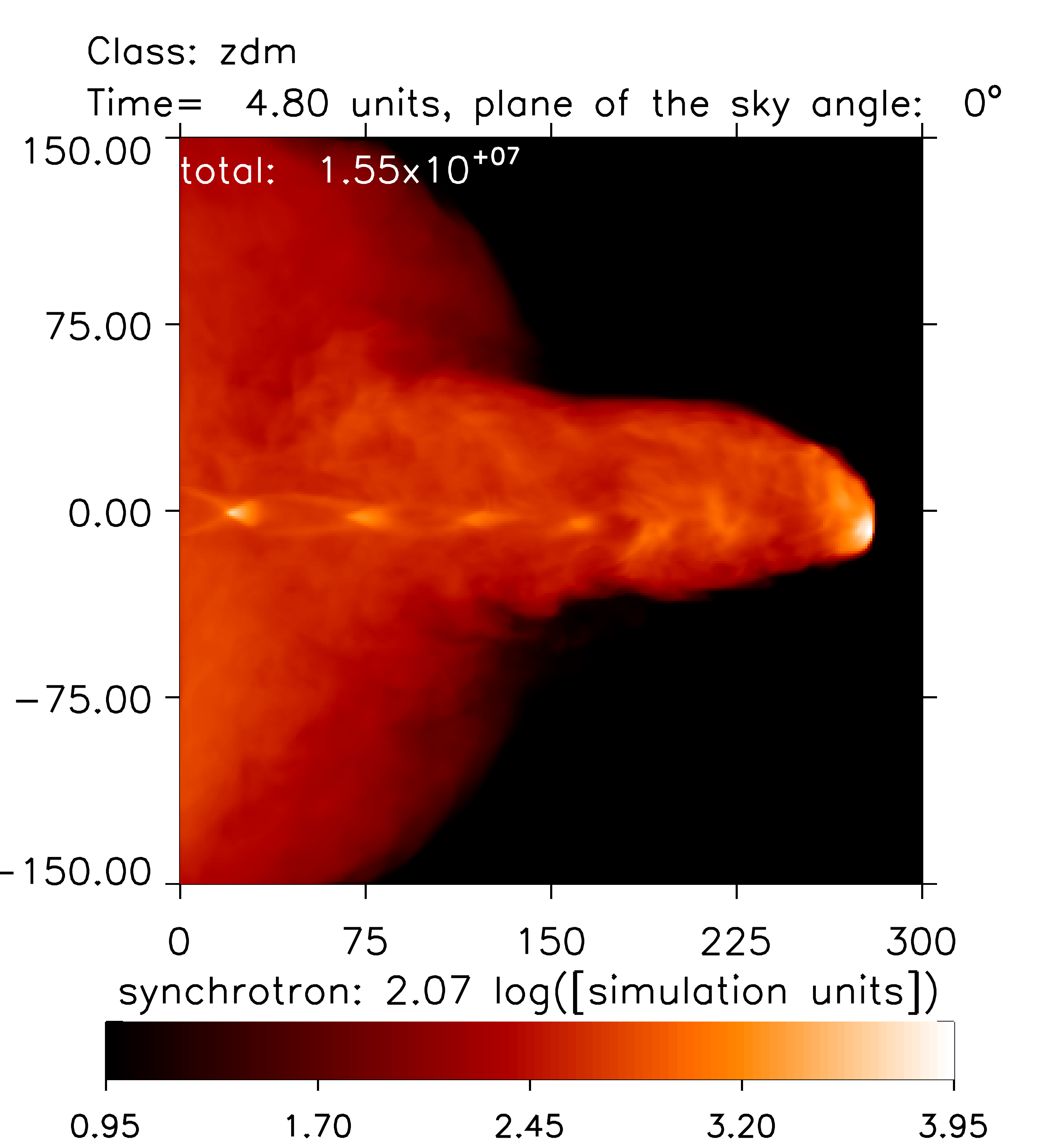} \\ 
(e) $\rho_j/\rho_a=0.1$ & (f) $\rho_j/\rho_a=0.01$ & (g) $\rho_j/\rho_a=0.001$ & (h) $\rho_j/\rho_a=0.0001$ \\ [6pt]
   \end{tabular}
\caption{ {\bf The density ratio.} The synthetic synchrotron emission maps as a function of the jet-ambient density ratio, $\rho_j/\rho_a$. Each simulation has the same conditions except the density ratio: 0.1, 0.01, 0.001 and 0.0001, from left to right. The first row locks the flux range to  within 100 and the second row locks it to 1,000  to highlight lobe emission.}
\label{fig-sync-tile-zda_zdb_zdc_zdm-subt_amb-oa_00}
\end{figure*}

\subsection{Precession}

The changing of the  jet direction is implemented  through a smooth precession. The precession is calculated in  the same way as covered in \citet{2016MNRAS.458..558D} by adding in a periodic function of the global time into the velocity components of the jet input. As seen from \citet{2016MNRAS.458..558D}, the precession spreads the jet energy perpendicular to the propagation axis. There is also a significant increase in the number of secondary shocks generated as the precessing jet catches up to the previous shock impact region. With these points outlined, we suspect that the radio emission is not only going to be focused at the head of the jet but will also highlight regions along the jet. 

\subsection{Classification: the Limb Brightening Index (LB Index)}

To round off this work, we introduce  a classification scheme based on the Fanaroff-Riley types in order to highlight how the starting parameters (density, Mach number, precession angle, precession rate) influence the observable properties as the radio galaxy evolves.  
To classify a radio galaxy, we take the following steps.
We assume that the radio galaxy is twin-lobed and is powered by symmetrical opposing jet and  counter-jet. The classification is determined after applying a Gaussian blur to the flattened image, which has a lower end emission cut at 5\% of the maximum emission for the specific time step (after the smoothing). The distance to the maximum emission and the maximum expansion is then obtained from the image giving us the Lobe/Limb Brightening Index. This may also be appropriately termed a Linear Brightness Index.

 The 'LB Index' system is applied here with the range lying between 0 and 1. It is the ratio of source length to the hot-spot to the entire length of the lobe.
   To summarise, any value that is less than 0.5 corresponds to a hotspot closer to the core, whereas values exceeding 0.5 will have the hotspot closer to the edge of the radio source. Note that in the graphs, none of the initial shock time steps have been omitted. The plots also generally reflect the jet evolution: the majority travelling in a predictable spline as it evolves.

\section{Results for radio emission} 

\subsection{Synchrotron radio maps}

 Here  we analyse the influence of density, precession and the Mach number on the radio surface brightness. The simulations covered in this paper are listed in Table \ref{simulation_name} in the Appendix.
Using the emission power-law indices as outlined in Sub-section \ref{synchrotron_emission}, the intensity maps for each of the runs have been created. These are integrated line of sight emission with the jet axis at various angles to the plane of the sky. Each simulation has been analysed at multiple viewing angles, from the first at 0\degsy, to 10\degsy, 30\degsy, 50\degsy, 70\degsy\,  and 90\degsy, in which the jet, respectively, is propagating closer to the observer's line of sight till the jet is propagating directly towards the observer. 
In addition to this, a Gaussian blur can be added to the emission map, emulating different resolutions or redshifts. This enables us to systematically investigate the effect of distance on the detectable structure.The radio galaxy produced by the straight jet
(1\degsy precession) in Fig.\,\ref{zda_power_law} is 
dominated by the hotspot with a weak and narrow cocoon of emission at the 1\% level. The six images demonstrate that a combination of a  shallow/flat spectrum  and a high dynamic range would be necessary to measure the trail of emission extending back to the nozzle.  

The opening angle as given by the perpendicular width of the hotspot is $\sim$ 3.0\degsy for the compact bright spot or  6.0\degsy for the full lateral extent, wider than the superimposed 
1\degsy precession. The cocoon contains both fine-scale structure and large-scale turbulence which leads to an asymmetric brightness distribution.

In contrast, the wide precessing jets displayed in Figs.\,\ref{zdg_power_law20}  and \ref{zda_zdd_zdg-subt_amb-radio-00_axis_1}  generate multiple hotspots. These  can appear
as two compact sub-spots, one moderately compact following spot  and one  weak diffuse region. The bright sub-spots are most obvious when a steep radio spectral index is employed (the right panels). On the other hand, as expected, the diffuse bridge is evident for the flatter spectrum with a high dynamic range (left and lower panels). 
 The jet also excites a broad channel with a significant hotpot of its own just behind the main impact hotspot. This is associated with feedback from the main impact which results in jet compression and partial disruption before the termination shock.

Relic lobes are produced in Fig.\,\ref{zdg_power_law20}. The lack of a hotspot  in the lobes towards the top of the displayed images suggests these could not be interpreted as X-shaped or double-double sources. 
 
 The wide precessing jet would clearly also be classified as FR\,II as illustrated across the panels of Fig.\,\ref{zda_zdd_zdg-subt_amb-radio-00_axis_1}. One difference is that the jet itself  is detectable being very distinct in  the high dynamic range maps. The higher pressure within the jets is caused by the  gradual bending of the supersonic flow which generates internal compression waves and weak oblique shocks.

\begin{figure*}
   \begin{tabular}{ccc}
        \includegraphics[width=0.34\textwidth]{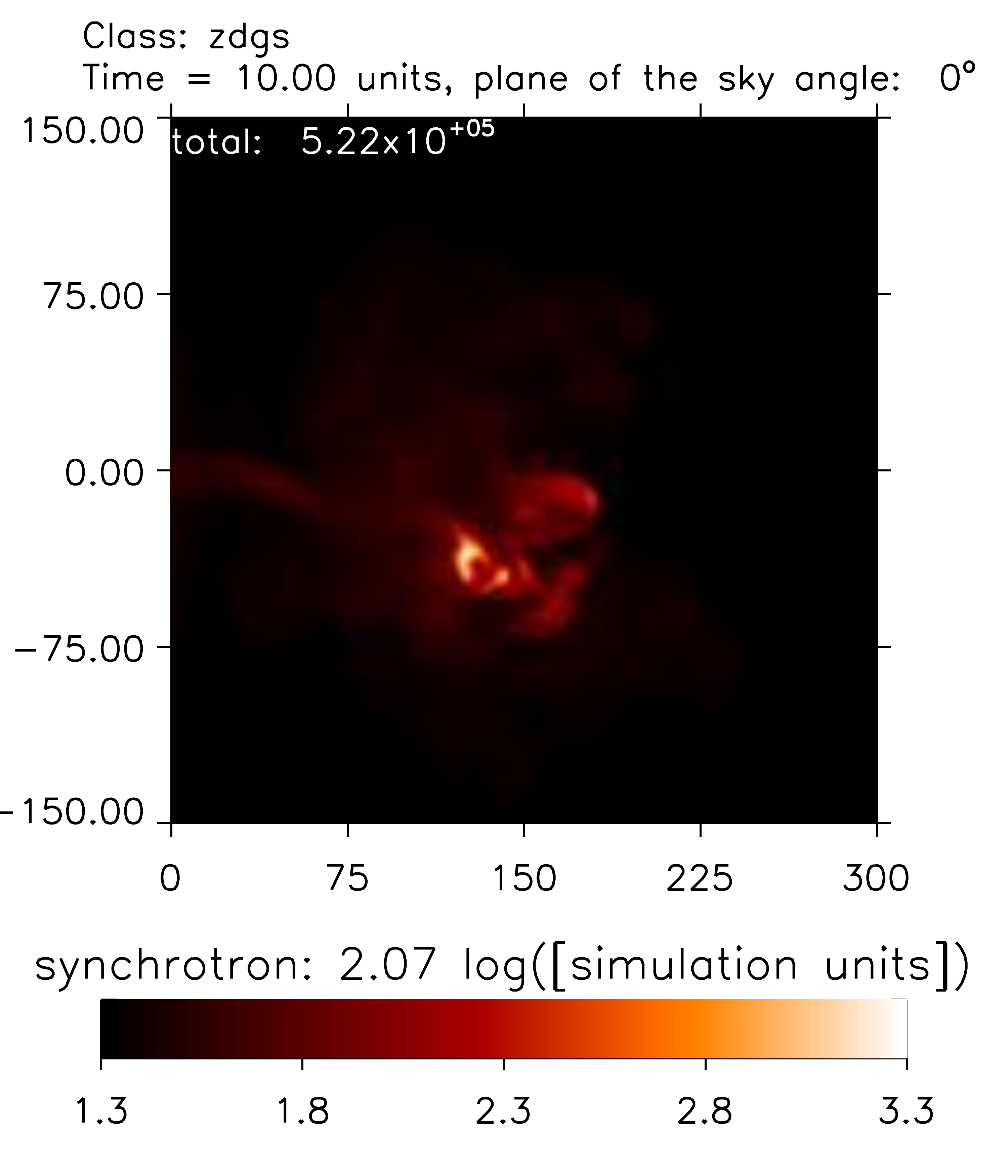} &  \hspace{-1.2cm} 
                \includegraphics[width=0.34\textwidth]{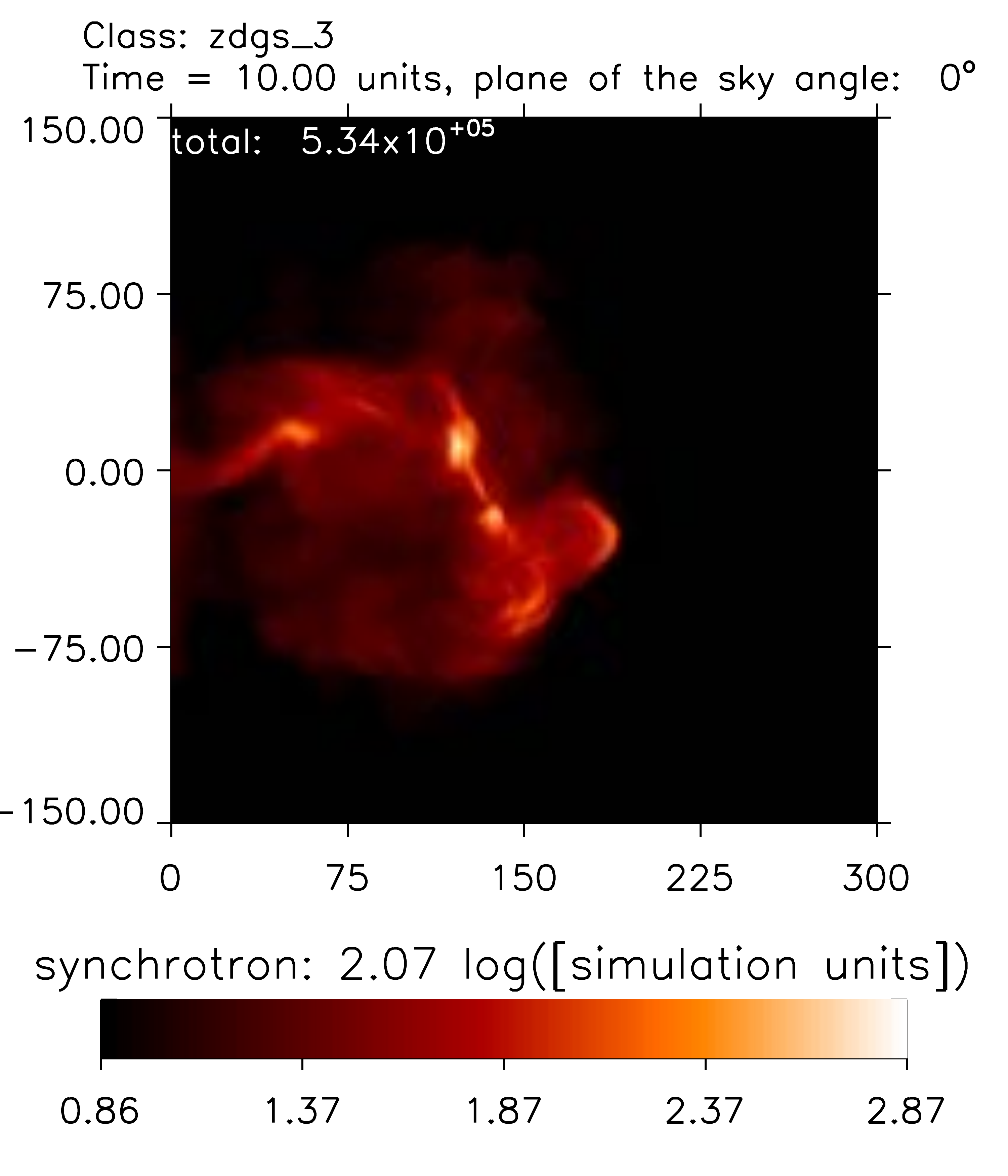} & \hspace{-1.2cm}   
         \includegraphics[width=0.34\textwidth]{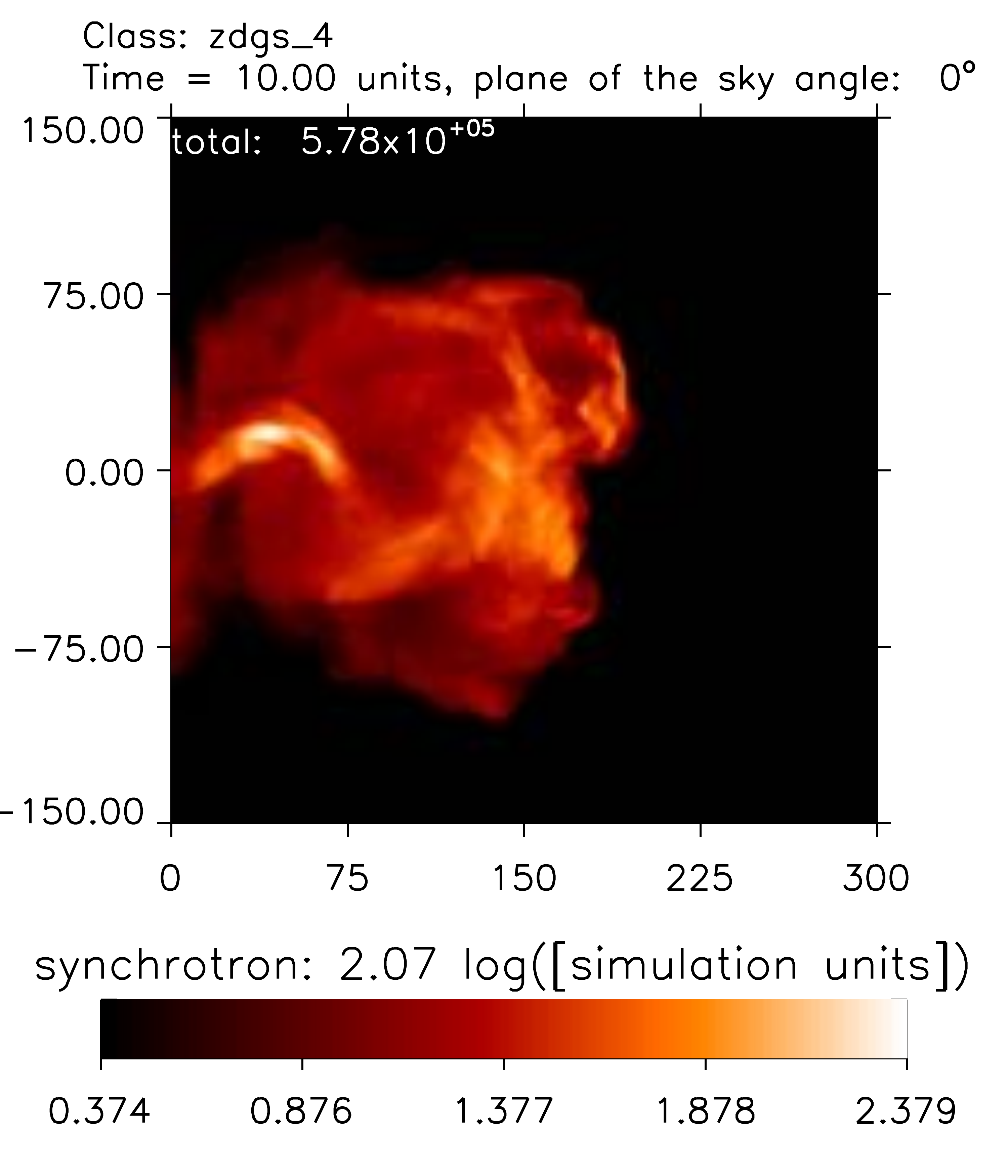}\\ 
 (a)  & (b)  & (c) \\ [6pt]
      \includegraphics[width=0.34\textwidth]{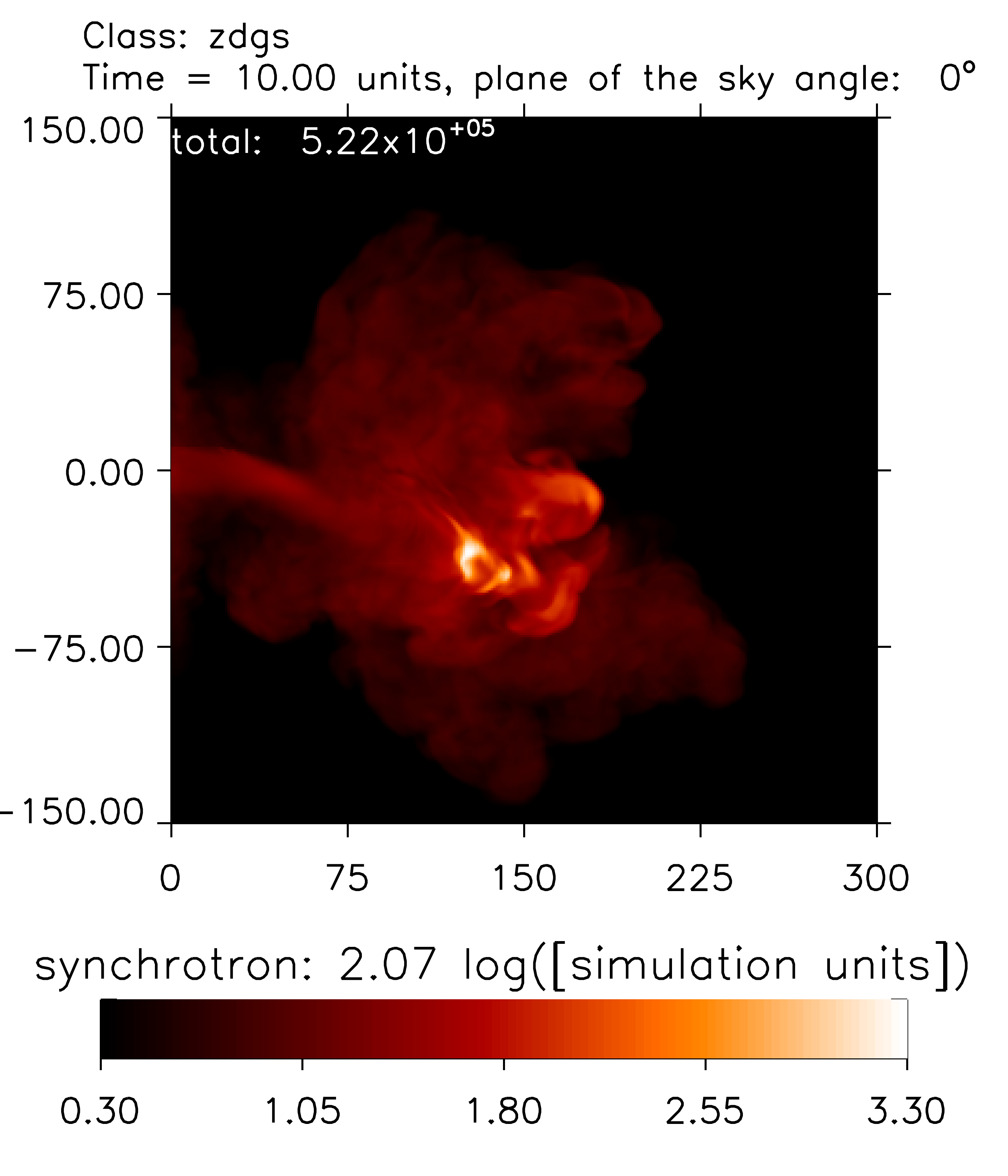} &   \hspace{-1.2cm} 
            \includegraphics[width=0.34\textwidth]{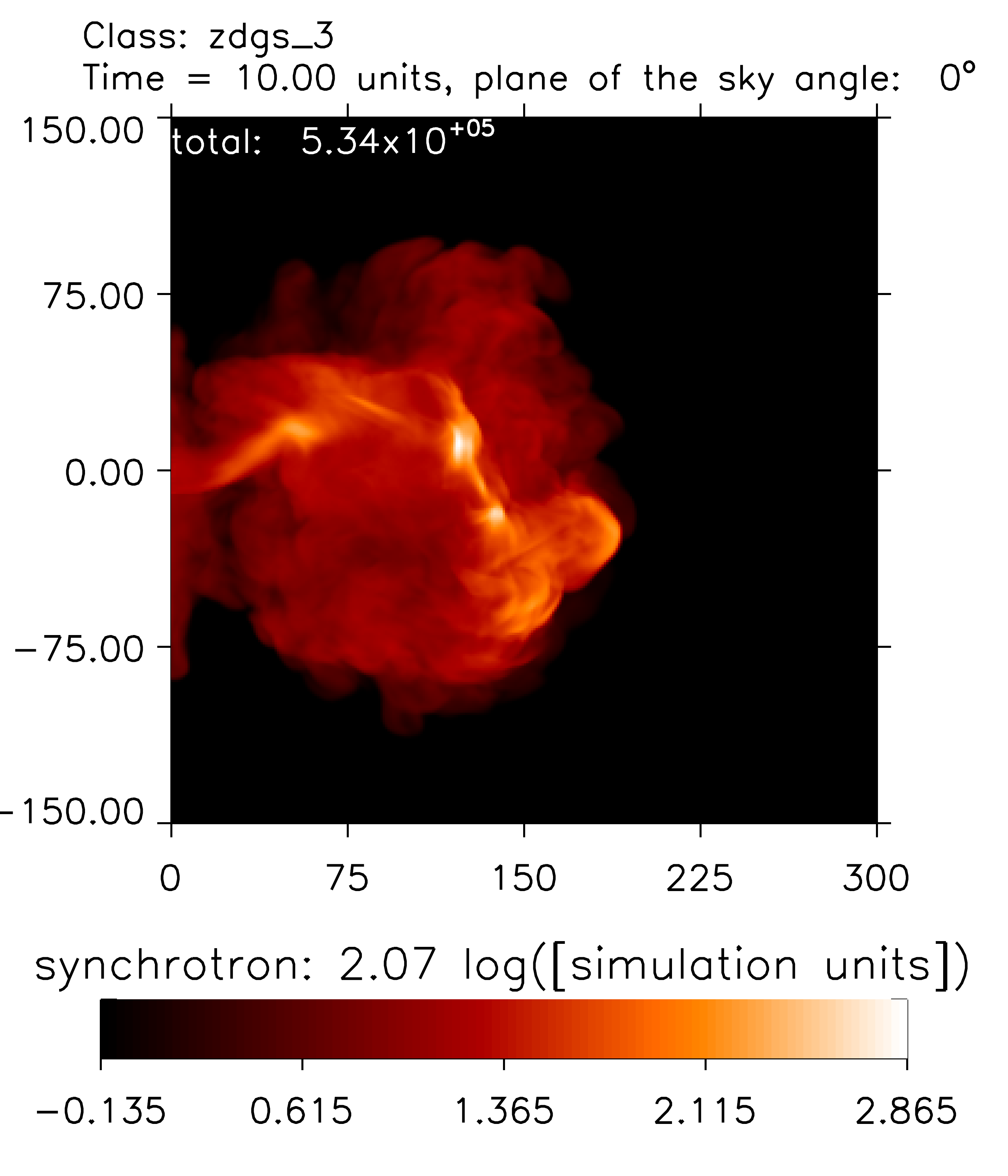} &   \hspace{-1.2cm}      
             \includegraphics[width=0.34\textwidth]{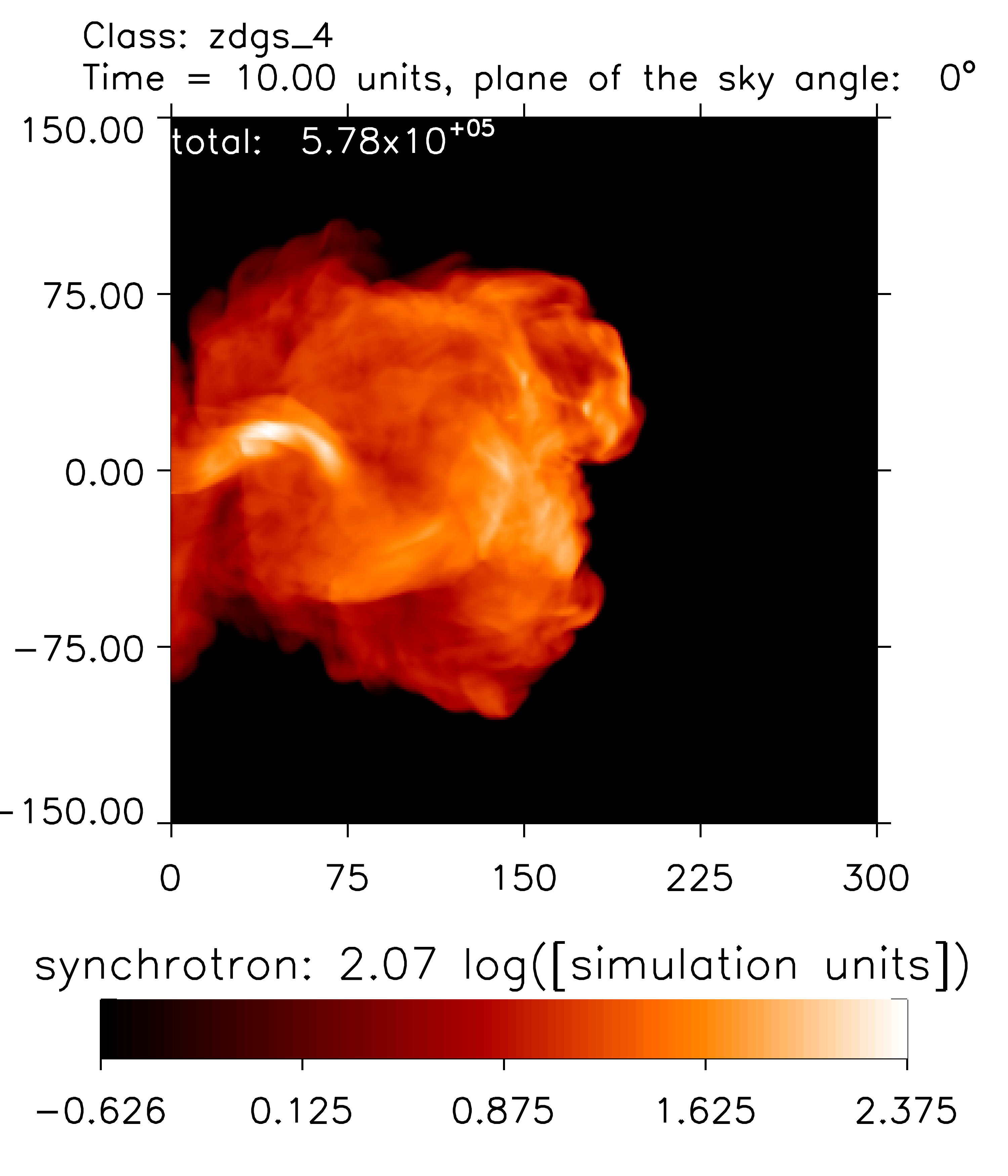} \\ (d) & (e) & (f) 
   \end{tabular}
\caption{ {\bf Rate of precession.} Three simulations that show how the rate of precession, $\omega$, is affecting the structure of the jet and cocoon system. All three have a precession angle of 20\degsy, with (b) having a precession rate 2x and (c) having a precession rate 4x  that of the slow precession model in (a).}
\label{zdg_precession_rate}
\end{figure*}

\begin{figure*}
   \begin{tabular}{ccc}
\includegraphics[width=0.335\textwidth]{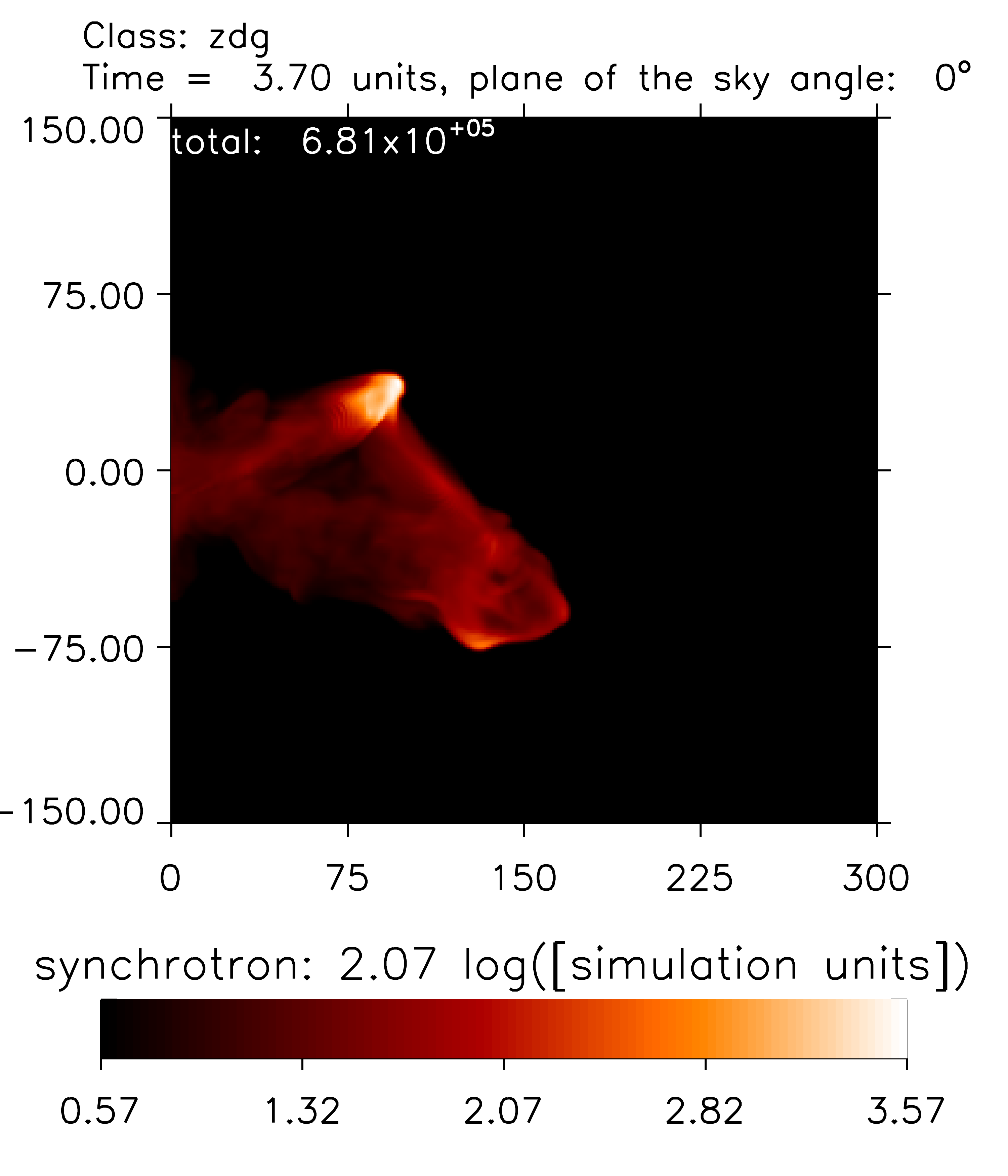} &   \hspace{-1.2cm}  
\includegraphics[width=0.335\textwidth]{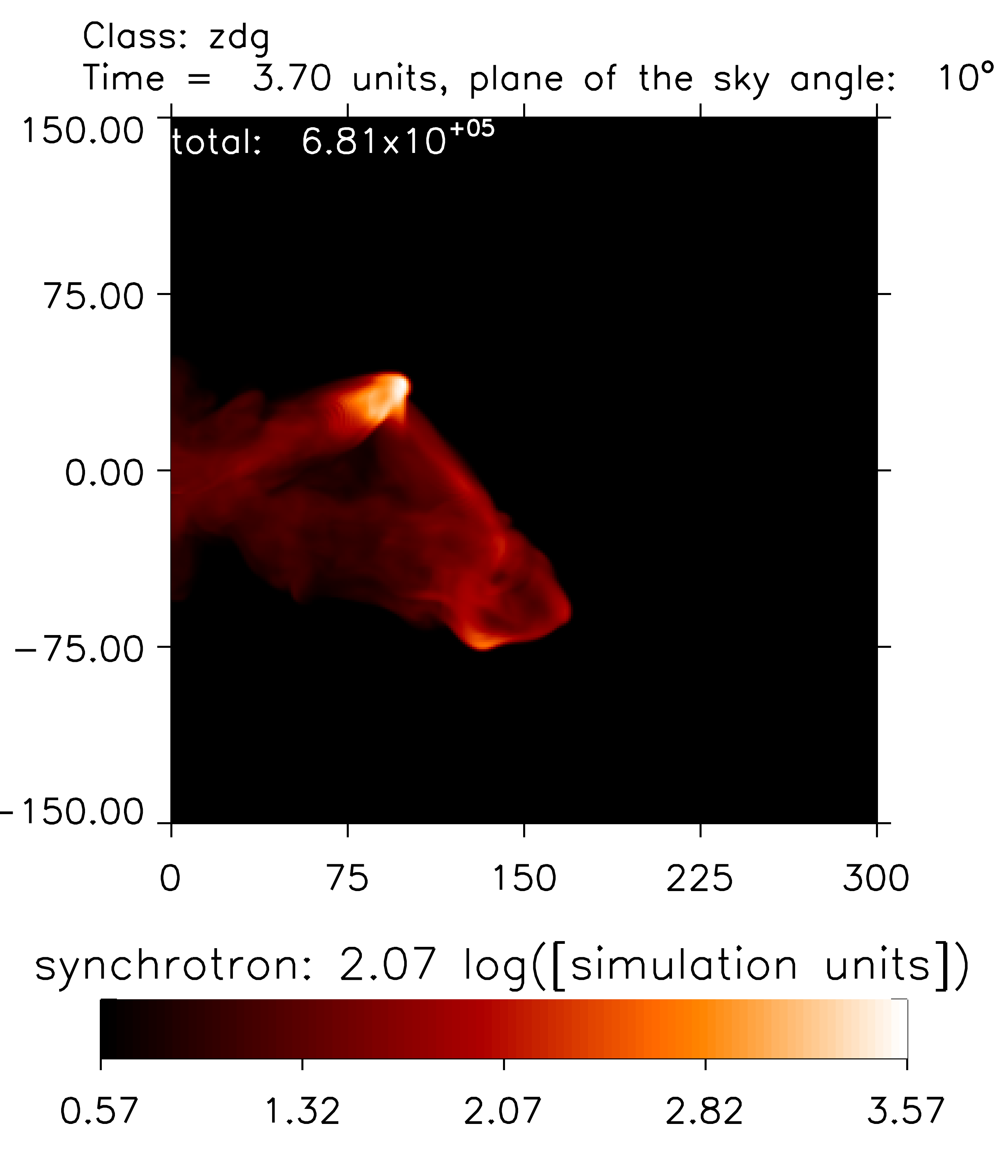} &  \hspace{-1.2cm}   
\includegraphics[width=0.335\textwidth]{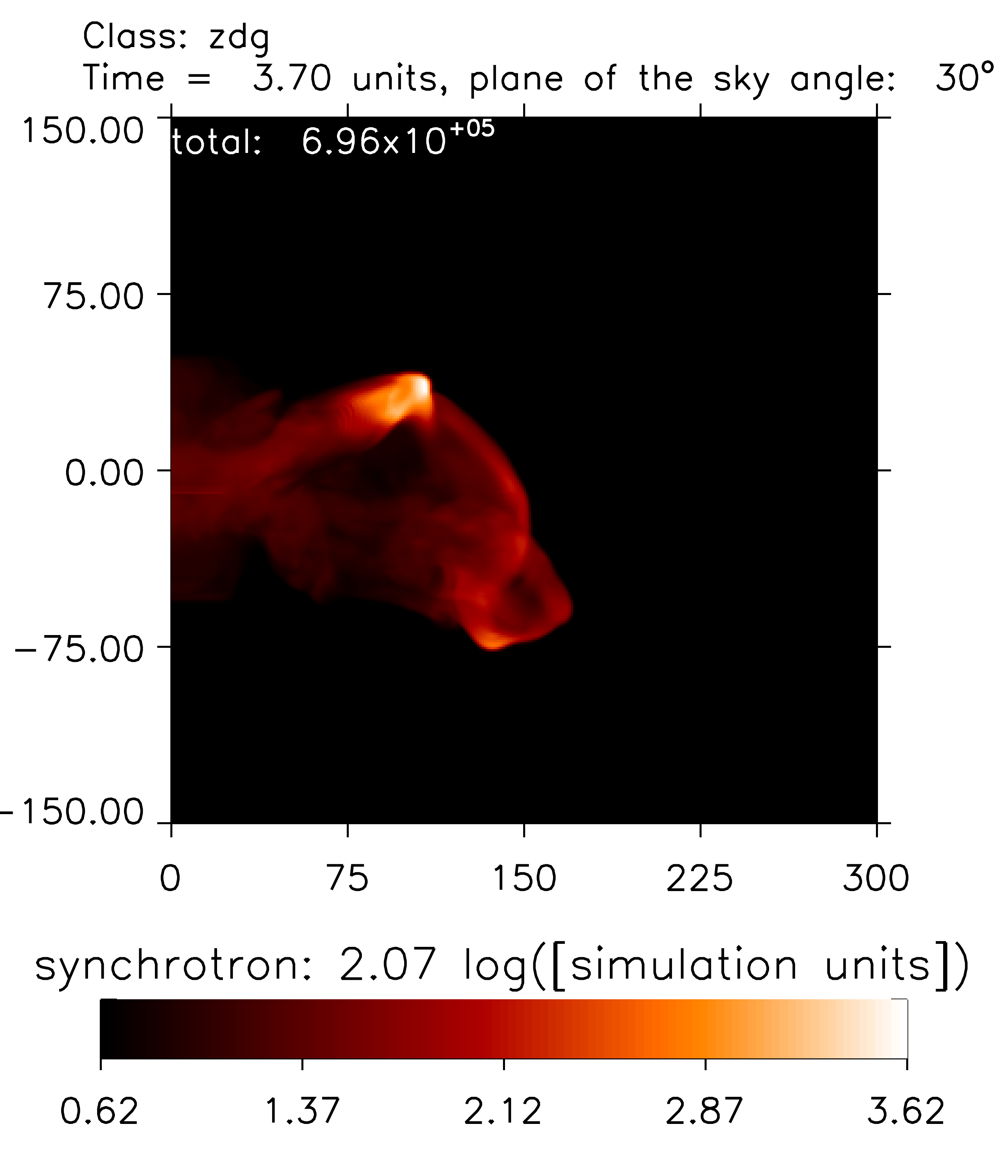}\\
(a) 0\degsy & (b) 10\degsy& (c) 30\degsy \\ [6pt]
\includegraphics[width=0.335\textwidth]{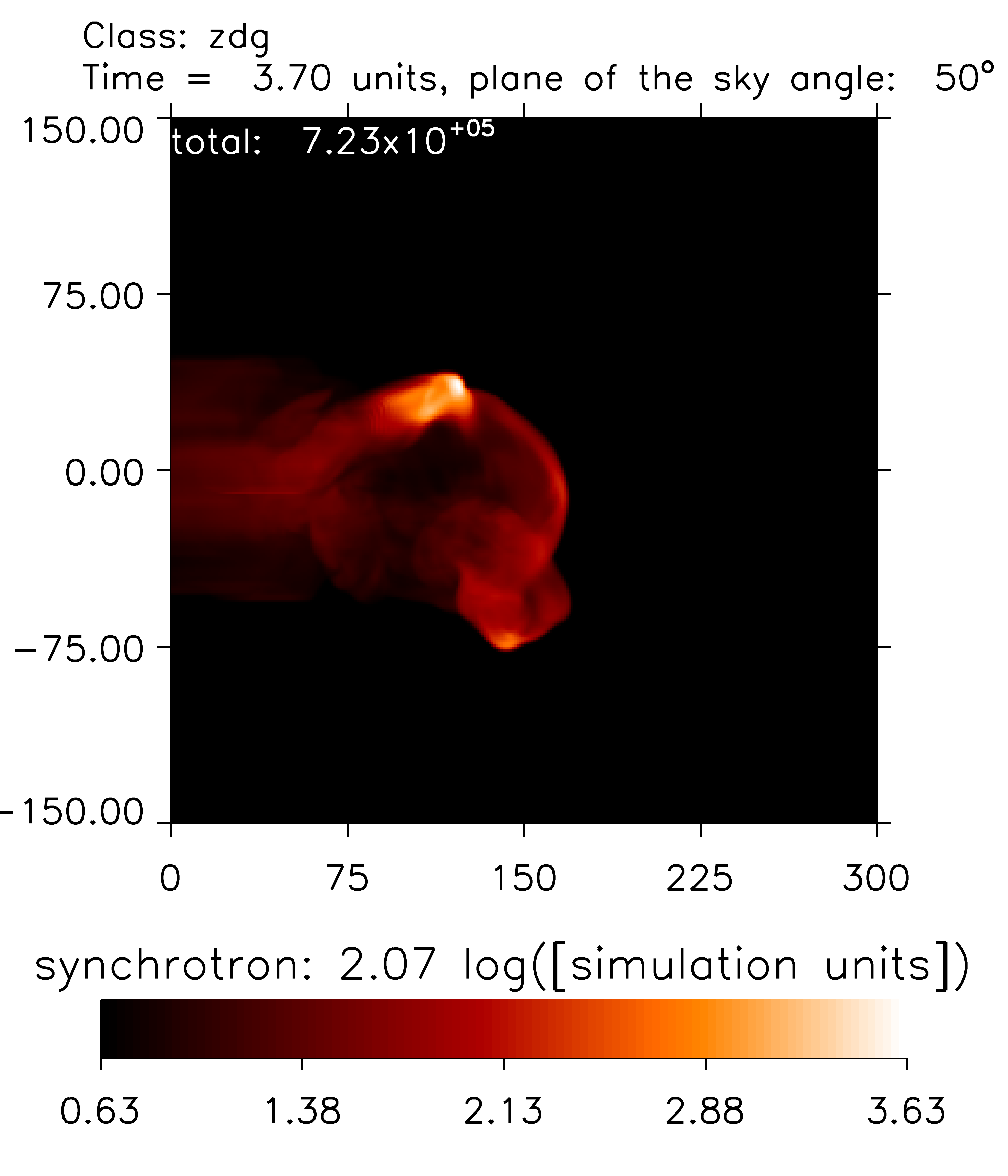} &  \hspace{-1.2cm} 
\includegraphics[width=0.335\textwidth]{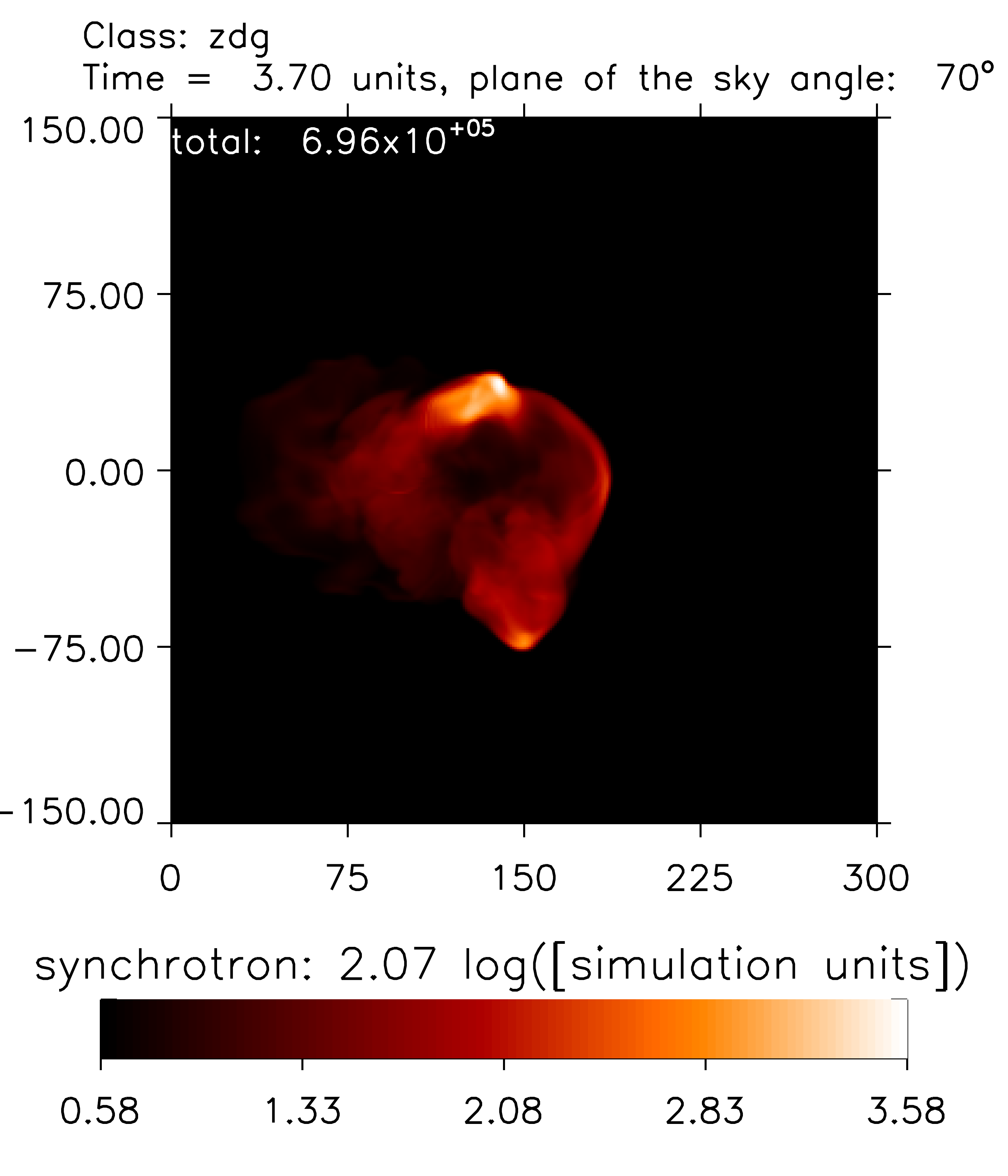} &   \hspace{-1.2cm}  
\includegraphics[width=0.335\textwidth]{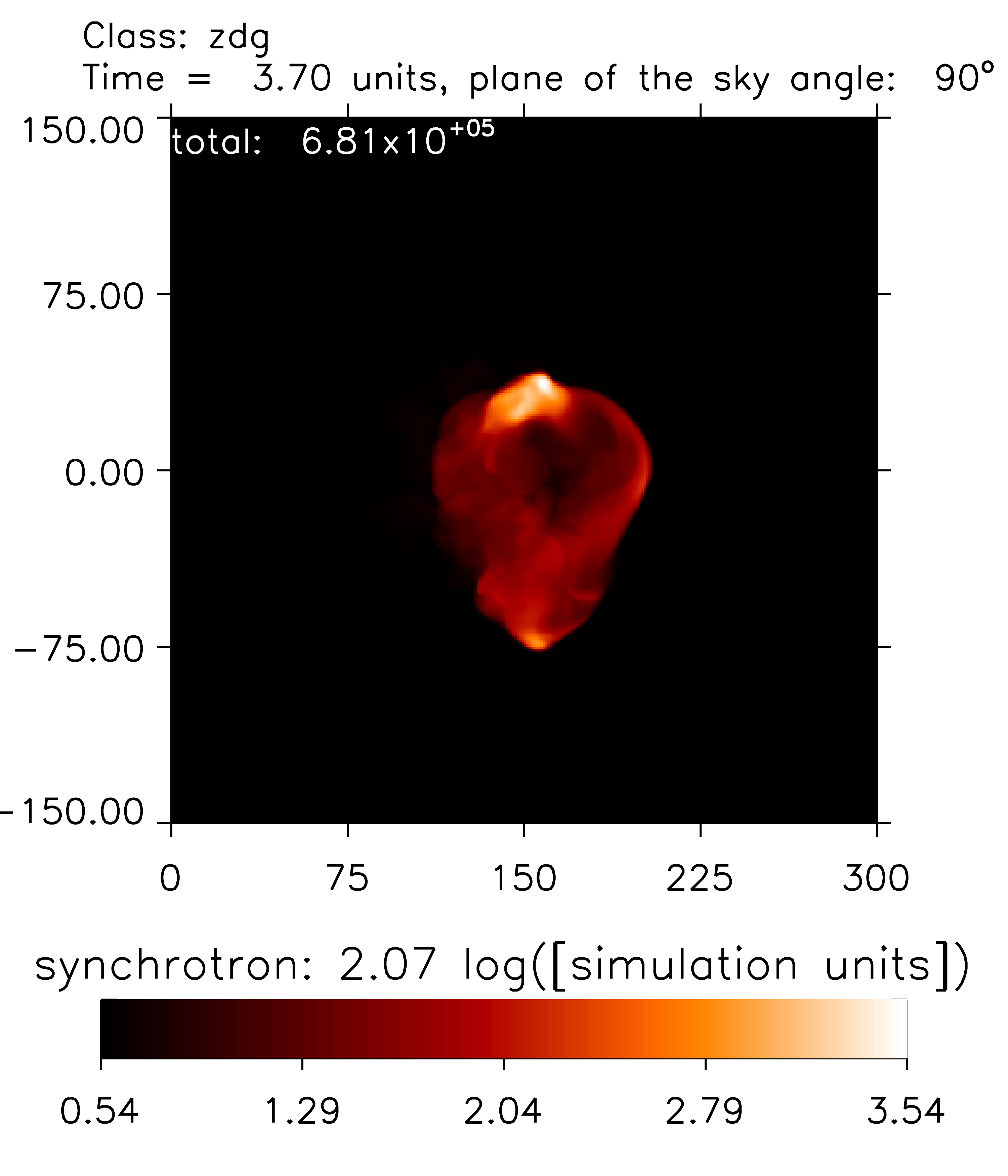}\\ 
(d) 50\degsy& (e) 70\degsy  & (f)  90\degsy  \\
   \end{tabular}
   \caption{{\bf Variation with increasing  angle out of the sky plane.} Simulated radio maps derived from integrated line of sight synchrotron emission. The angle indicated is relative to the sky plane: 0\degsy, \ 10\degsy, \ 30\degsy, \ 50\degsy, \ 70\degsy \ \& 90\degsy \ (a) to (f) respectively. The maps are derived by rotating the cube of data about the centre which, for this example, keeps the emission on the square detector displayed. The run with 20$^\circ$ (denoted zdg) is taken.}   
   \label{fig-sync-tile-zdg-oa_000_010_030_050_070_090}  
\end{figure*}

\begin{figure*}
   \begin{tabular}{cc}
 \hspace*{-0.6cm}   \includegraphics[width=0.4\textwidth]{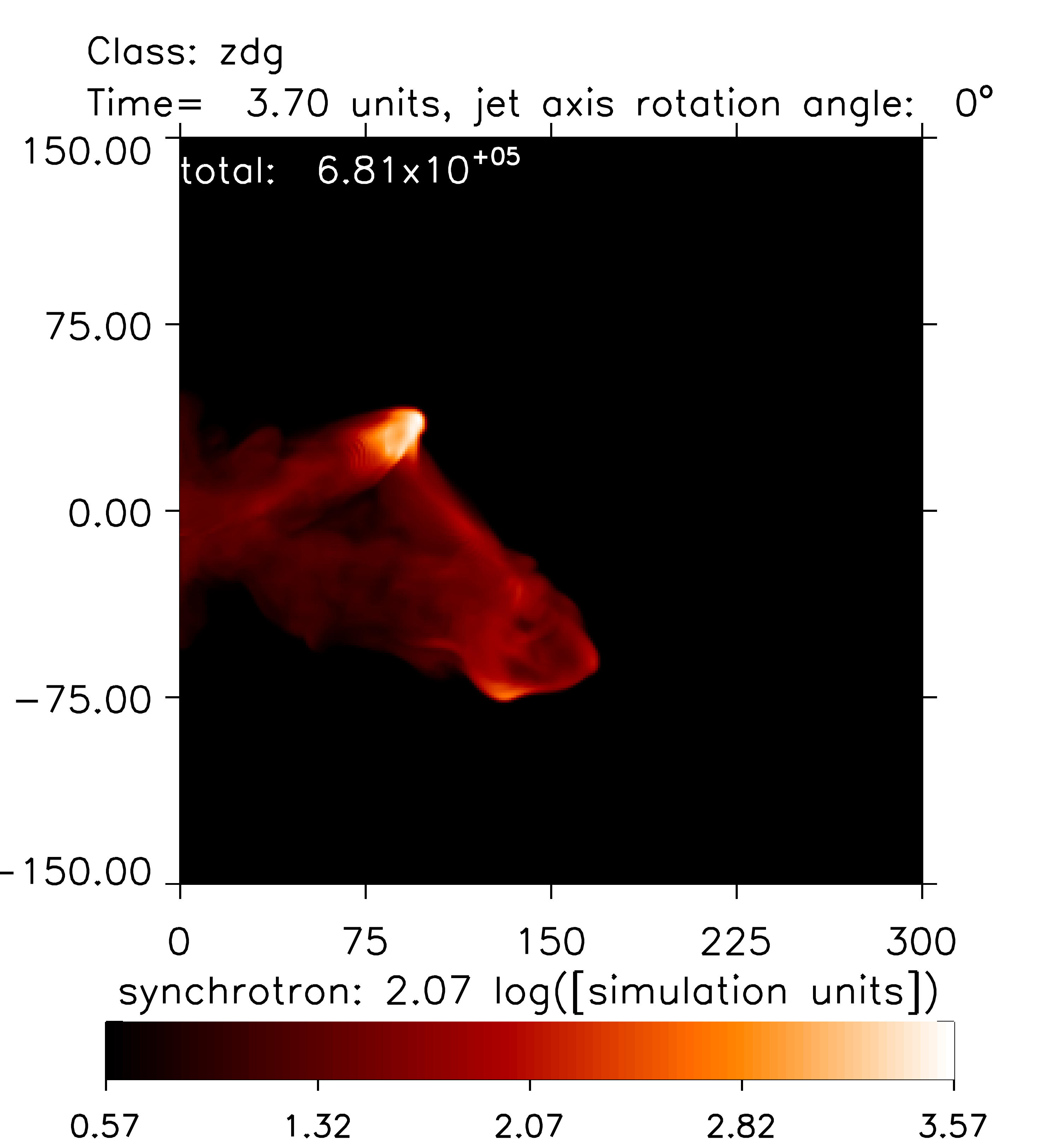} & \hspace*{-1.3cm}
\includegraphics[width=0.4\textwidth]{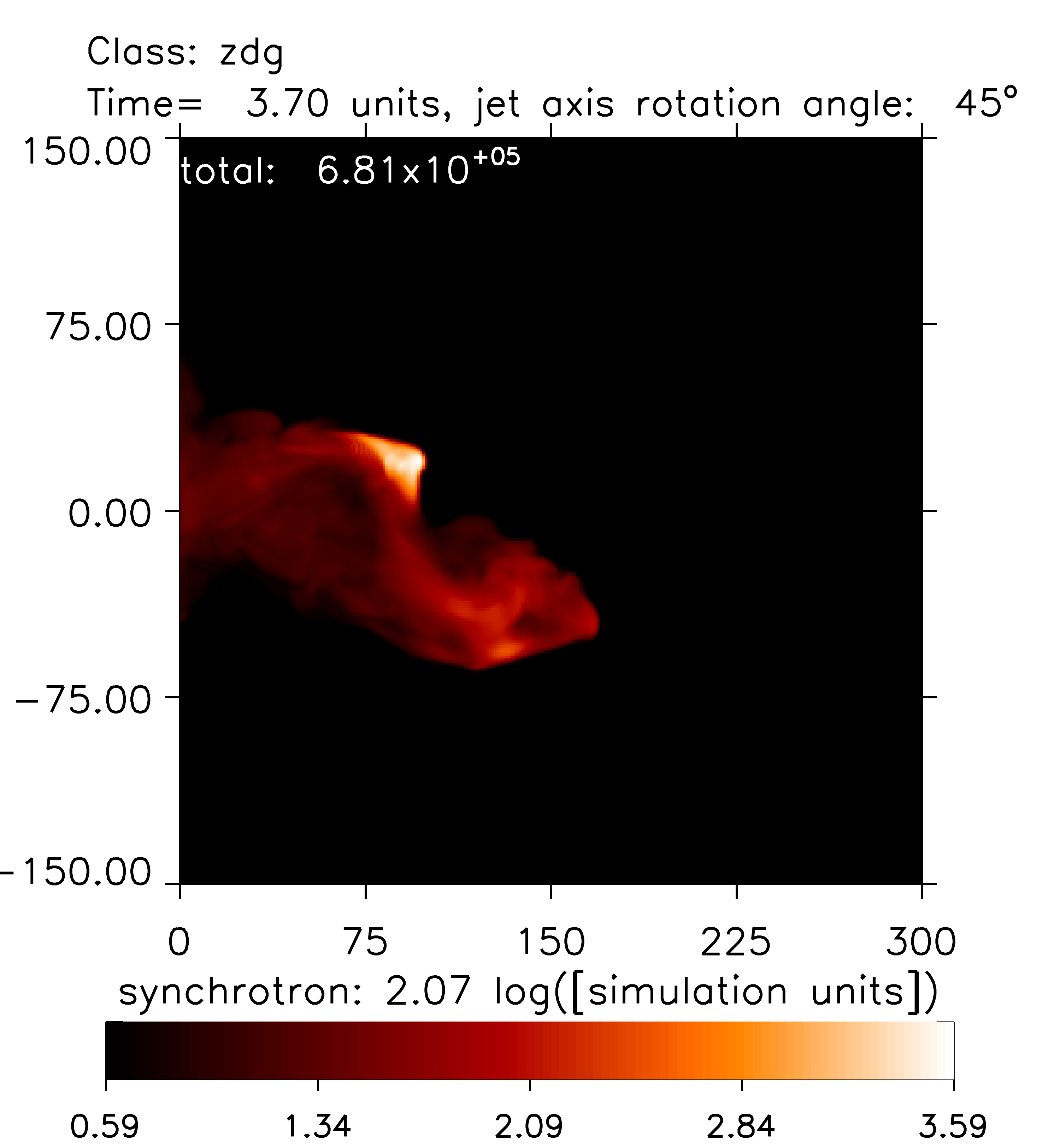} \\
(a) 0\degsy& (b) 45\degsy \\
\includegraphics[width=0.4\textwidth]{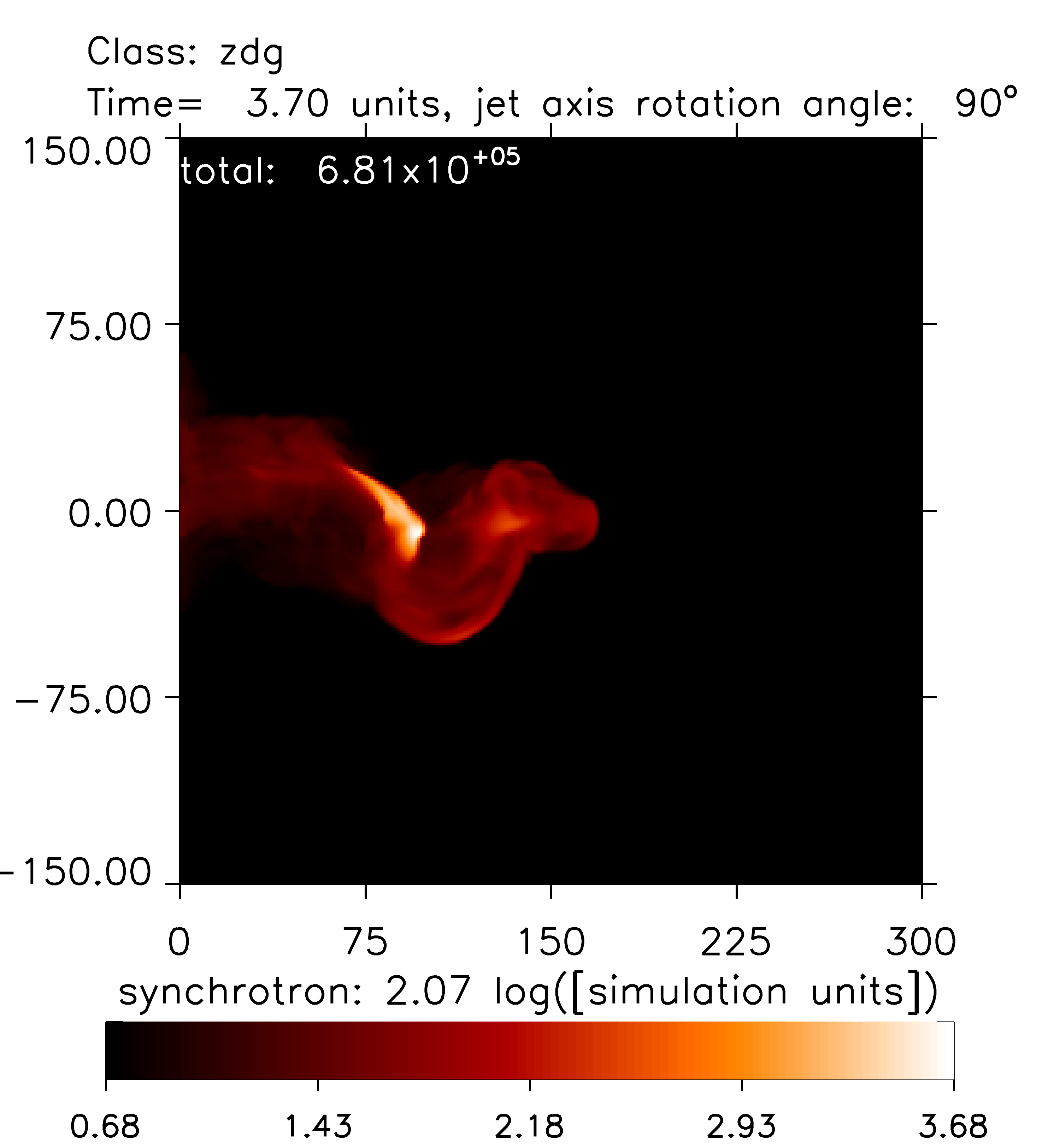} &  \hspace*{-1.3cm} 
\includegraphics[width=0.4\textwidth]{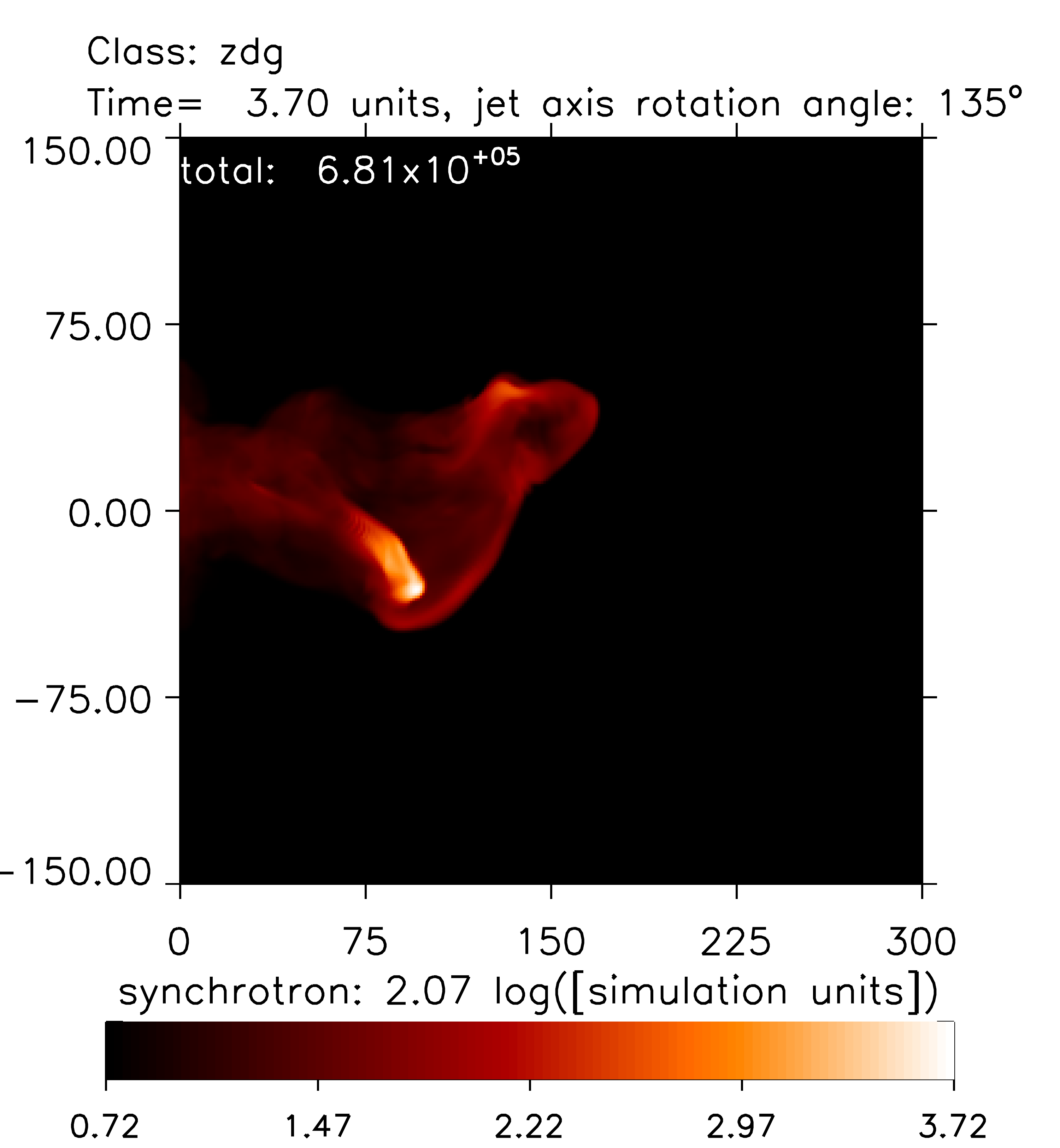}\\ 
 (c) 90\degsy & (d) 90\degsy \\ [6pt]
   \end{tabular}
    \caption{{\bf Rotation: viewing at different angles of the same intrinsic lobe with the jet axis  in the sky plane.} Simulated radio maps derived from integrated line of sight synchrotron emission. The angle indicated  is the  jet rotation angle about the axis: 0\degsy, \ 45\degsy, \ 90\degsy,   (a) to (c), respectively. The run with 20$^\circ$ (denoted zdg) is taken}
  \label{fig-sync-tile-zdg-ja_000_045_090_135_180_225_270_315}
  \end{figure*}
  
\begin{figure}
     \hspace*{-2.6cm} \includegraphics[width=0.7\textwidth]{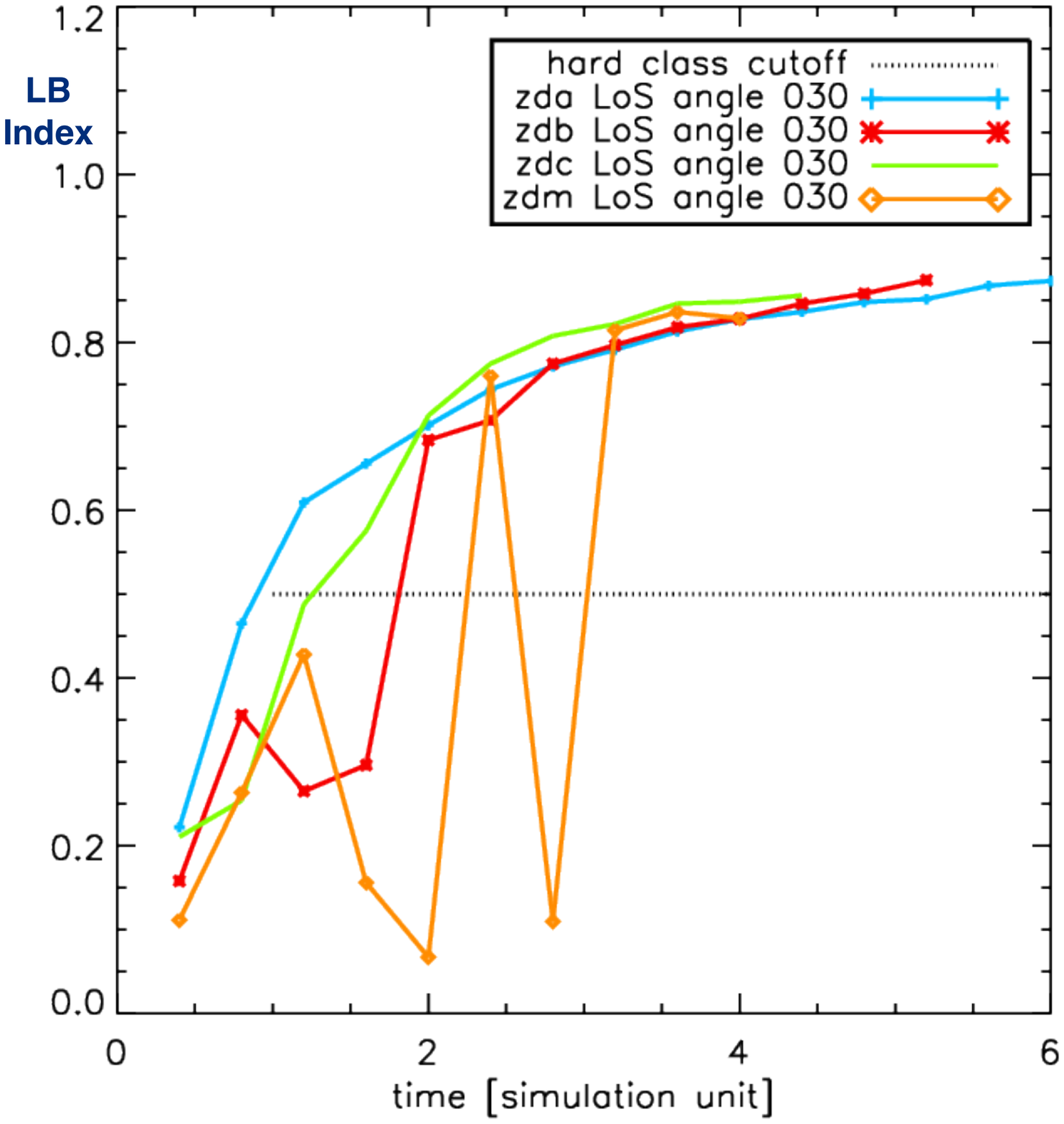} 
  
 \caption{The LBI Index versus time for the four   jet-ambient density ratios running from 0.1 (zda) to 0.0001 (zdo). Here M=6 and 
 a fixed viewing angle out of the plane of 30$^\circ$ is taken.}
 \label{frtype-density}
\end{figure}

\begin{figure*}
   \begin{tabular}{cc}
\includegraphics[height=0.3\textheight]{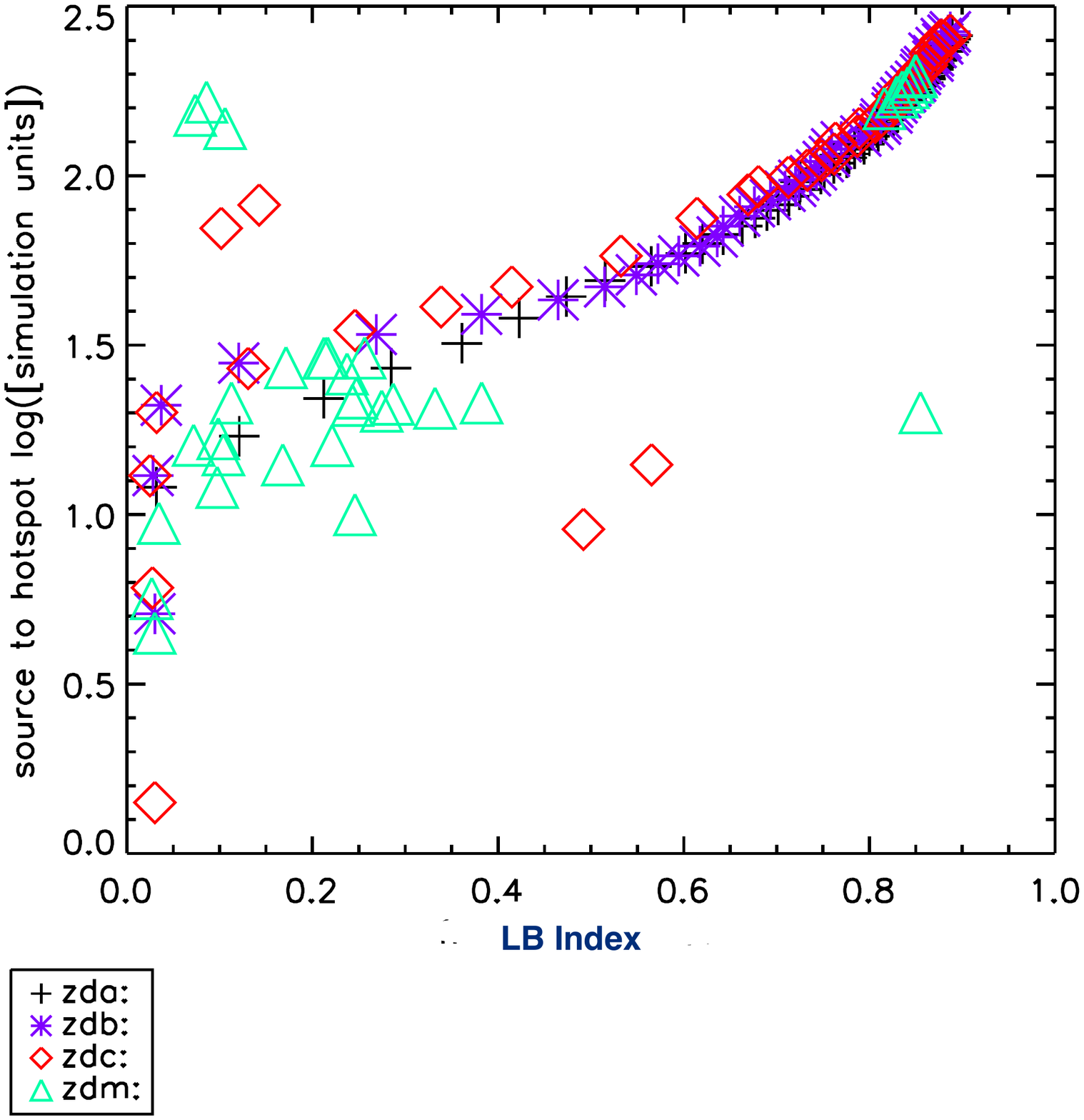} & \hspace{-3cm}
\includegraphics[height=0.3\textheight]{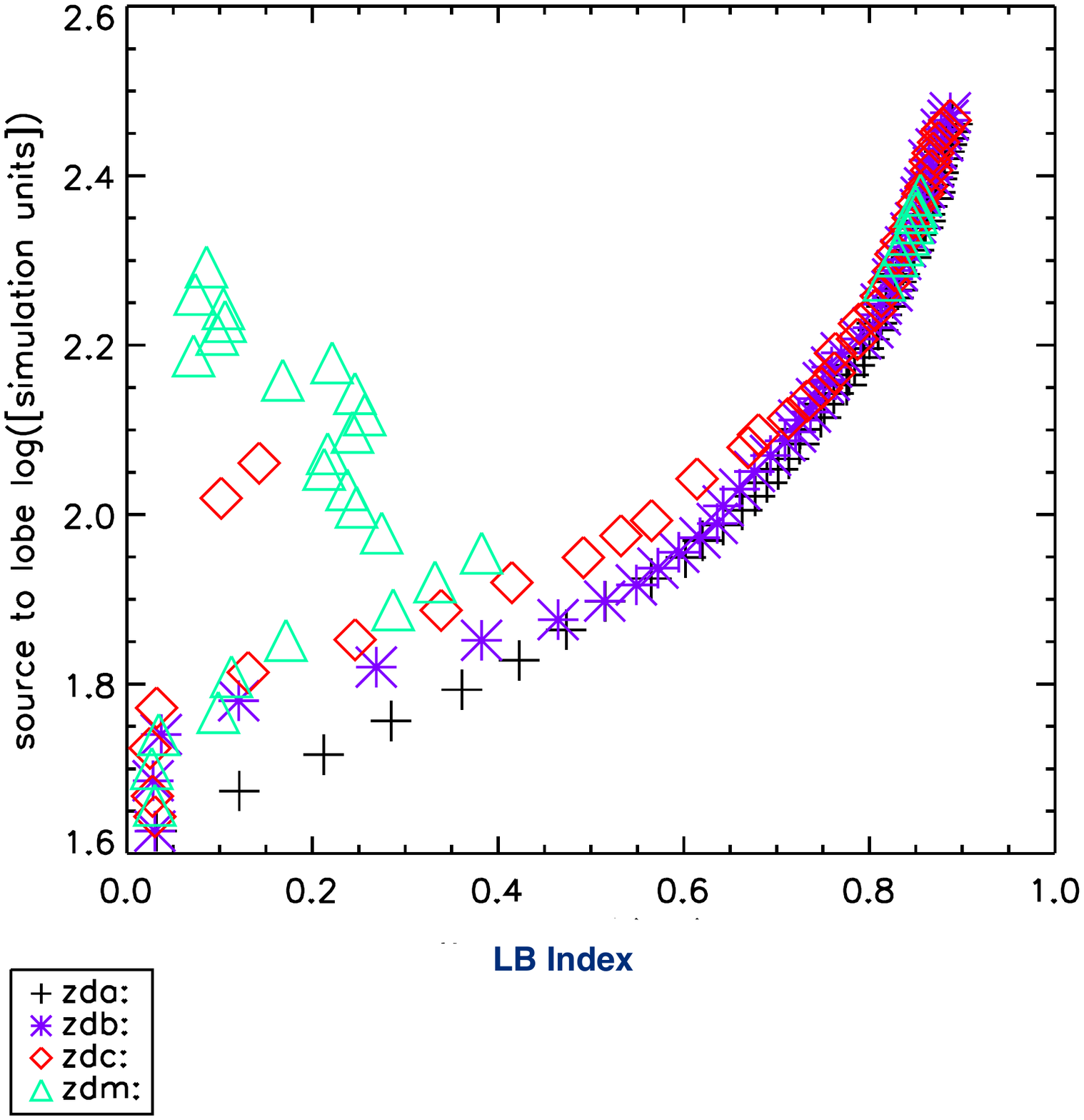} \\ (a) & (b) \\ [6pt]
\includegraphics[height=0.3\textheight]{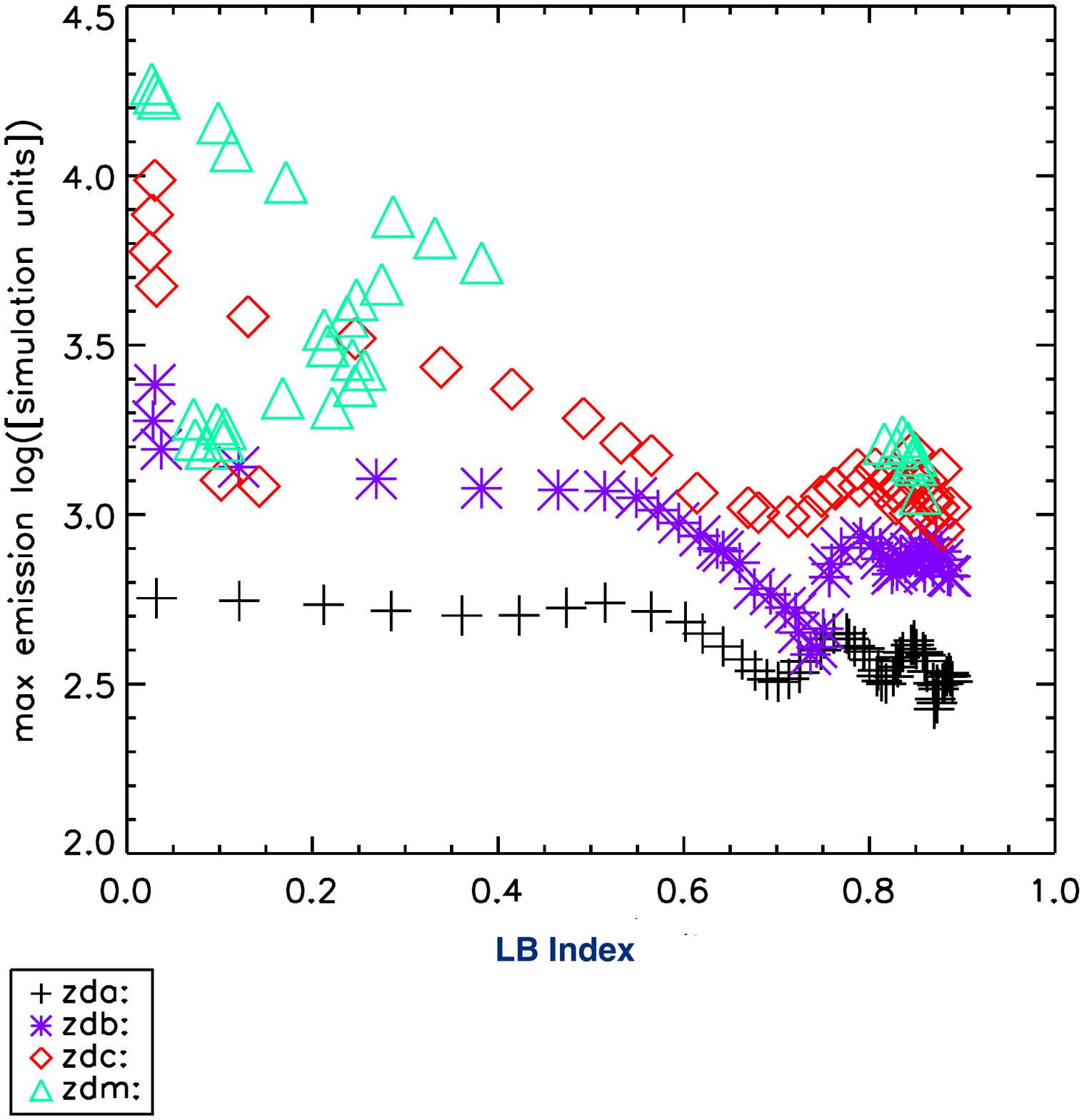}  & \hspace{-3cm} 
\includegraphics[height=0.3\textheight]{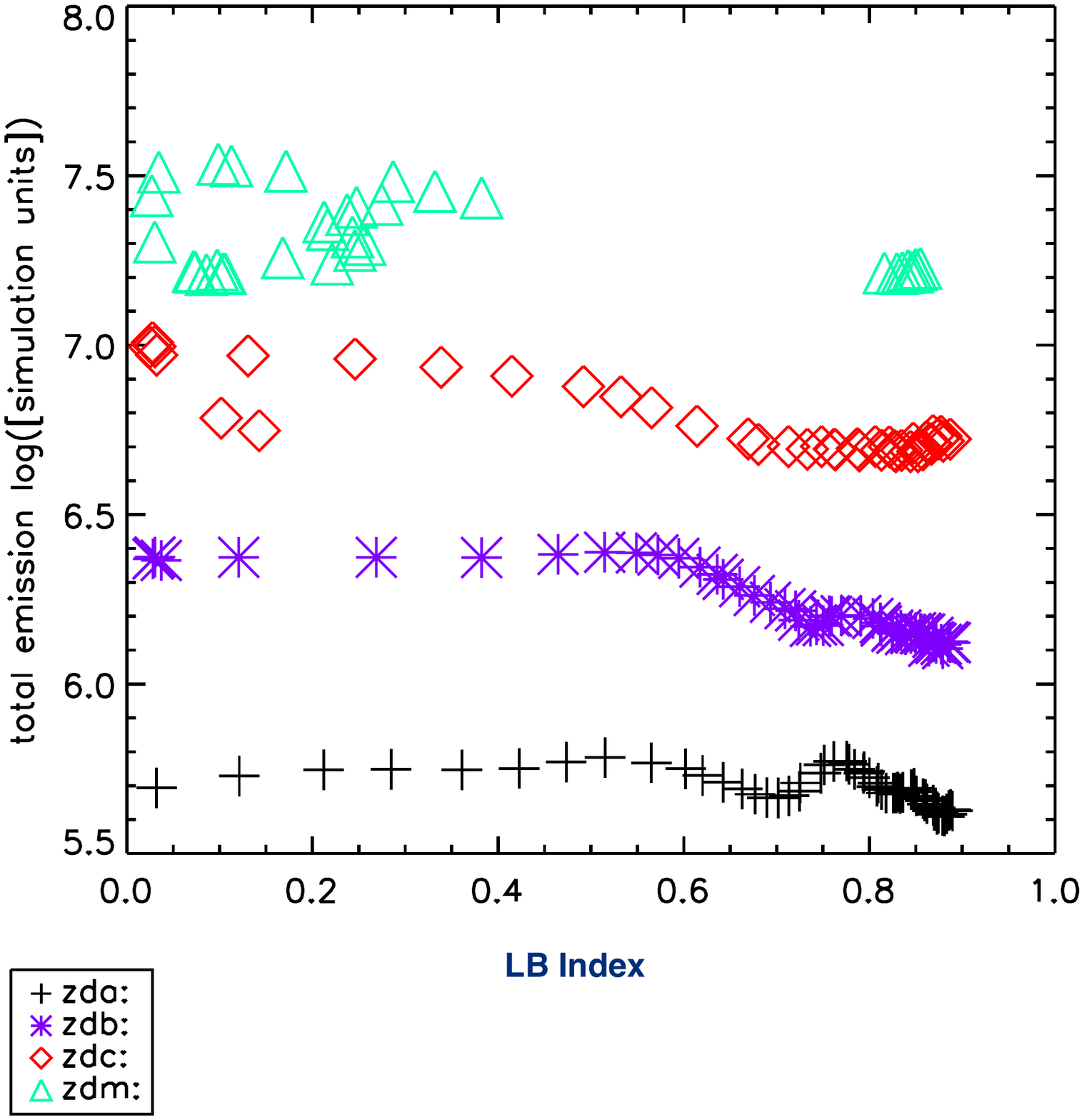} \\ (c) & (d) \\ [6pt] 
  \end{tabular}
   \caption{Relation of the   Limb Brightening Index to the  major evolving properties of the straight propagating jets. 
   The four jet-ambient density ratios are displayed on each panel from 0.1 (plus signs), 0.01 (asterisks), 0.001 (diamonds) to 
   0.0001 (triangles).  The panels show (a) radio hotspot distance from source, (b) maximum expansion distance from source with a  5\% hotspot cut off, (c) maximum emission and (d) total radio emission. The legends also provide the rub designations which correspond to those of Table~\ref{simulation_name}. }
   \label{fig_zda_zdb_zdc_zdm_emission-class}
\end{figure*}
\begin{figure*}
   \begin{tabular}{cc}
\includegraphics[width=0.42\textwidth]{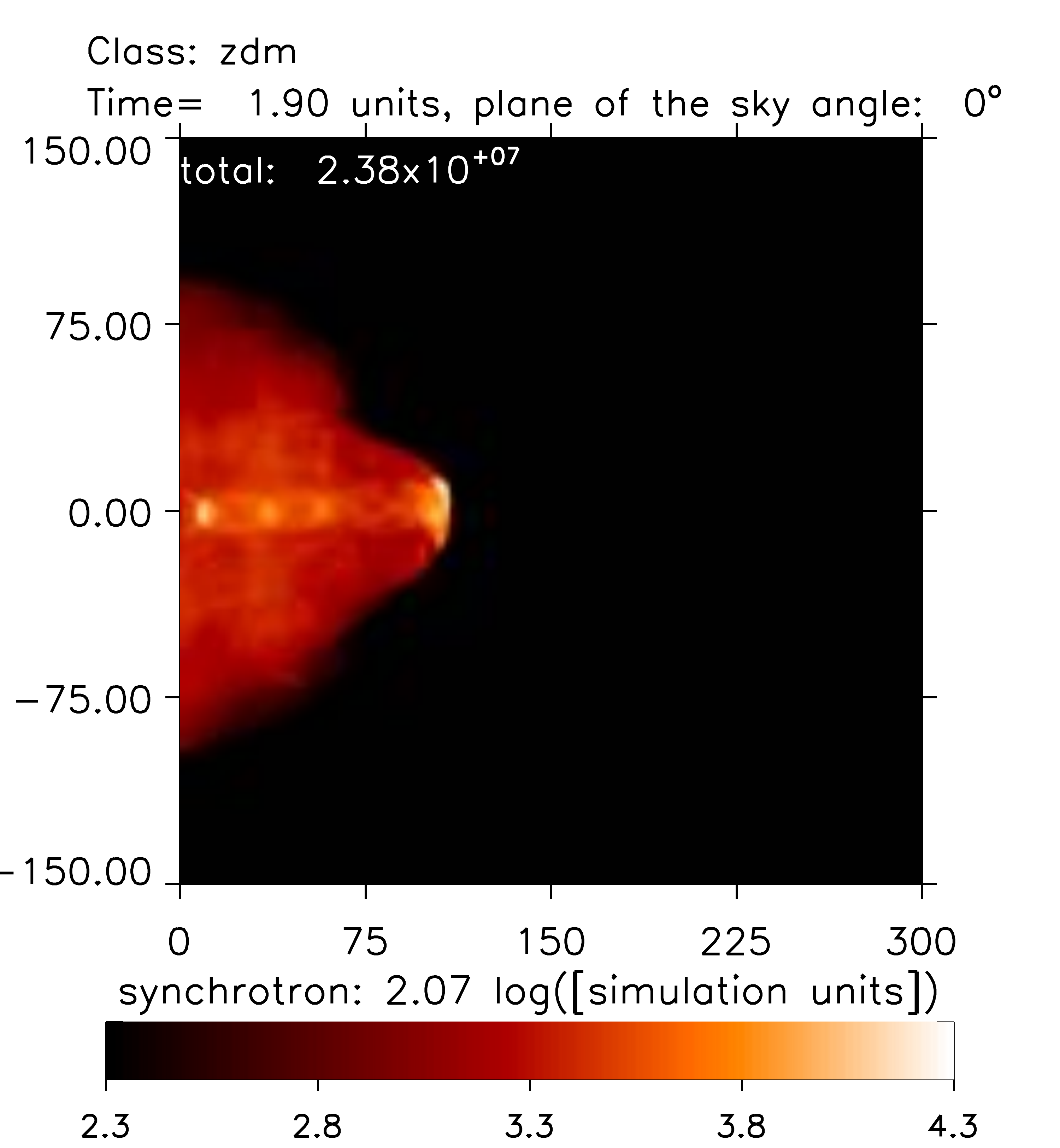} &
\includegraphics[width=0.42\textwidth]{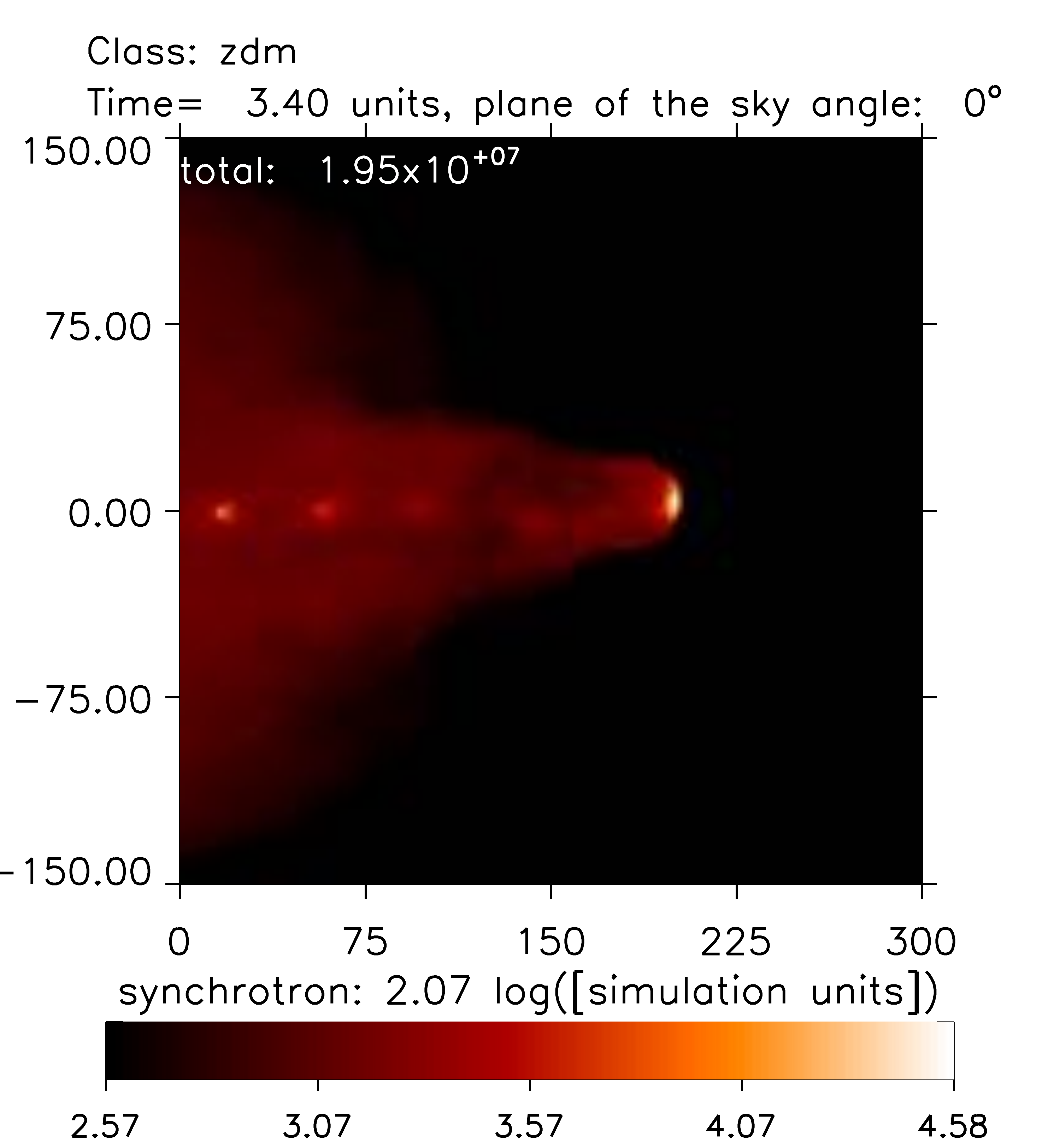} \\ (a) & (b) \\ [6pt]
\includegraphics[width=0.42\textwidth]{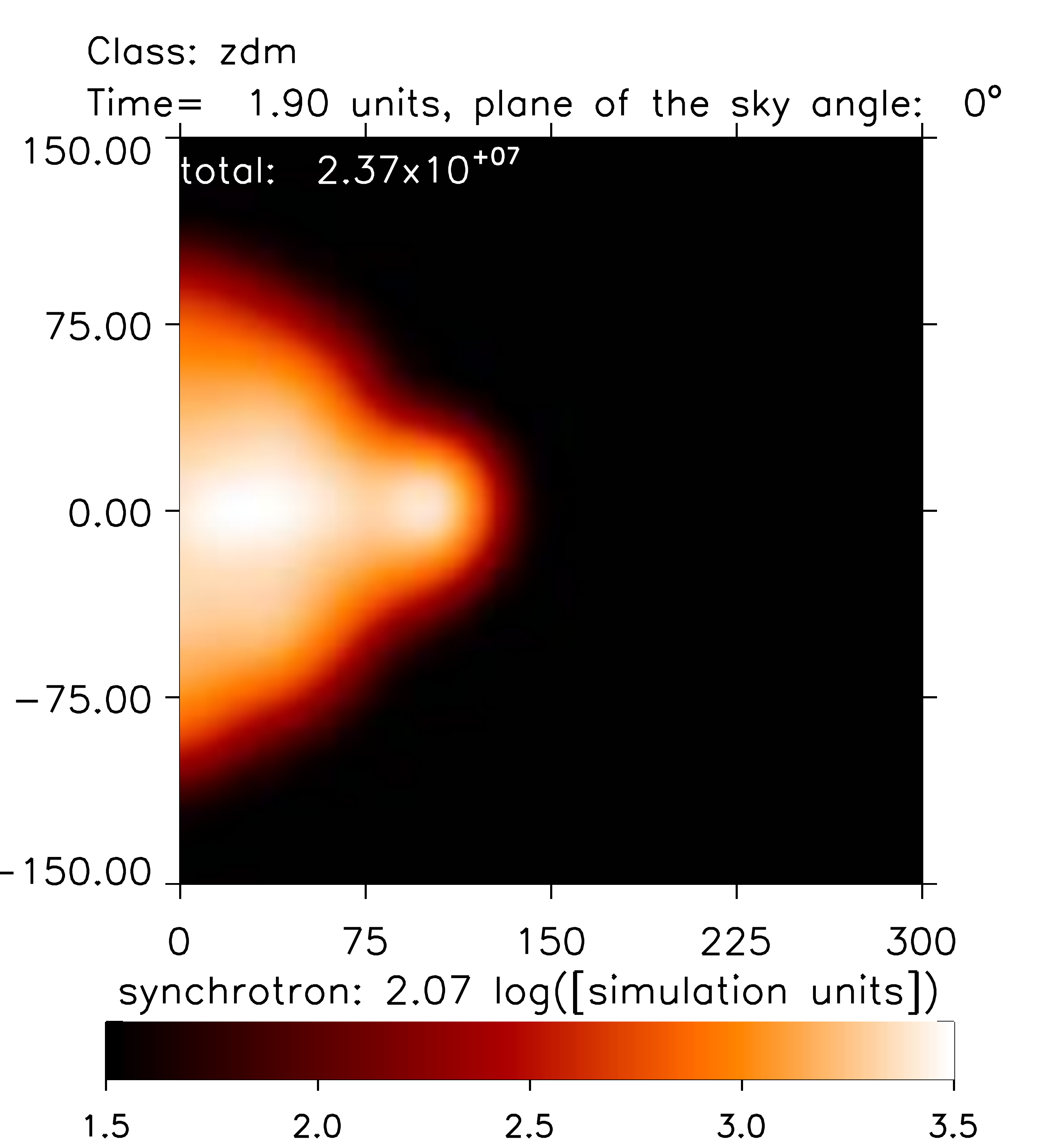} &
\includegraphics[width=0.42\textwidth]{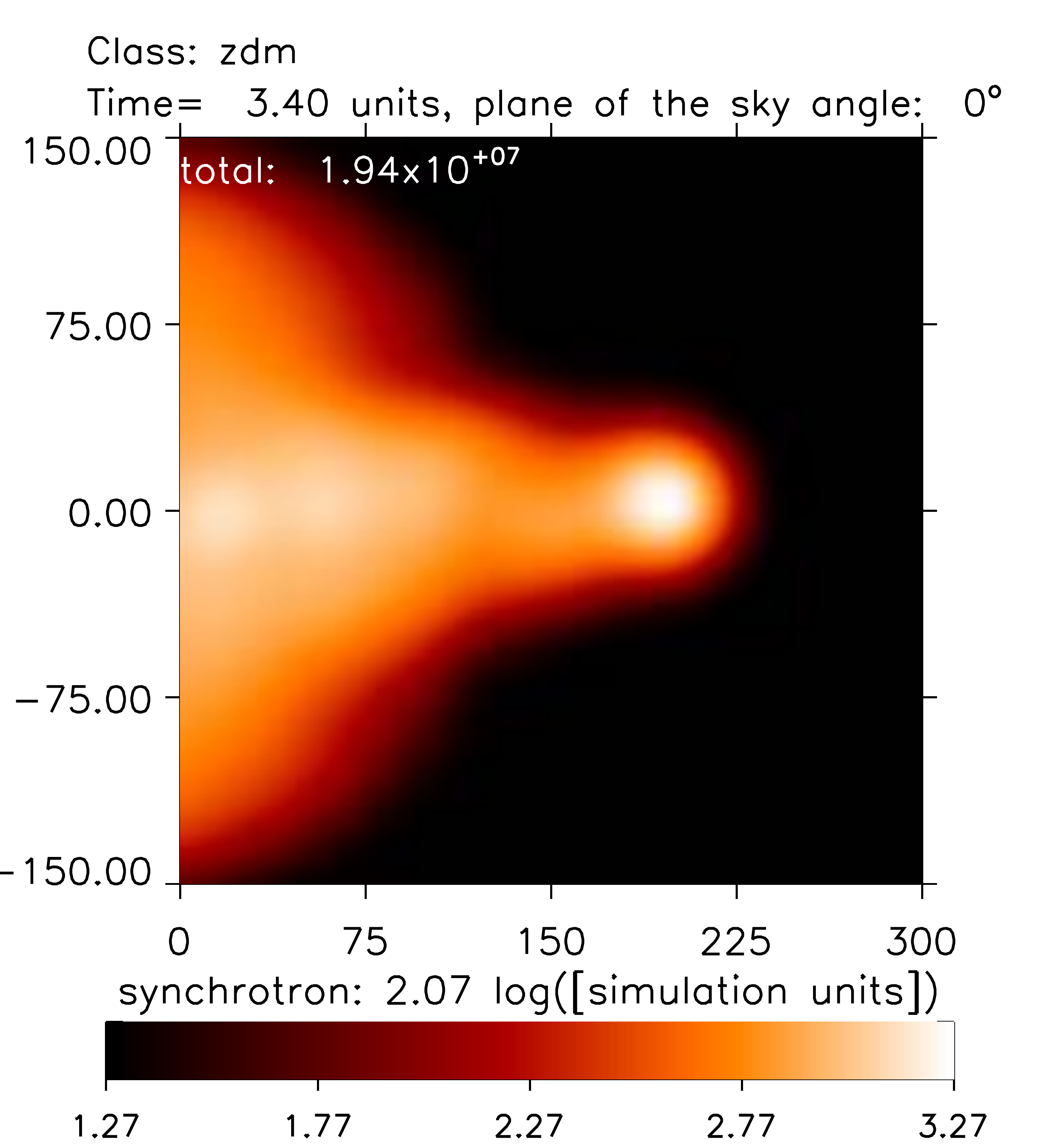} \\ (c) & (d) \\ [6pt]
   \end{tabular}
\caption{Radio images generated from the  single very low density simulation (0.0001, denoted zdm) at the two indicated times in (a) and (b). Below these, in panels (c) and (d), are the Gaussian blurred images that have been used to calculate the corresponding graphs. The jet axis is in the plane of the sky.}
\label{fig-zdm-sync_tile-019_034}
\end{figure*}

\begin{figure*}
   \begin{tabular}{cc}
\hspace{-1.2cm}   \includegraphics[height=0.33\textheight]{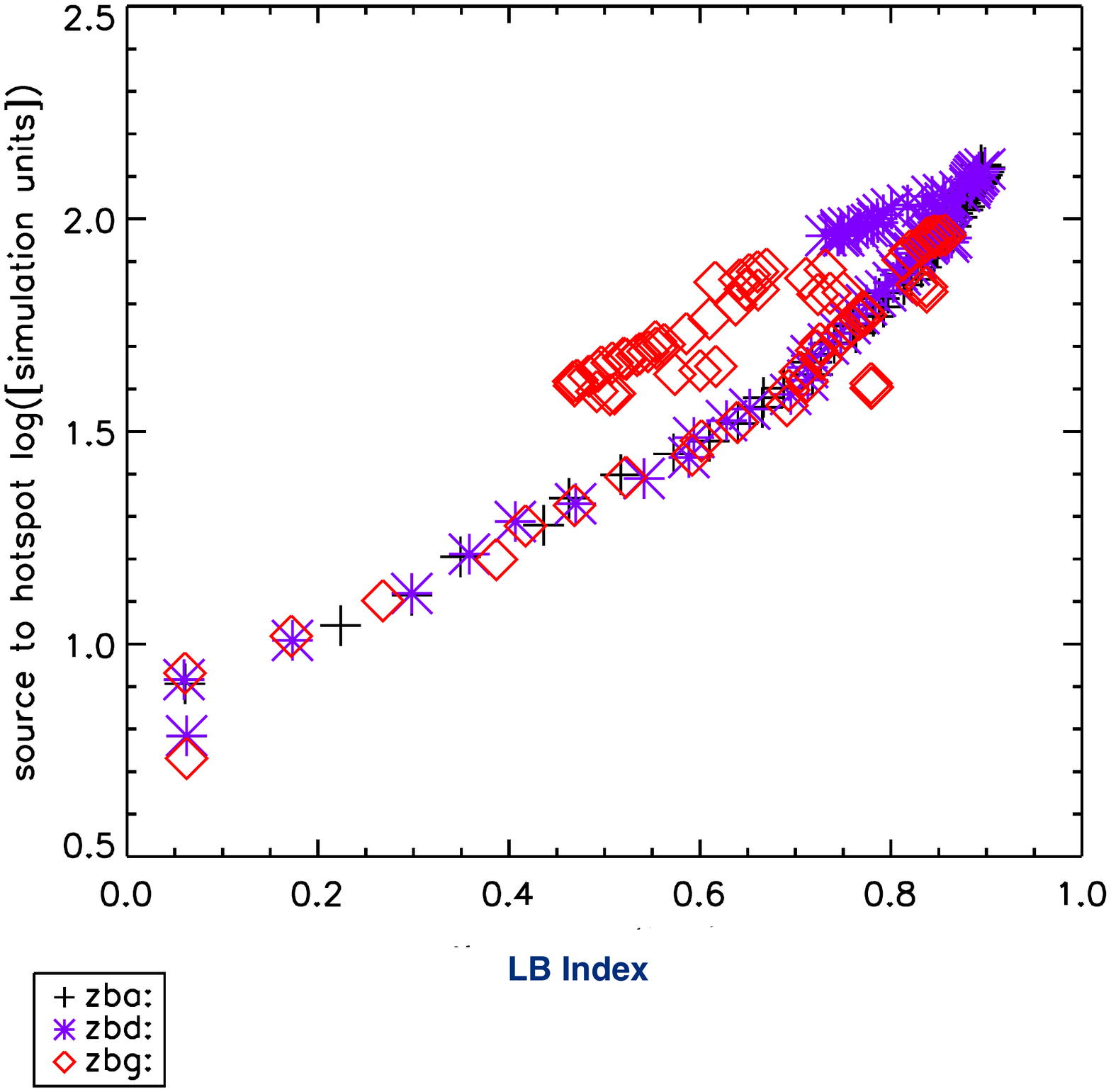} & \hspace{-3cm}
\includegraphics[height=0.33\textheight]{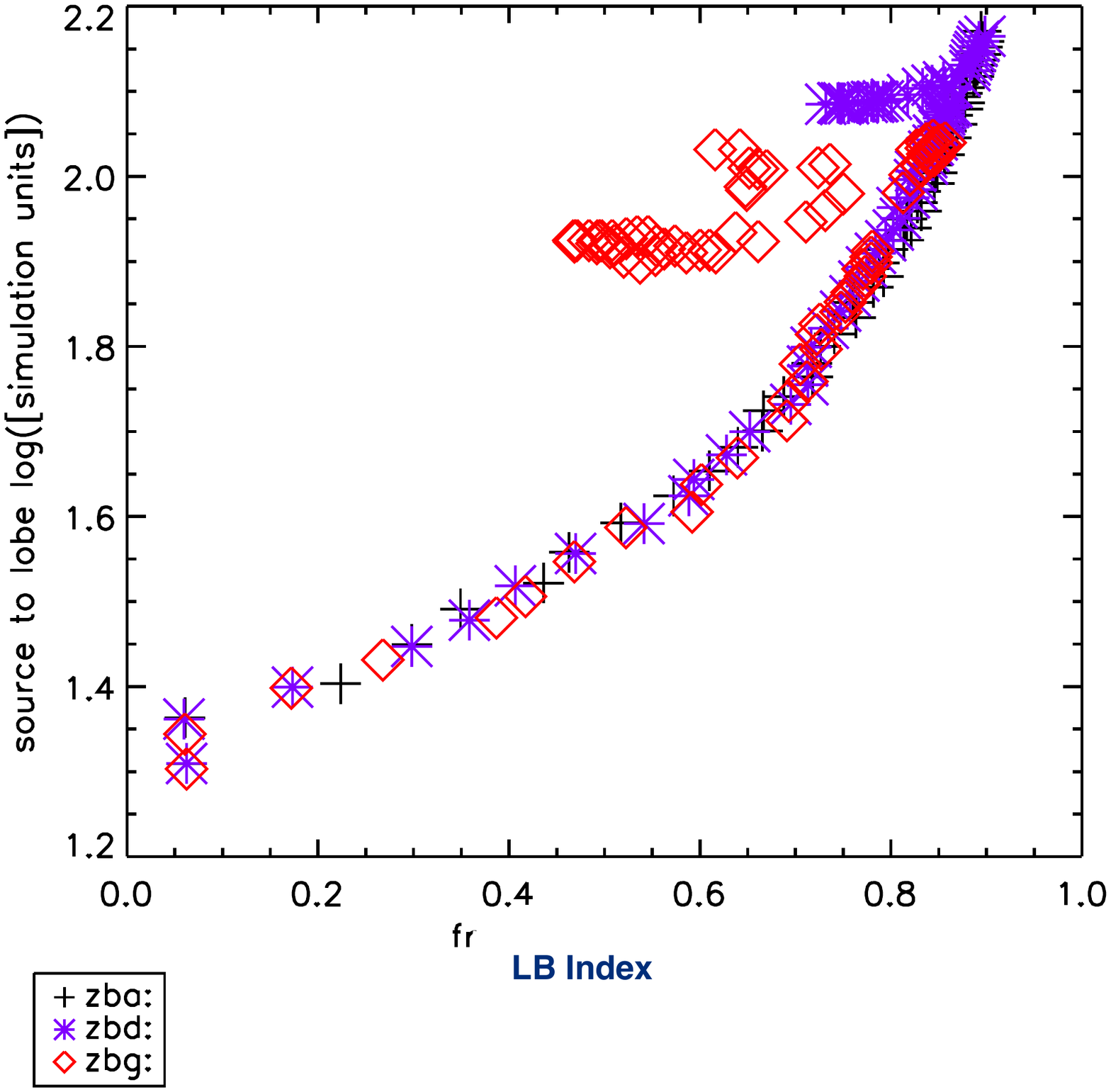} \\ (a) & (b) \\ [6pt]
\hspace{-1.2cm} \includegraphics[height=0.33\textheight]{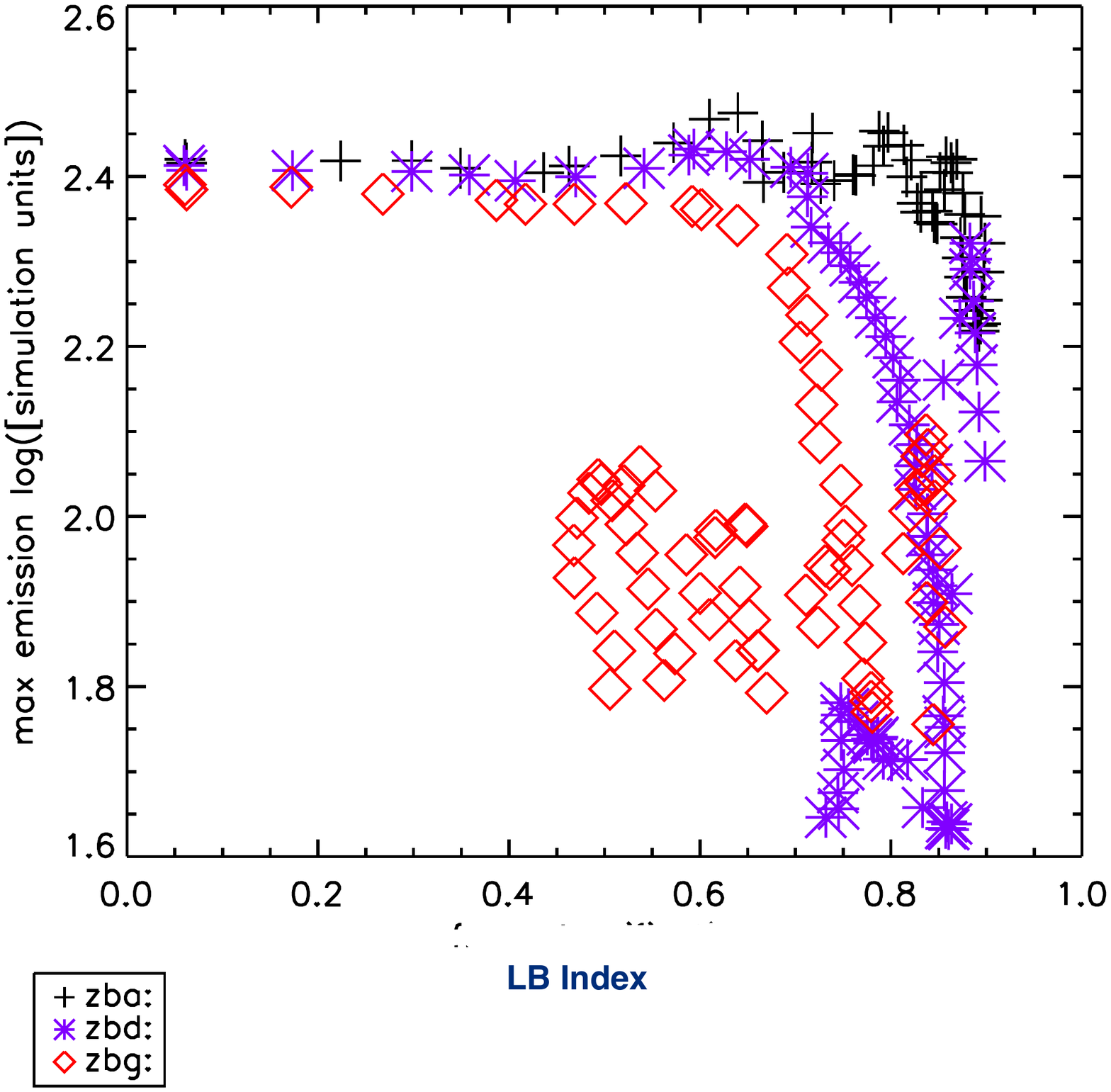}     & \hspace{-3cm}
\includegraphics[height=0.33\textheight]{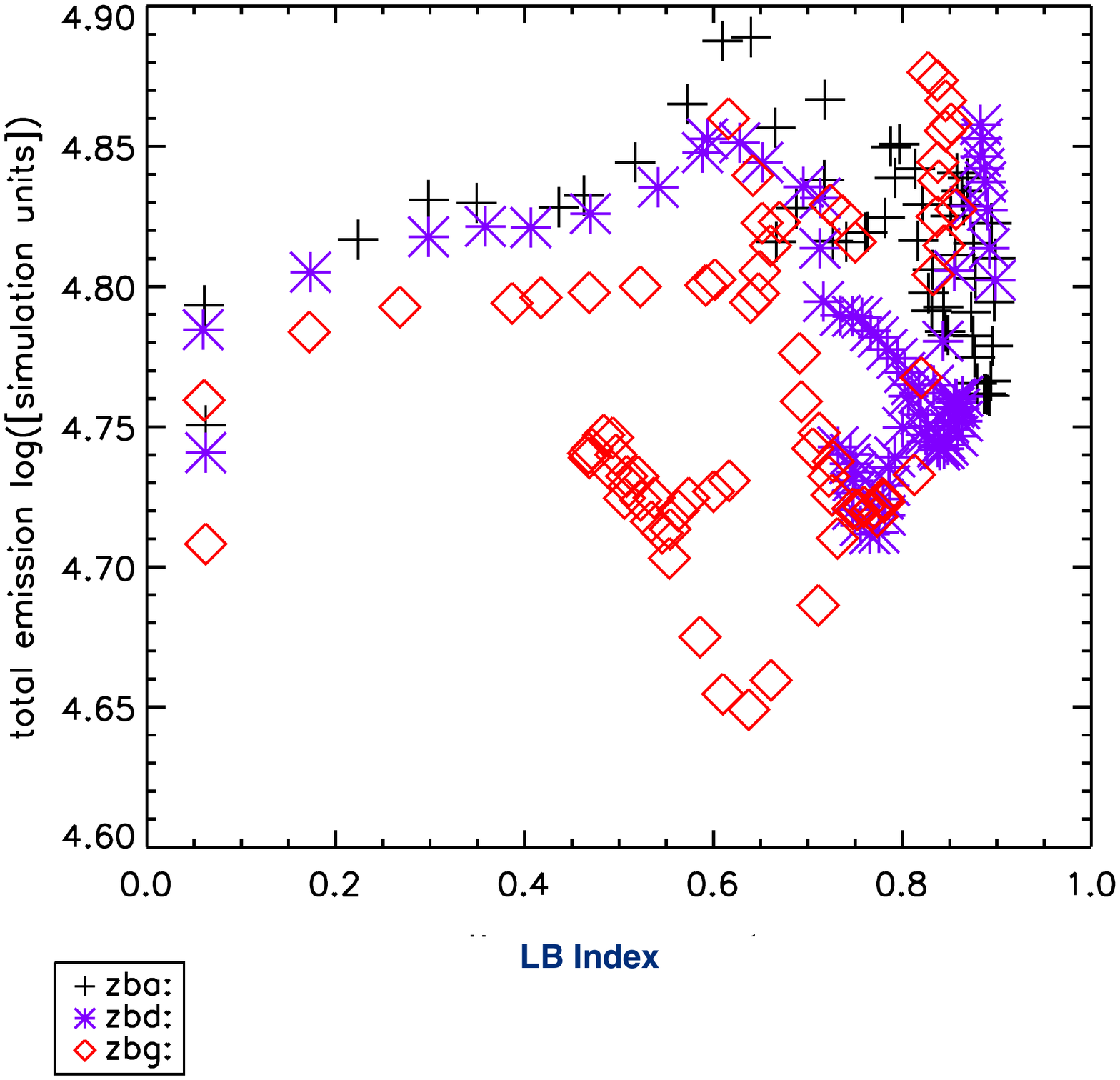} \\ (c) & (d) \\ [6pt]
 \end{tabular}
   \caption{The properties of the radio galaxies for three precession angles demonstrating  trends with the Limb Brightening Index.
   The precession opening angles for runs zba (1\degsy), zbd (10\degsy) \& zbg (20\degsy) are shown . The data  logged on the y-axis are (a) radio hotspot distance from source, (b) maximum expansion distance from source, (c) maximum emission and (d) the  total radio emission.}
   \label{fig_zba_zbd_zbg_emission-class}
\end{figure*}

The dependence on the density ratio is probed in Fig.\,\ref{fig-sync-tile-zda_zdb_zdc_zdm-subt_amb-oa_00}. 
Again, two maps have been supplied for each density ratio. The differences between the two radio maps is the limits that are used to highlight the hotspot or the cocoon material.
For these straight jets, there is no great difference in the maximum surface brightness. This is partly because the Mach number is held constant so that the lower density jets have a correspondingly higher speed and, hence, the same momentum flow rate. 

A broader lobe is generated as the density ratio is reduced as expected from the analysis of Paper 1. It should be noted that in three dimensions, the forced symmetry is broken and the cocoon does not expand transversely  in the same manner as the two dimensional equivalent. 

There is a remarkable transition in the visible jet structure at density ratios lower than 0.001. The channel itself can be clearly traced in the
very low density ratio simulation on the right with a string of oblique shocks.
Here, the wider cocoon protects the jet, keeping the pressure high. This squeezes the jet which consequently maintains a convergent-divergent configuration along its entire length. This can be considered as a potential diagnostic for jet density. 

The jet precession rate can be expected to be a major influence.  
This is seen from  the precession equations for the transverse speed components
\begin{align}\label{eq_velocity}
     &v_{\text{y}}=  v_{\text{jet}} \cdot \sin(\theta) \cdot \sin(\omega_{\text{l}} \cdot t) \nonumber \\ 
     &  \\
     &v_{\text{z}}=  v_{\text{jet}} \cdot \sin(\theta) \cdot \cos(\omega_{\text{l}} \cdot t) \nonumber ,
\end{align}
where $\omega_{\text{l}}$ determines the angular frequency of precession.  

Faster precession has a significant effect  on the radio lobe structure as shown in Fig.~\ref{zdg_precession_rate} where the last two columns correspond to precession rates of $2 \cdot \omega_{\text{l}}$ \& $4 \cdot \omega_{\text{l}}$, respectively. Here, $\omega_{\text{l}}$ 
corresponds to a period of two time units or 58.9 Myr for a giant radio galaxy.
The result is that the helical dynamics becomes apparent in the radio structure  of the inner lobes. In addition, the hotspot surface brightness falls as the visible lobe enlarges. The hotspot location also changes even in these projections where the jet rotates around an axis in the plane-of-sky.

\subsection{Viewing angles}

 There is very little information concerning the orientation of an observed radio galaxy that can be derived from the 
 synchrotron emission and its distribution. However, simulations can help connect the orientation to the structure to some extent. Each simulation at each time step then yields a diverse range of possible structures as a source is rotated. This is particularly true
for wide precessing flows, as now discussed.

The influence of the orientation of the jet axis for the 20$^\circ$ run  is shown 
in Fig. \ref{fig-sync-tile-zdg-oa_000_010_030_050_070_090}. In this case, a pronounced dog-leg in one lobe or Z-shaped structure (across two lobes with point symmetry) is apparent at intermediate angles reminiscent of the giant radio galaxy
NGC\,315 \citep{1976Natur.262..179B}. Although the simulation can generate double-lobed structures with inner and outer sets of lobes, it is clear that the outer lobes do not leave a trail or bridge away from the origin but connect to the inner hotspot instead. We thus exclude this particular model for X-shaped sources but favour it for Z and S-shaped varieties, consistent with recent observations and their interpretation  \citep{2018ApJ...852...48S}.

The viewing angle can also be changed  by rotating the radio source about the jet axis, keeping the axis fixed in the sky plane, as shown in Fig.\,\ref{fig-sync-tile-zdg-ja_000_045_090_135_180_225_270_315}. This figure demonstrates how a Z-shaped edge-darkened source at one angle (a) can appear as an S-shaped edge-darkened source when rotated by 90 degrees, as shown in (c). This raises serious questions about how to classify radio galaxies.

\subsection{The LB Index as a source evolves }

We determine the position of maximum intensity,
the maximum lobe size and, hence, the LB Index by passing the  maps for synchrotron emission  through an IDL  script. 
The maps are derived from the 3D data cubes, stored at time steps of 0.4 in simulation units. These are the time steps plotted in the following figures.

The Index evolution is shown in Fig\,\ref{frtype-density} for each of  the four jet-ambient density ratios (models denoted zda to zdm, all 300$^3$).
 For these straight jets which propagate with little hinderance, we obtain straightforward results. We show this in the top two panels of
Fig.\,\ref{fig_zda_zdb_zdc_zdm_emission-class}  where we 
plot separately the location of the peak surface brightness and  the total source length against the LB Index.

Fig\,\ref{frtype-density} displays smooth evolutionary tracks for the lobes of radio galaxies with one major exception.  For the lowest jet-ambient density ratio, the radio galaxy can switch
between FR\,I and FR\,II, corresponding to an LB Index of 0.5 (the dotted horizontal line). This is caused by the significant variations in jet brightening at these low densities due to induced pressure variations from the  interaction with the hot high-pressure  lobe.
Thus at very low jet densities there is a  distinct LB Index regime between 0.1 and  0.2 which persists until finally disappearing as the jet  approached the end of the grid and the feedback weakens. 

Apart from this exception, for  jets that have little to no precession, the hotspot is located at the leading edge of the jet. So the recorded distances will then be consistently increasing with time.
The LB Index gradually increases as both the length to the hotspot (Panel a) and total lobe length (Panel b) increase with the switch from FR\,I to FR\,II occurring after just $\sim$ 70 of the 300 zones.   

The lengths of the lobes are almost identical after the initial settling period. This is because we maintain jet-ambient pressure equilibrium while the jet density is varied. This means that the jet sound speed increases as the density decreases, which, since the Mach number is held constant,  raises the jet speed in proportion. Hence, the momentum flow rate is constant and hence, provided the jet opening and aerodynamic drag are similar, the propagation speed is approximately constant. 

The simulated luminosities are, however, very different. The hotspot and total luminosities are displayed in Panel (c) and Panel (d).  First, it is clear that, while the momentum flow rate is fixed,  the beam power increases proportional to the jet speed. Therefore the power is inversely proportional to the square root of the input jet density.  The simulated total radio emission, quite surprisingly at first, also increases with density in the same manner. However, the total luminosity does not increase with time for each run.
Even though the radio power is derived by summing the squares of the pressure in each cell, beam power and radio power are roughly proportional. Therefore, in these FR\,II types, the luminosity is dominated by the high brightness region and the material accumulated in the diffuse lobe contributes little to the total.
There is a small tendency for the radio power to diminish with time probably related to the gradual fall in maximum pressure as the jet propagates further into the growing radio lobe and expands somewhat. 

The region which generates the extra emission in the lower density runs is related to the lobe extent. As the input density of the jet is reduced, there is an increase in the width of the base of the cocoon where the galaxy is located  (note the reflective boundary). This larger cocoon is highlighted in Fig.\,\ref{fig-sync-tile-zda_zdb_zdc_zdm-subt_amb-oa_00}, and results in a higher total emission and broader distribution. This is further demonstrated in
Fig.\,\ref{fig-zdm-sync_tile-019_034} which
 also illustrates how a distant radio galaxy could be re-classified as an FR\,I due to the lower resolution (lower panels) as compared to the high resolution images in the top row. 
   
Of further  interest  from this density analysis is  that the maximum emission converges with time as the LB Index increases. Panel (c) of Fig.\,\ref{fig_zda_zdb_zdc_zdm_emission-class} predicts that the hotspot flux itself is purely dependent on the pressure as given by the jet momentum flow rate per unit area and that the volume occupied by the gas generating the hotspot is roughly constant. Therefore, care has to be applied when employing total  emission of a distant radio galaxy that  sufficiently sensitive measurements have been made to include the entire lobe. Nevertheless, a moderate increase in surface brightness of the hotspot as the jet density decreases is predicted.
   
The precession angle has a significant effect on the evolutionary tracks, as shown in Fig.\,\ref{fig_zba_zbd_zbg_emission-class}. 
 Panels (a) and (b) demonstrate that the length  indicators, as well as the LB Index, reverse in value once the precession has significantly spread  the beam momentum.  In fact, the brightest region becomes associated with the location where the jet is impacting upon itself. 
Rather than advancing, the hotspot recedes and the LB Index may even fall below 0.5 for the highest precession angle.

\begin{figure*}
   \begin{tabular}{cc}
\hspace{-1.2cm} \includegraphics[height=0.32\textheight]{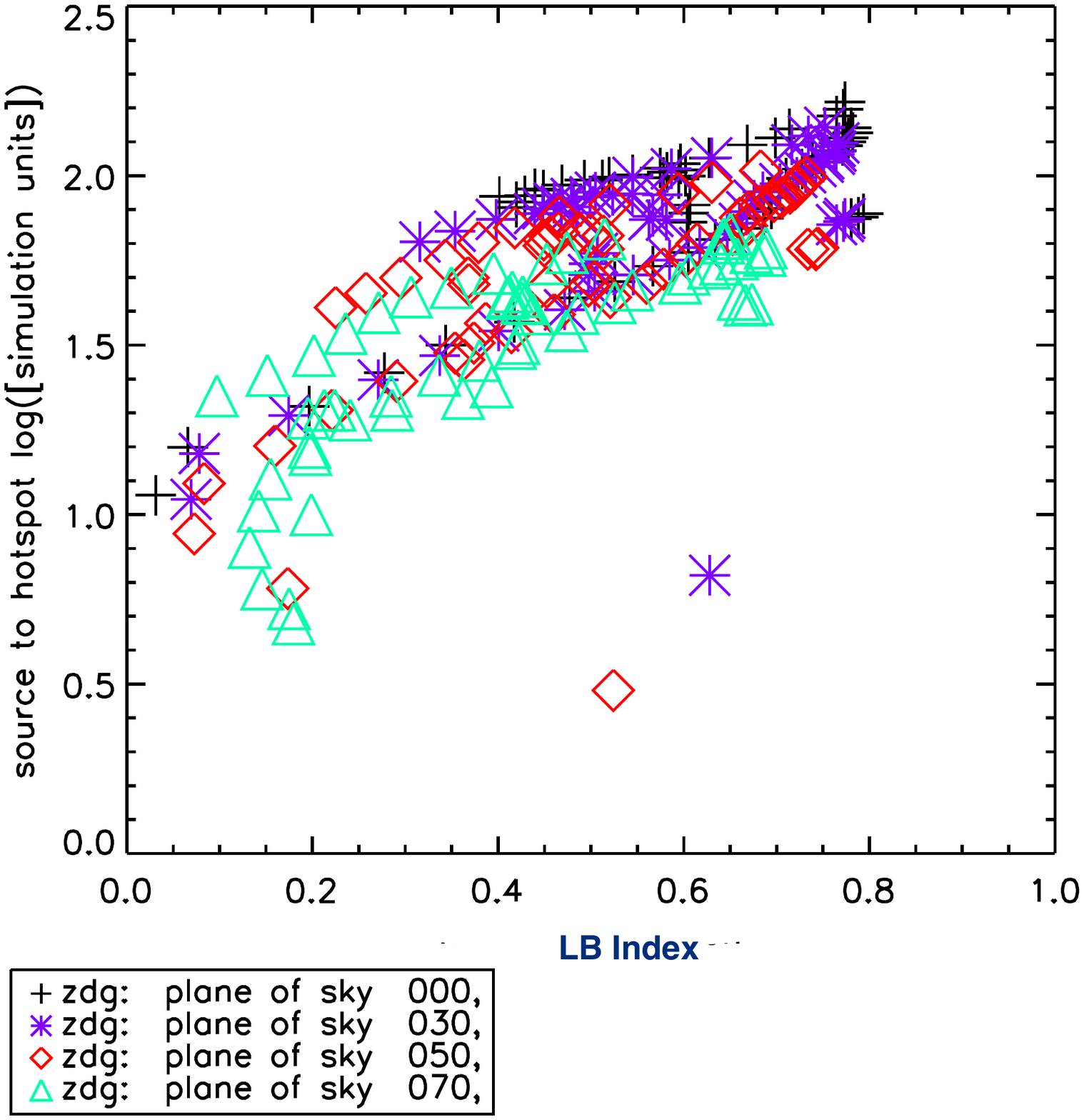} & \hspace{-3cm}
\includegraphics[height=0.32\textheight]{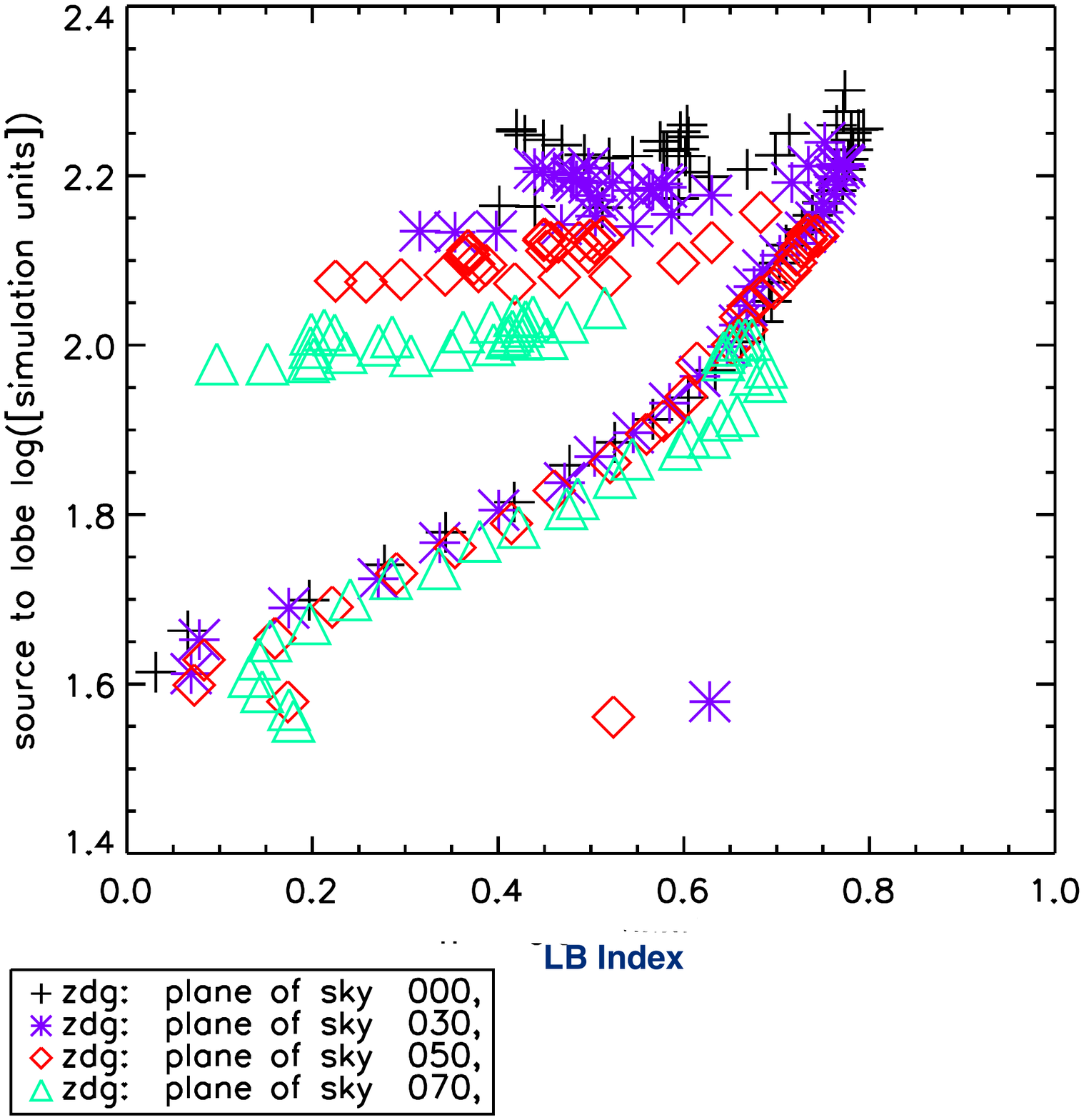} \\ (a) & (b) \\ [6pt]
\hspace{-1.2cm}   \includegraphics[height=0.32\textheight]{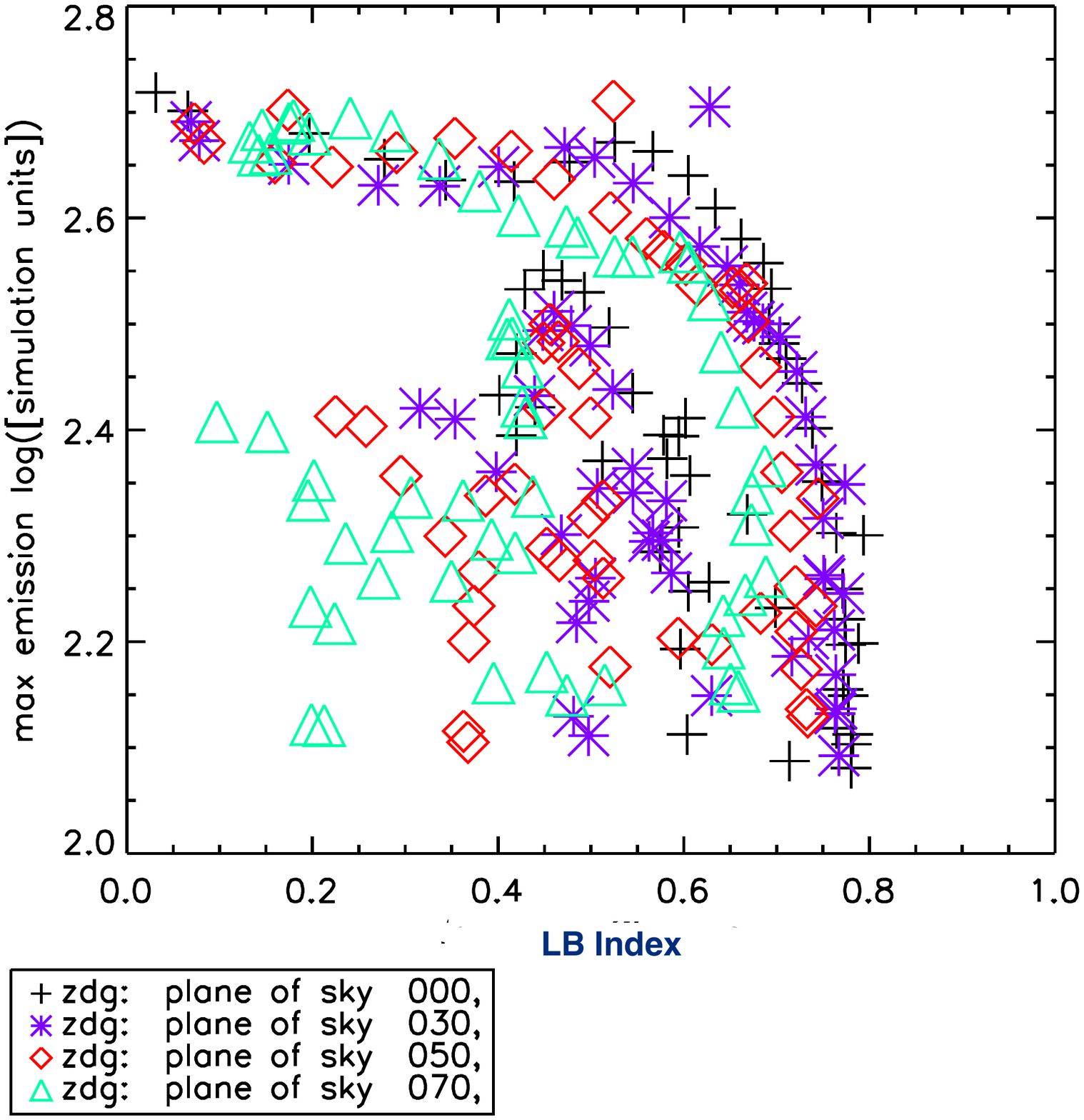}  & \hspace{-3cm}
\includegraphics[height=0.32\textheight]{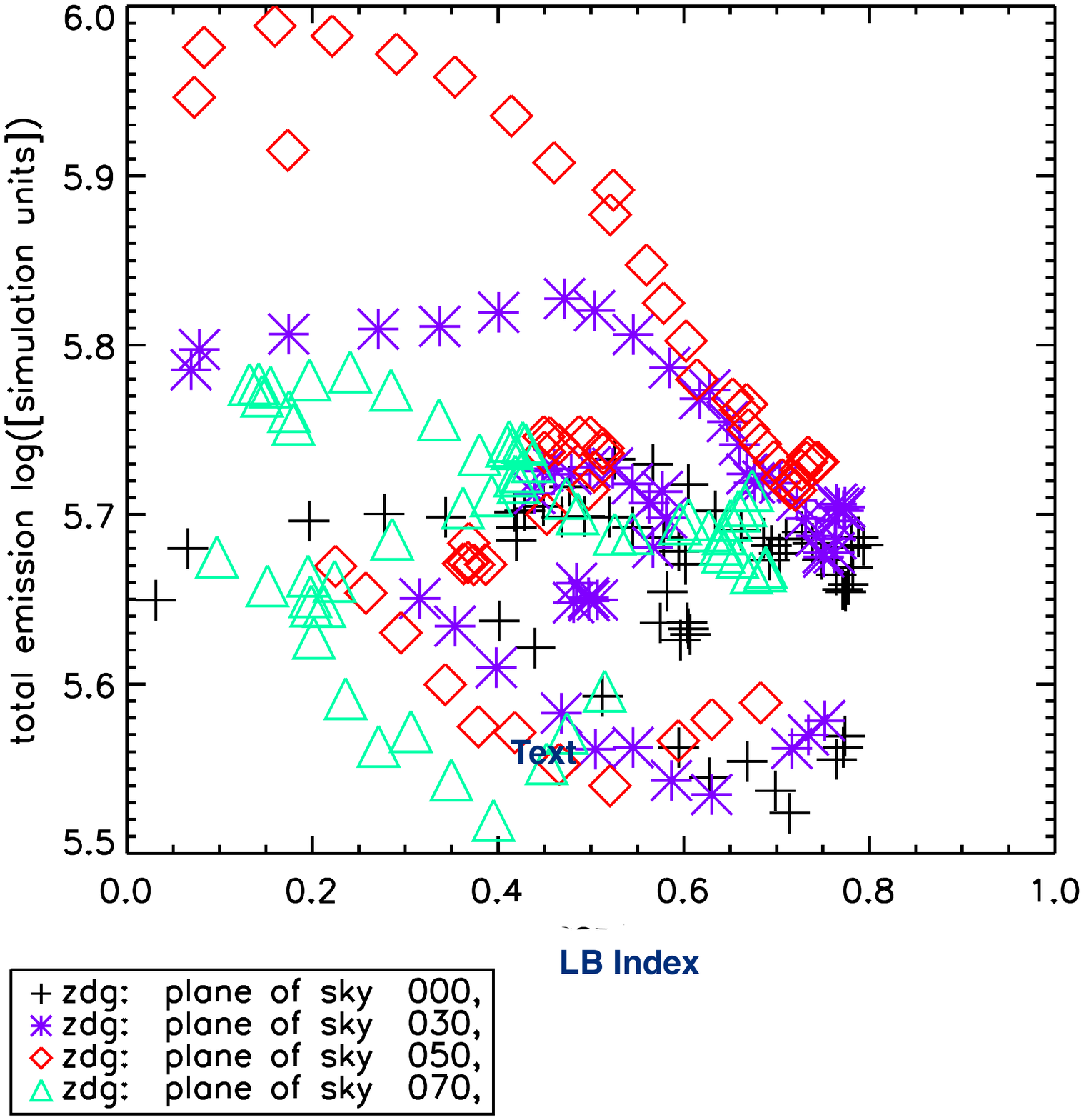} \\ (c) & (d) \\ [6pt]   \end{tabular}
  \caption{Effects of the viewing angle on the four major diagnostics for the 20\degsy precessing jet. The indicated viewing angles are those between the jet axis and the plane of the sky. So 0\degsy is perpendicular to the observer and 90\degsy would be propagating to the observer. Panel (a) radio hotspot distance from source,  (b) maximum expansion distance from source, (c) maximum emission and (d) total radio emission with 5\% hotspot cut off.}
   \label{fig_zbgs_emission-angle}
\end{figure*}

The LB Index is very sensitive to the angle of the axis to the plane of the sky. This can be seen qualitatively in the radio emission tiles in Fig.\,\ref{fig-sync-tile-zdg-oa_000_010_030_050_070_090}. Fig.\,\ref{fig_zbgs_emission-angle} shows the breakdown of the emission properties. It is notable that the LB Index can take values as low as 0.2 for angles within 40\degsy  of the line of sight. Note also that the variation in radio brightness and power remains within narrow bands. They are indeed influenced by the intrinsic three dimensional nature of the hotspot and the 5\% cut-off imposed as the detection limit.

\begin{figure*}
   \begin{tabular}{cc}
\includegraphics[height=0.31\textheight]{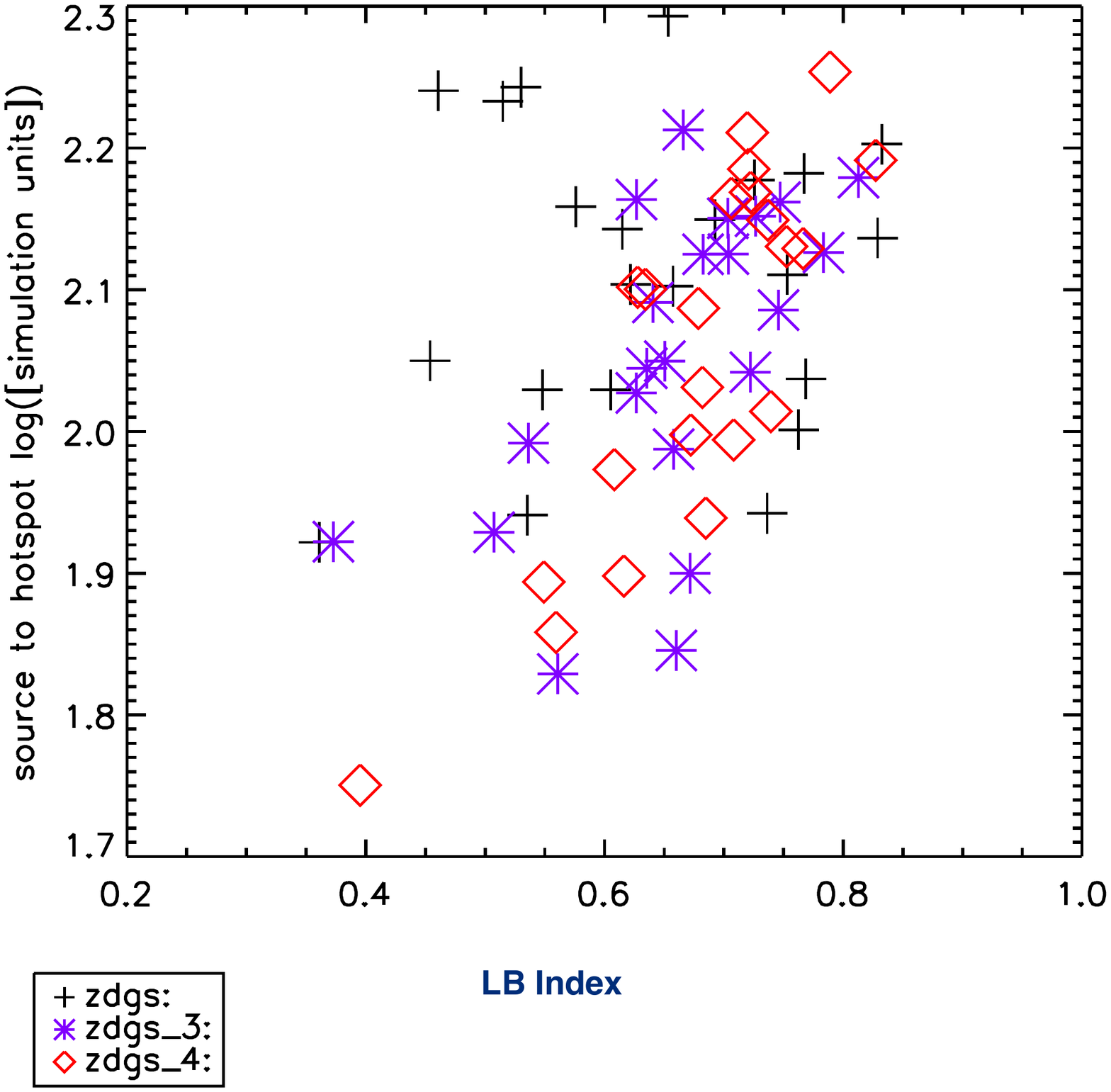} & \hspace{-3cm}\includegraphics[height=0.31\textheight]{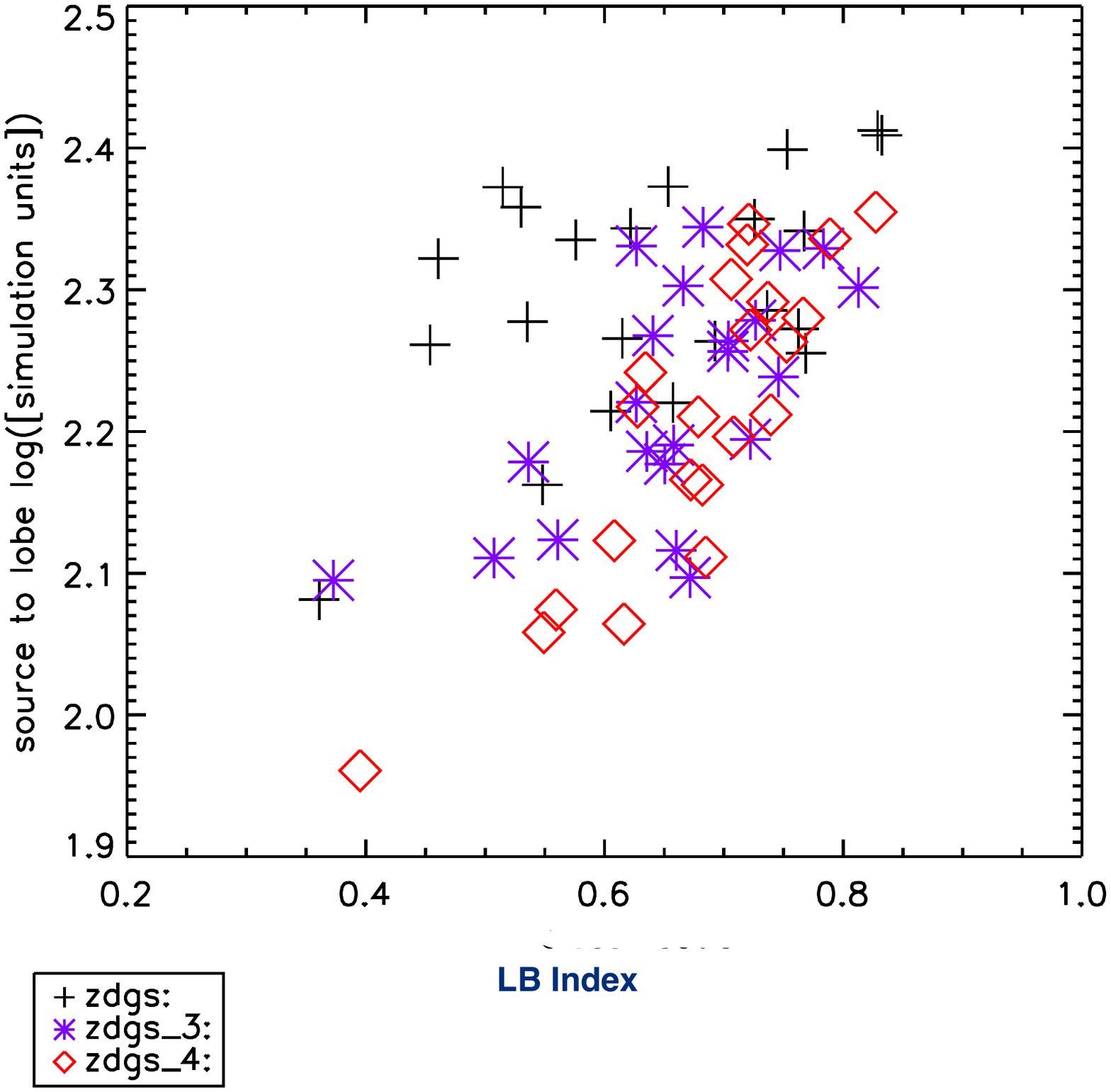} \\ (a) & (b) \\ [6pt]
\includegraphics[height=0.31\textheight]{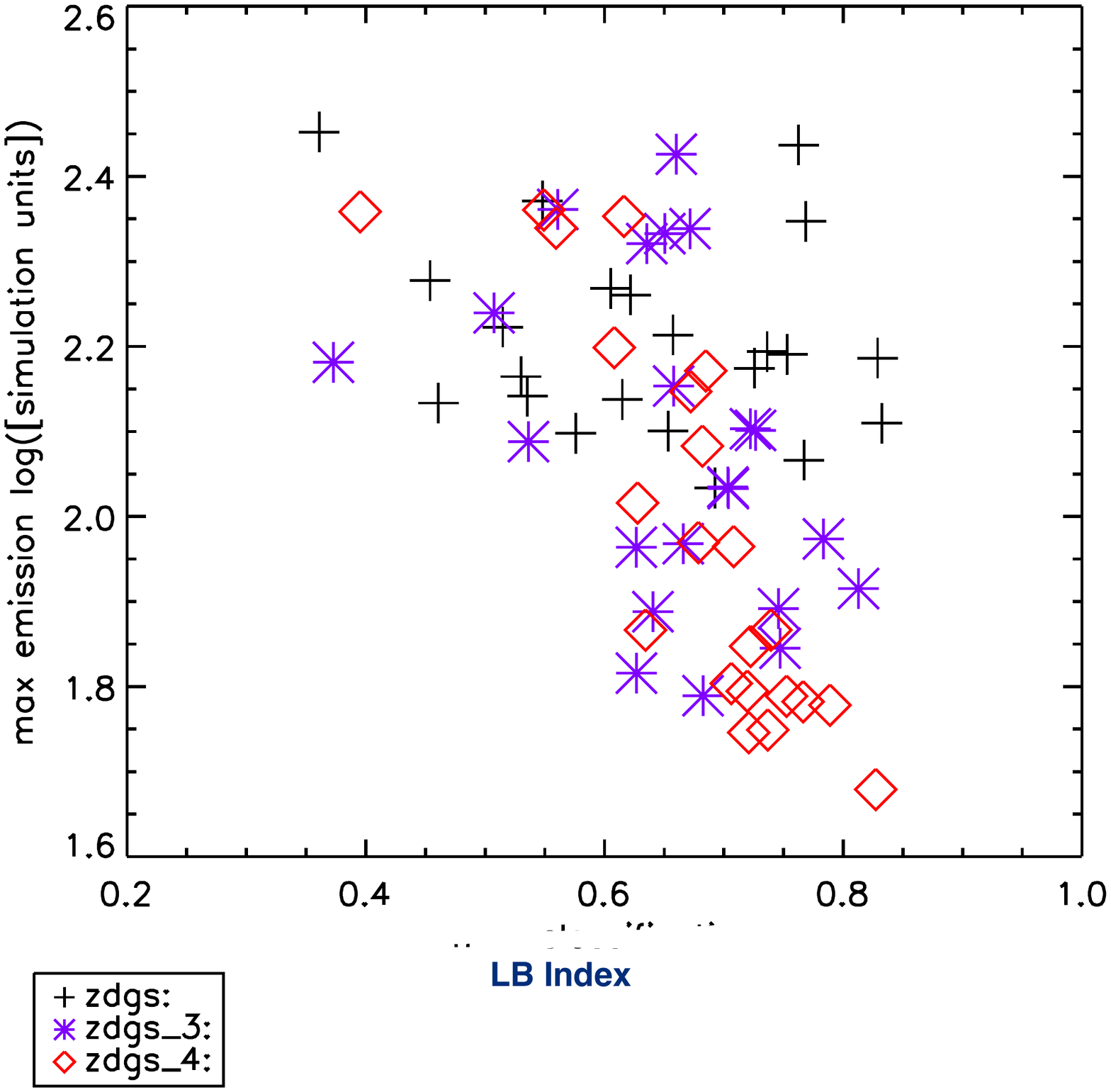} & \hspace{-3cm}\includegraphics[height=0.31\textheight]{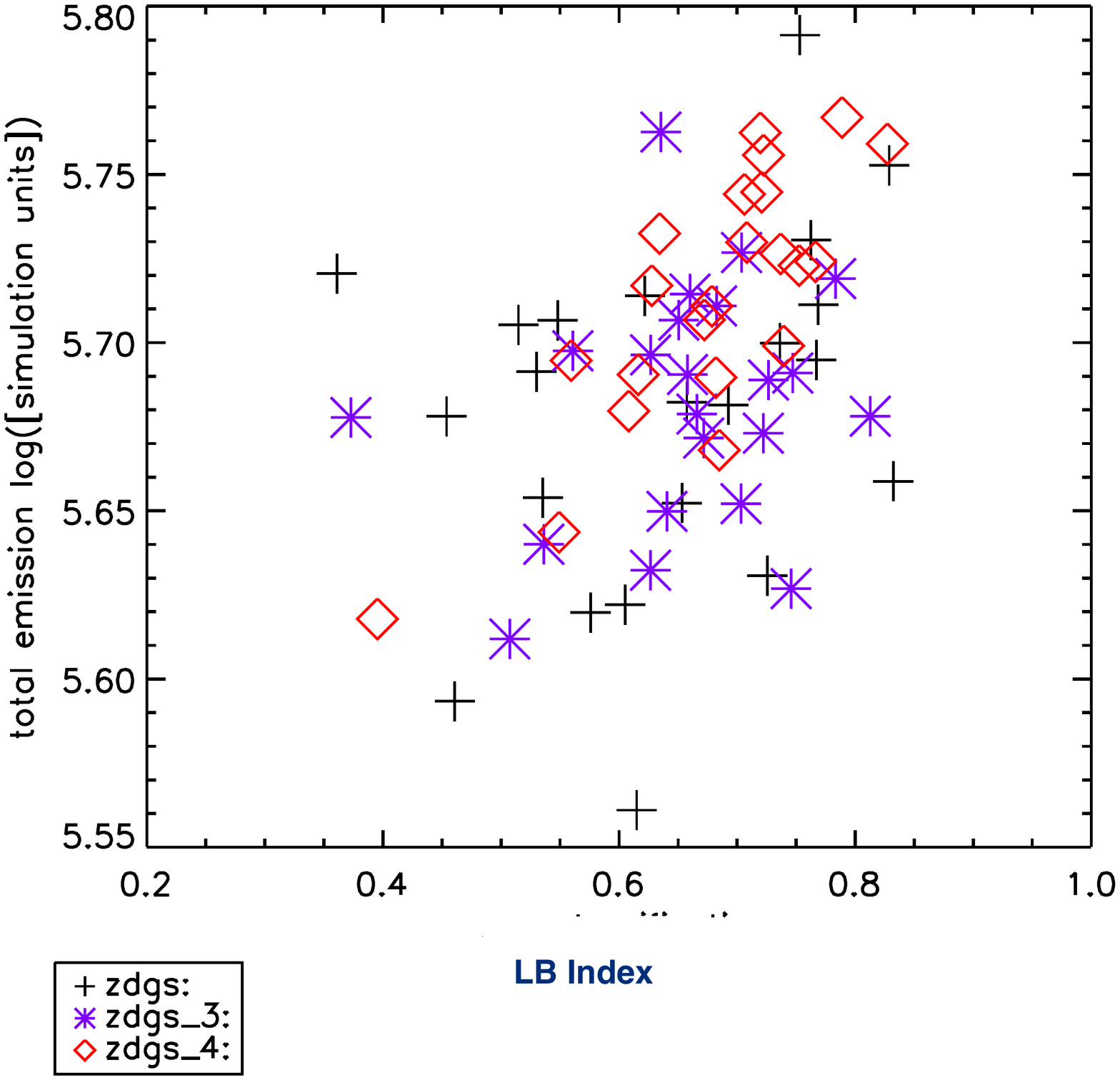} \\ (c) & (d) \\ [6pt]   \end{tabular}
  \caption{The rate of precession, from the default (zdgs),  twice the rate (zdgs\_3) and four times the rate (zdgs\_4) are displayed which demonstrate the
  scattered distributions of the properties and the Limb Brightening Index.
  Panels display  (a)  the radio hotspot distance from the source, (b) maximum expansion distance from source with 5\% hotspot cut off,
  (c) the maximum surface brightness and   and (d) the total radio emission. These data were derived from a 300$^3$ run with an extended staggered grid (run zdgs) to check for boundary issues.}
   \label{fig_zdgs_zdgs_3_zdgs_4-emission-class}
\end{figure*}

To this point, the greatest effect on the LB Index is the precession angle. The next step is to study how the rate of precession changes this, as was illustrated
  in the images of Fig.\,\ref{zdg_precession_rate}.    Increasing the rate of precession causes the LB Index to become highly variable as shown in 
  Fig.\,\ref{fig_zdgs_zdgs_3_zdgs_4-emission-class} (a) \& (b).  The classification is not predictable although lying in the range 
  0.4\,--\,0.8.
     The hotspot brightness can also vary considerably for high precession rates   with a trend for a brighter hotspot when it is located nearer the source. 
   However, the total emission, as seen from Fig.\,\ref{fig_zdgs_zdgs_3_zdgs_4-emission-class}(d), is quite uniform during the evolution on noting the narrow range in values in this panel.

The appearance of a remote radio galaxy of the same size can be simulated by smoothing the images. We add a Gaussian blur to the calculated emission maps. This compensates for the lower spatial resolution while maintaining the same angular resolution but with no account taken for the lower received flux.
Moreover, a Gaussian blur does not take into consideration the interferometric constraints on  short and long baseline spacings.

\begin{figure*}
   \begin{tabular}{cc}
\hspace{-1.2cm} \includegraphics[height=0.33\textheight]{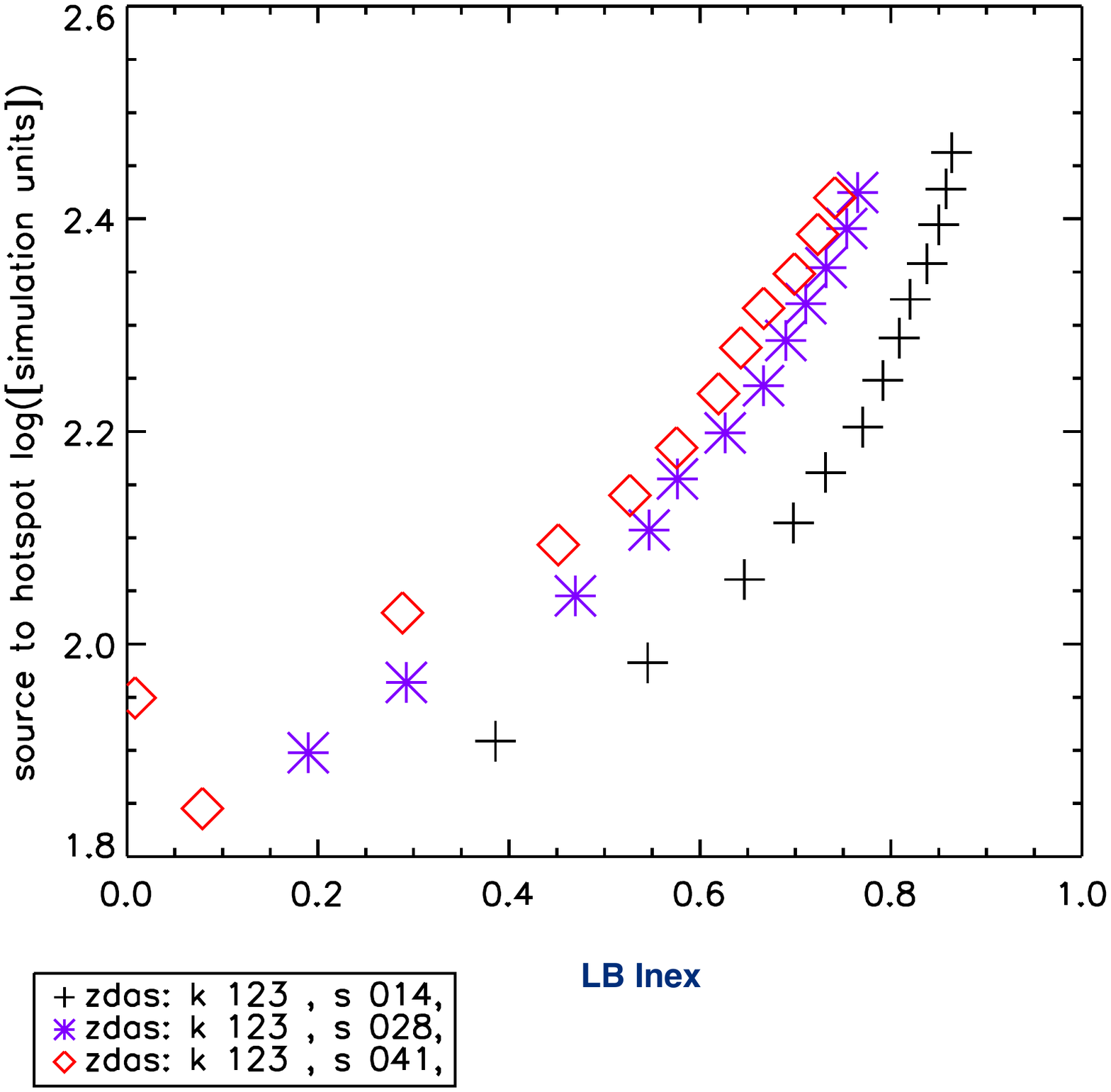} &  \hspace{-3cm}  
\includegraphics[height=0.33\textheight]{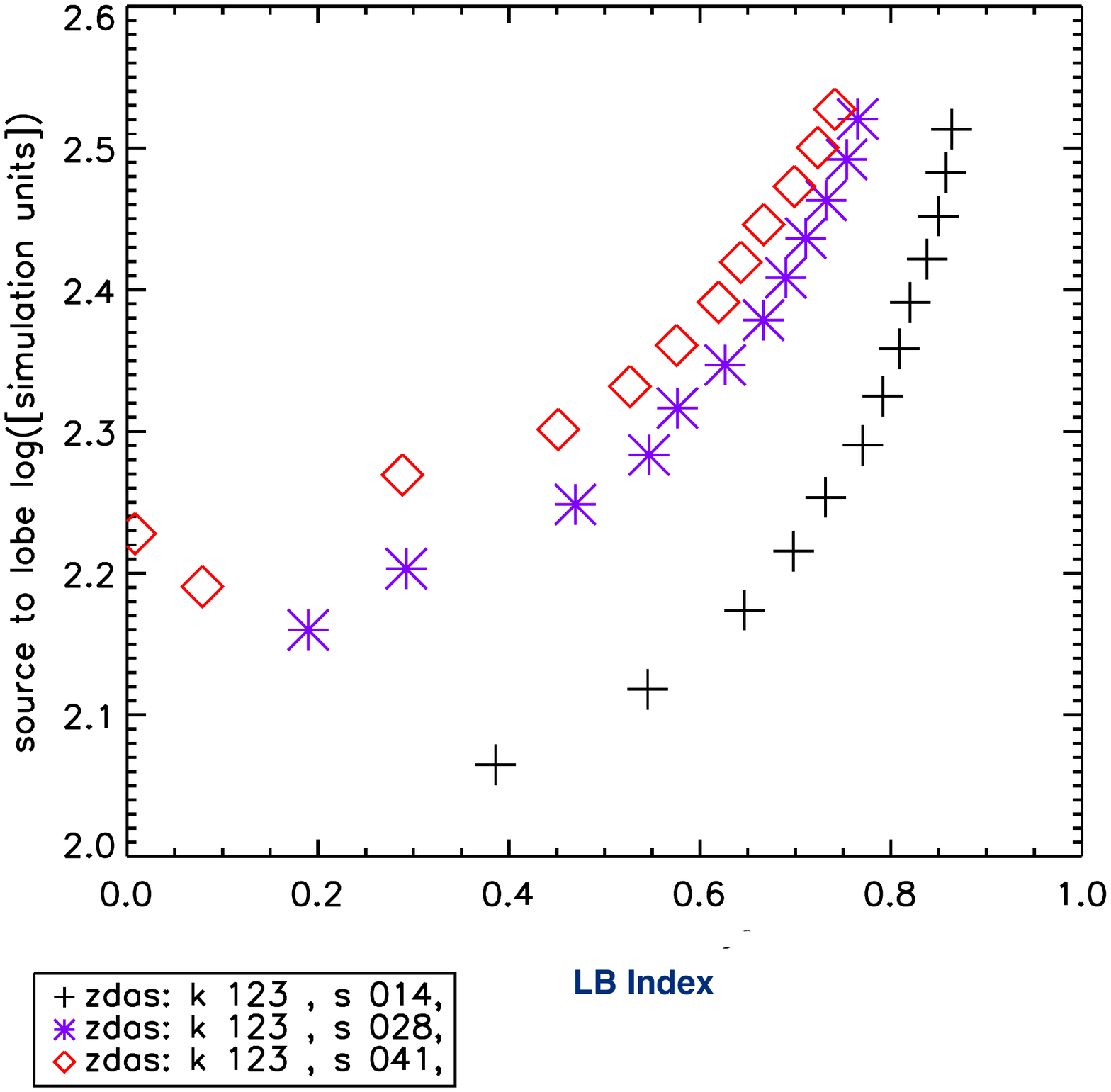} \\ (a) & (b) \\ [6pt]
\hspace{-1.2cm}  \includegraphics[height=0.33\textheight]{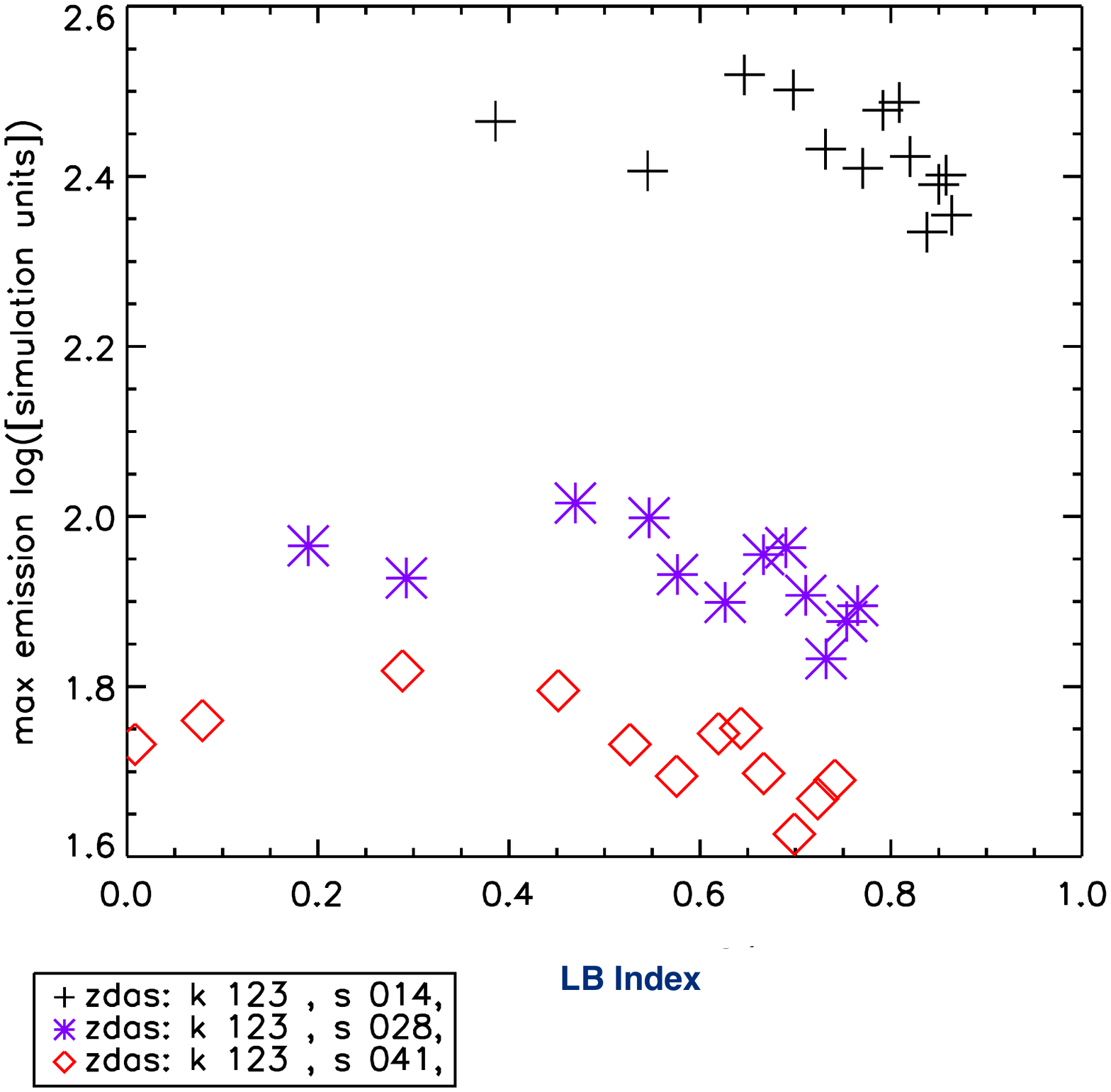}   &   \hspace{-3cm}  
\includegraphics[height=0.33\textheight]{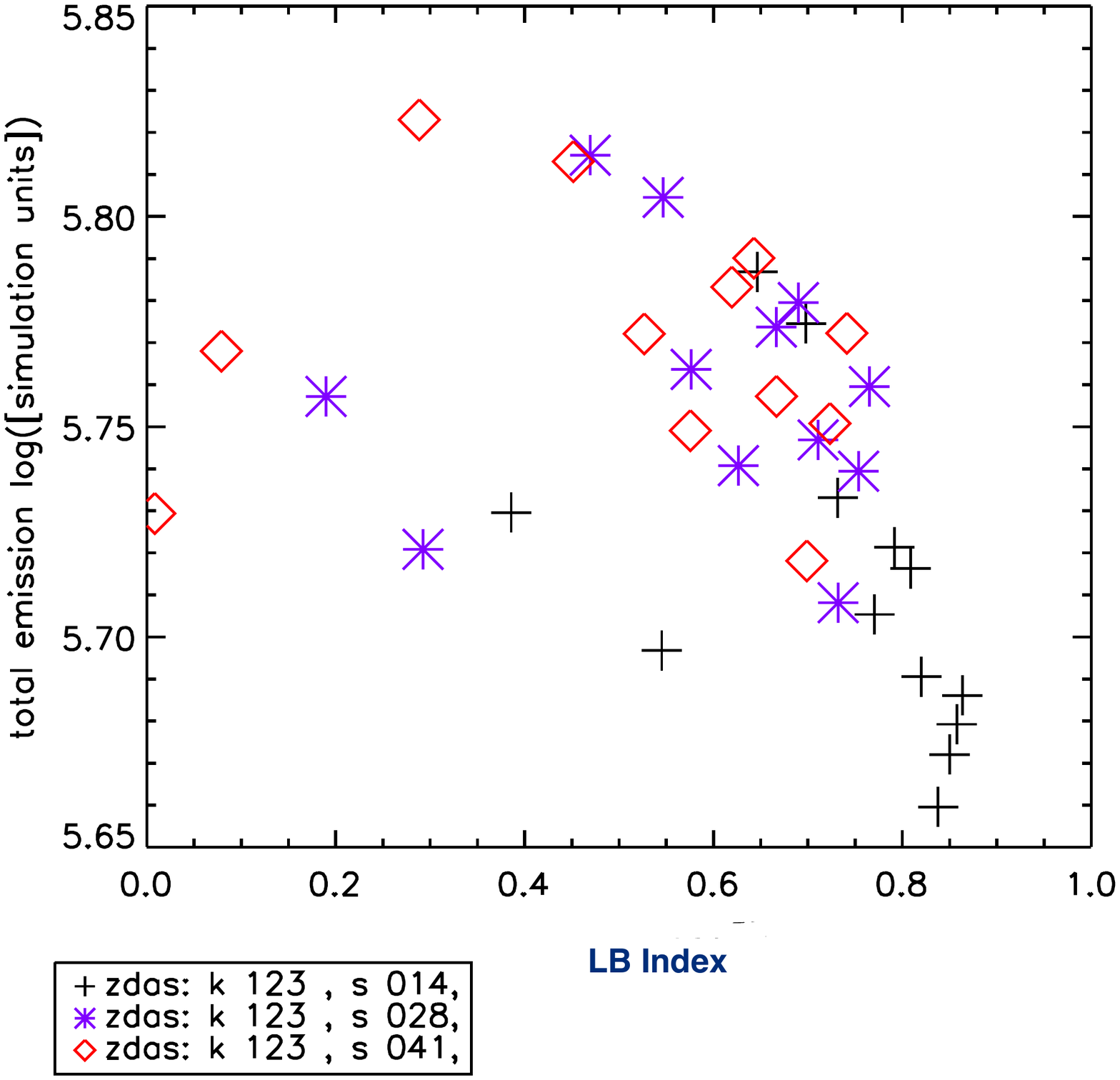}    \\ (c) & (d) \\ [6pt]   \end{tabular} 
 \caption{Effects on classification due to increase in Gaussian smoothing (to simulate redshift) for the straight jet with jet density 0.1 of the ambient density. The smoothing indices are in simulation units with standard deviation s and kernel k. The panels show (a) radio hotspot distance from source, (b) maximum expansion distance from source with 5\% hotspot cut off, (c) peak radio emission, (d) total radio emission.}
   \label{fig_zdas_redshift-emission-graph}
\end{figure*}

In Fig.\, \ref{fig_zdas_redshift-emission-graph}, we have added Gaussian smoothing to the emission for the standard run of a Mach 6  straight jet (zda) with the indicated standard deviation and kernal sizes.
 The first result is that the LB Index is reduced typically by a factor of 0.2, an effect which can transform the FR type in the early expansion phase. The hotspot distance is reduced but the source size is unchanged.
The peak emission is significantly reduced by the smoothing, as expected. In turn, this raises the total emission somewhat because the dynamic range has been held constant, allowing regions of diffuse emission to now contribute.

\begin{figure*}
   \begin{tabular}{cc}
\hspace{-1.2cm} \includegraphics[height=0.33\textheight]{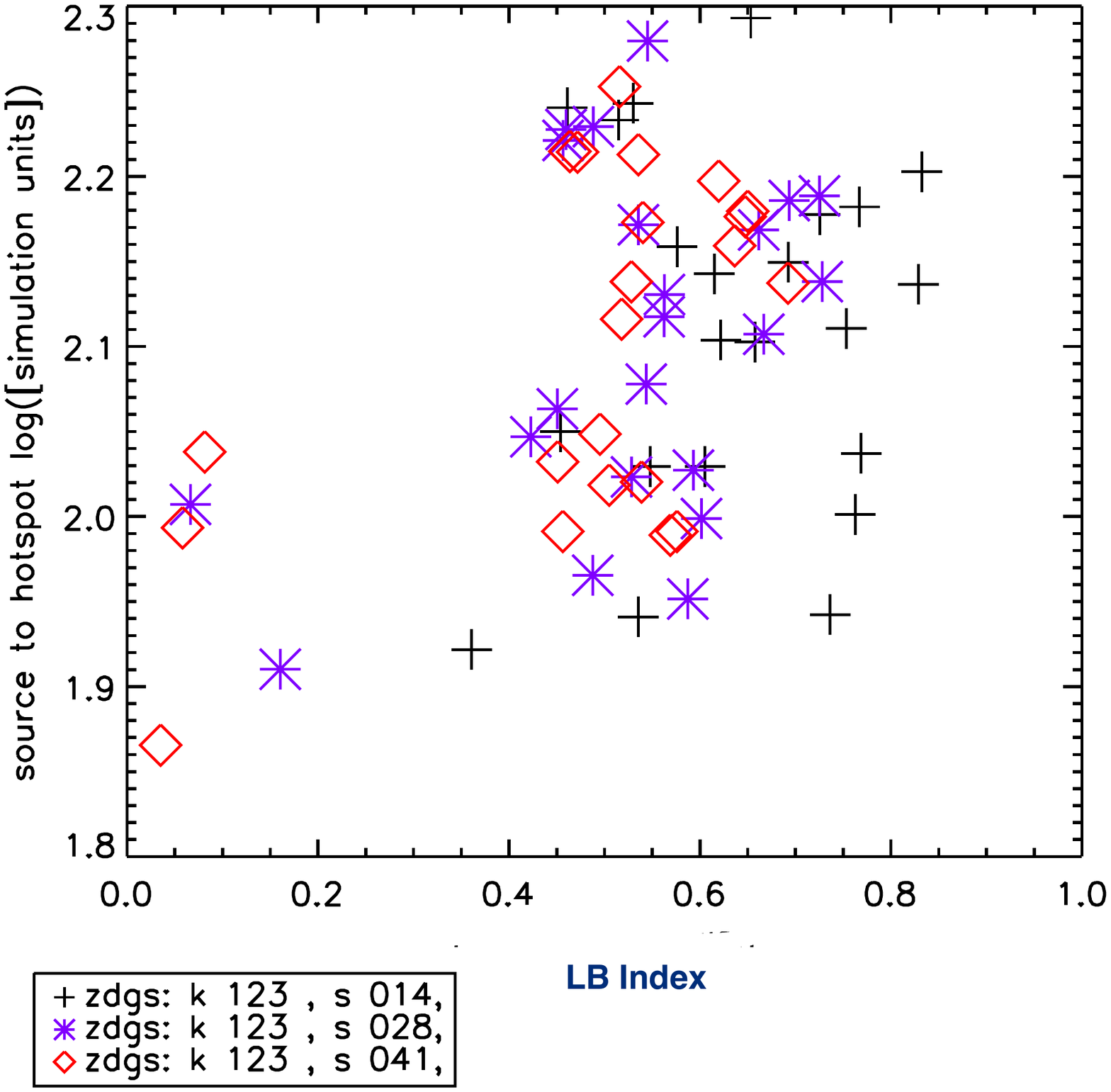} & \hspace{-3cm}  
\includegraphics[height=0.33\textheight]{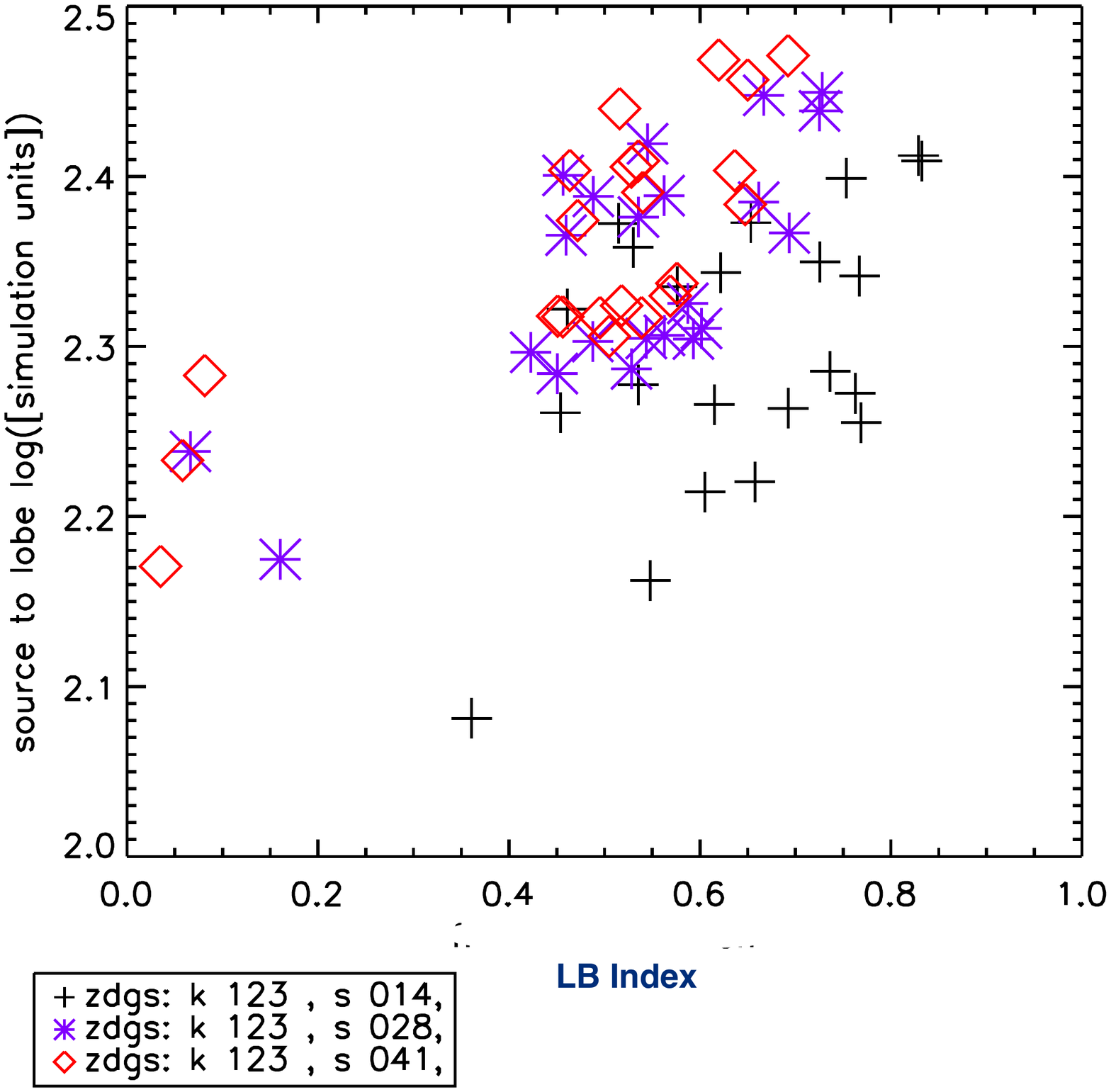} \\ (a) & (b) \\ [6pt]
\hspace{-1.2cm}  \includegraphics[height=0.33\textheight]{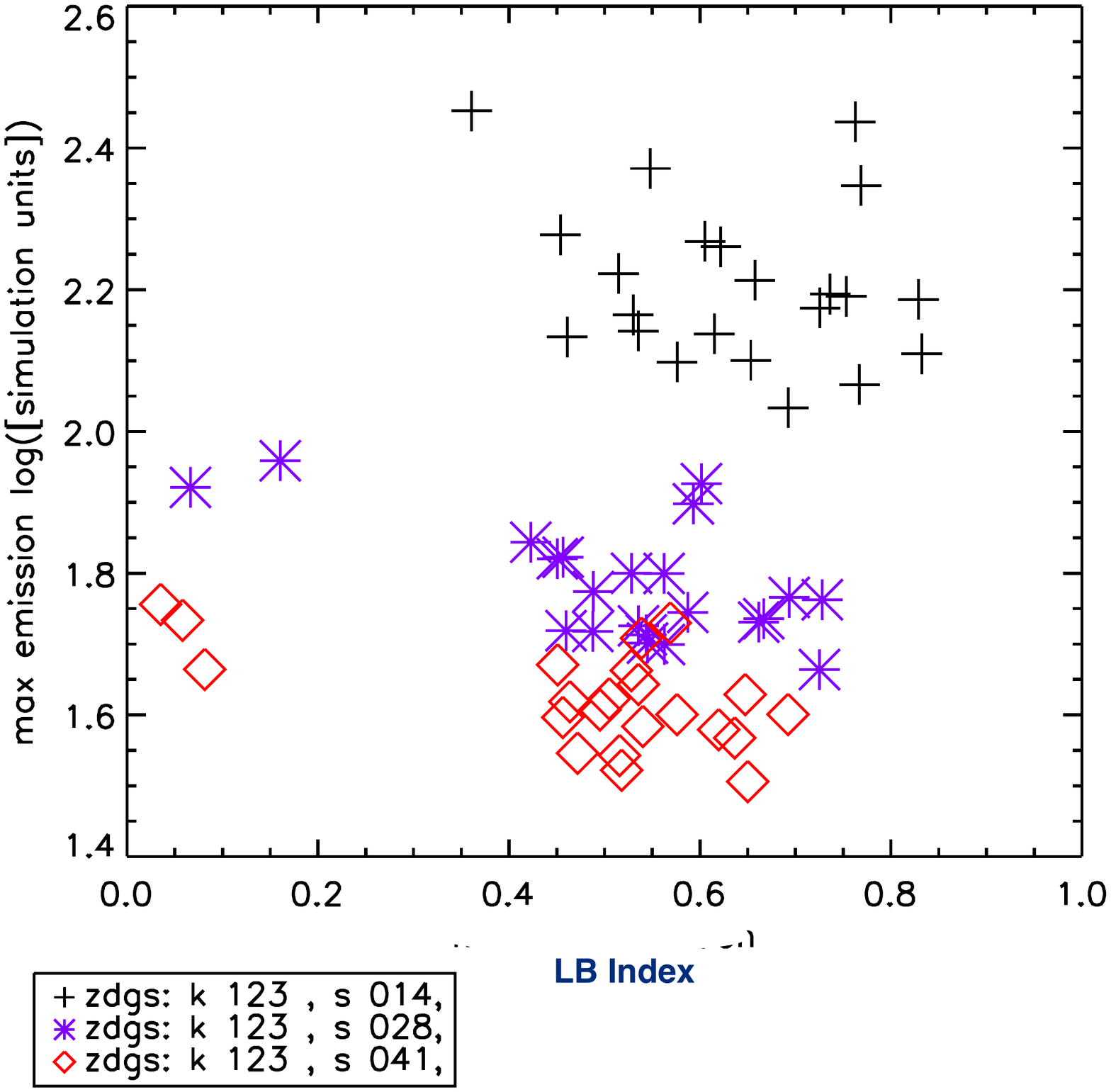} &  \hspace{-3cm}  
\includegraphics[height=0.33\textheight]{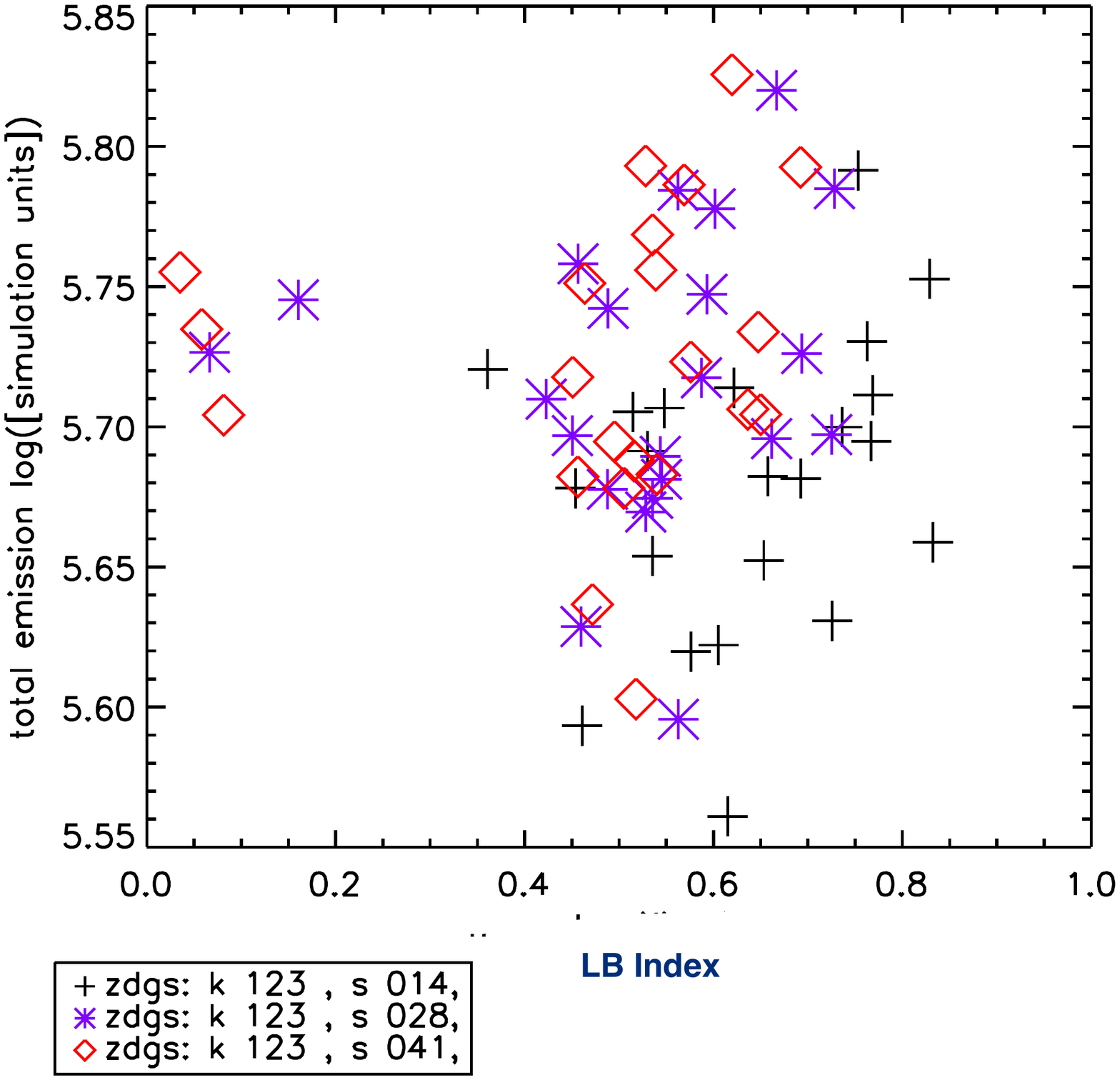}    \\ (c) & (d) \\ [6pt]   \end{tabular}
\caption{Effects on classification due to increase in Gaussian smoothing (to simulate redshift) for the 20$^\circ$ precessing jet with jet density 0.1 of the ambient density. The smoothing indices are in simulation units with standard deviation s and kernel k. The panels show (a) radio hotspot distance from source, (b) maximum expansion distance from source with 5\% hotspot cut off, (c) peak radio emission, (d) total radio emission.}   
\label{fig_zdgs_redshift-emission-graph}
\end{figure*}

  Fig.\,\ref{fig_zdgs_redshift-emission-graph}  displays the same results for the wide precessing jet derived from tun zdg. This demonstrates the large reductions possible in the LB Index   and potential switch in FR type.
  Large variations in source dimensions also occur as shown in Panels (a) and (b). The peak emission shown in panel (c) displays a more systematic trend, which indicates that the smoothing alters the classification for specific times rather than overall. 

\begin{figure*}
   \begin{tabular}{cc}
 \hspace{-1.8cm}   
 \includegraphics[height=0.35\textheight]{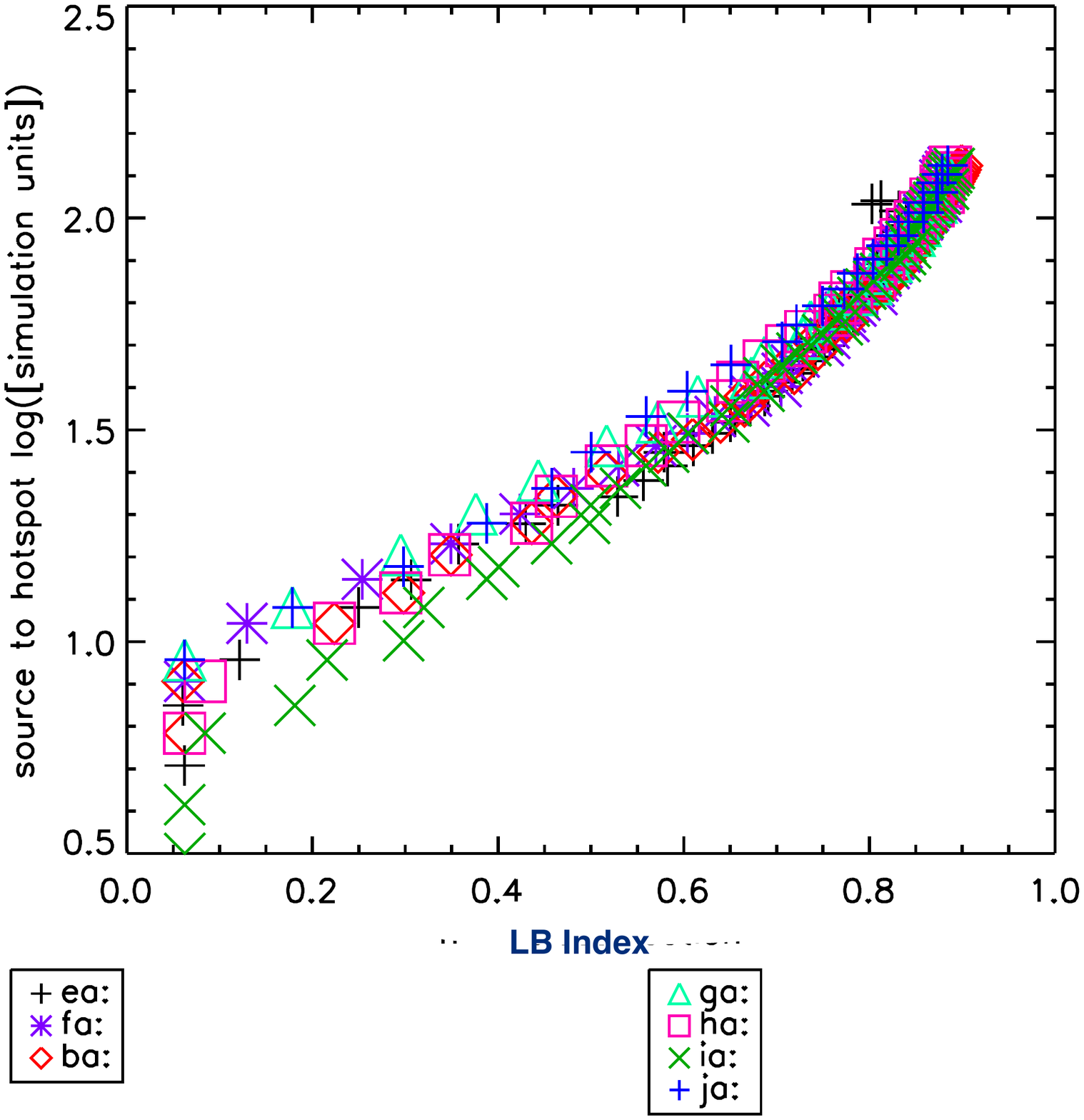} &  \hspace{-3cm}
 \includegraphics[height=0.35\textheight]{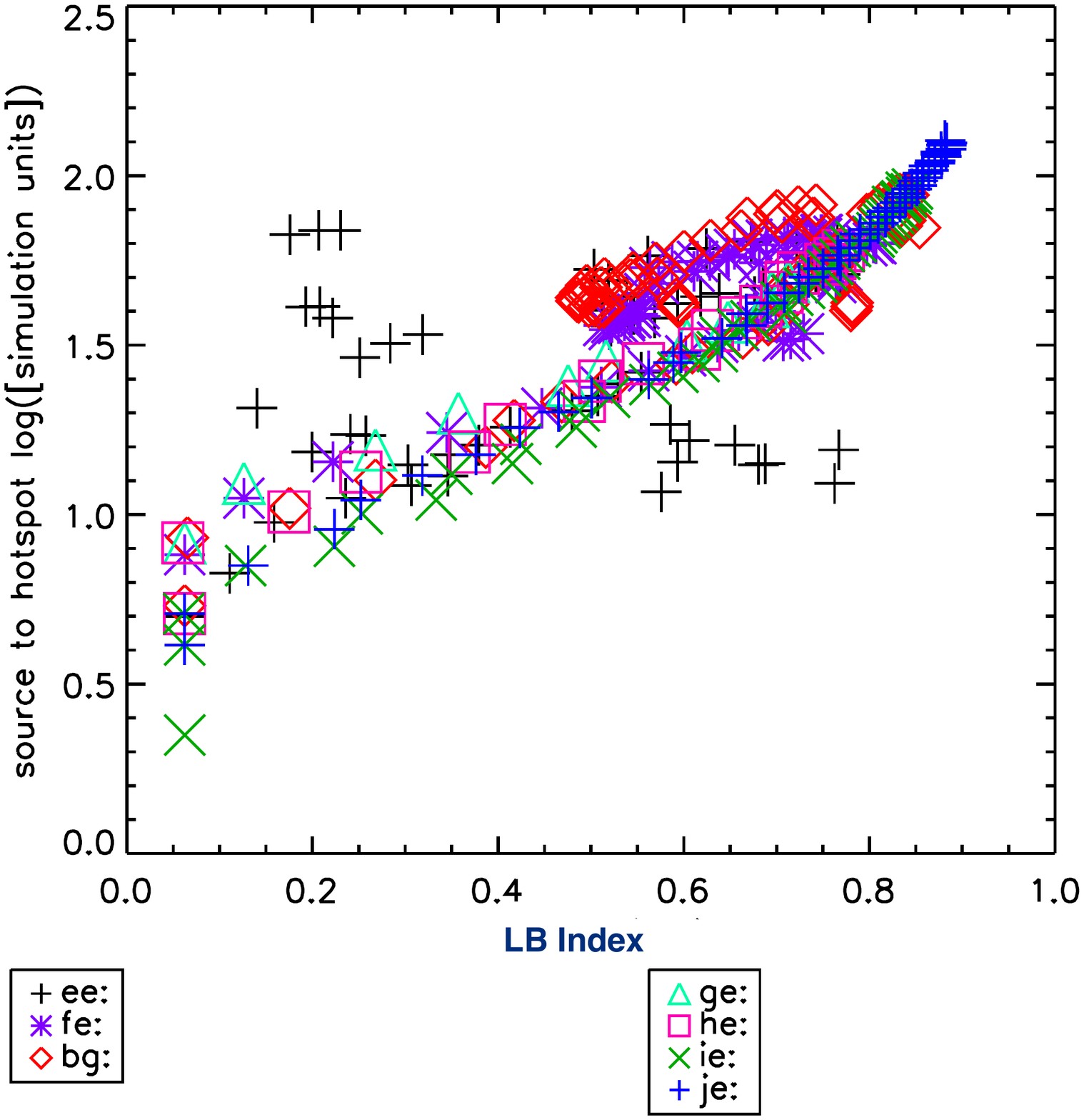} \\ 
(a) M=2 (ea) to M=48 (ja)  & (b)  M=2 (ee) to M=48 (je)  \\ [6pt]
 \end{tabular}
 \caption{{\bf Mach number.} Comparisons of increasing Mach number with 1\degsy precession (left  panel) and 20\degsy precession (right panel). The Mach number runs from M=2 (ea), M=4 (fa), M=6 (ba),
   M=8 (ga), M=12 (ha), M=24 (ia) to M=48 (ja) with similar notation from ee to je for the precessing run. This highlights the source to hotspot distance.}
   \label{fig_mach_1pre_20pre_emission-class}
\end{figure*}

 The dependence on the jet  Mach number is one of the most studied features of extragalactic jets. It is one of the leading explanations of the different type of radio galaxies. Fig.\,\ref{fig_mach_1pre_20pre_emission-class} has a side by side comparison of the increasing Mach number at a precession angle of 1\degsy\,  and 20\degsy. The straight jets simply drill their way through the ambient medium at all Mach numbers and we find no particular trend in classification. 
 It should be remarked that the present set of simulations is restricted in domain size; the development of surface instabilities requires much longer propagation lengths. 
 
 Since all the simulations have the same spatial dimensions of 300$^3$, the faster that the jet propagates, the less time the jet has on the grid to complete or even start precessing. This is evident from the first three 20\degsy \ simulations (ee, fe \& bg), where the location of the hotspot and the maximum lobe distance are erratic compared to their straight jet counterparts. All 20\degsy \ simulations do show some degree of stunted growth while at higher Mach numbers the propagation speed is starting to overshadow any affect that precession would add.

\begin{figure*}
 \begin{tabular}{ccc} 
   \hspace*{-2.3cm} \includegraphics[width=0.52\textwidth]{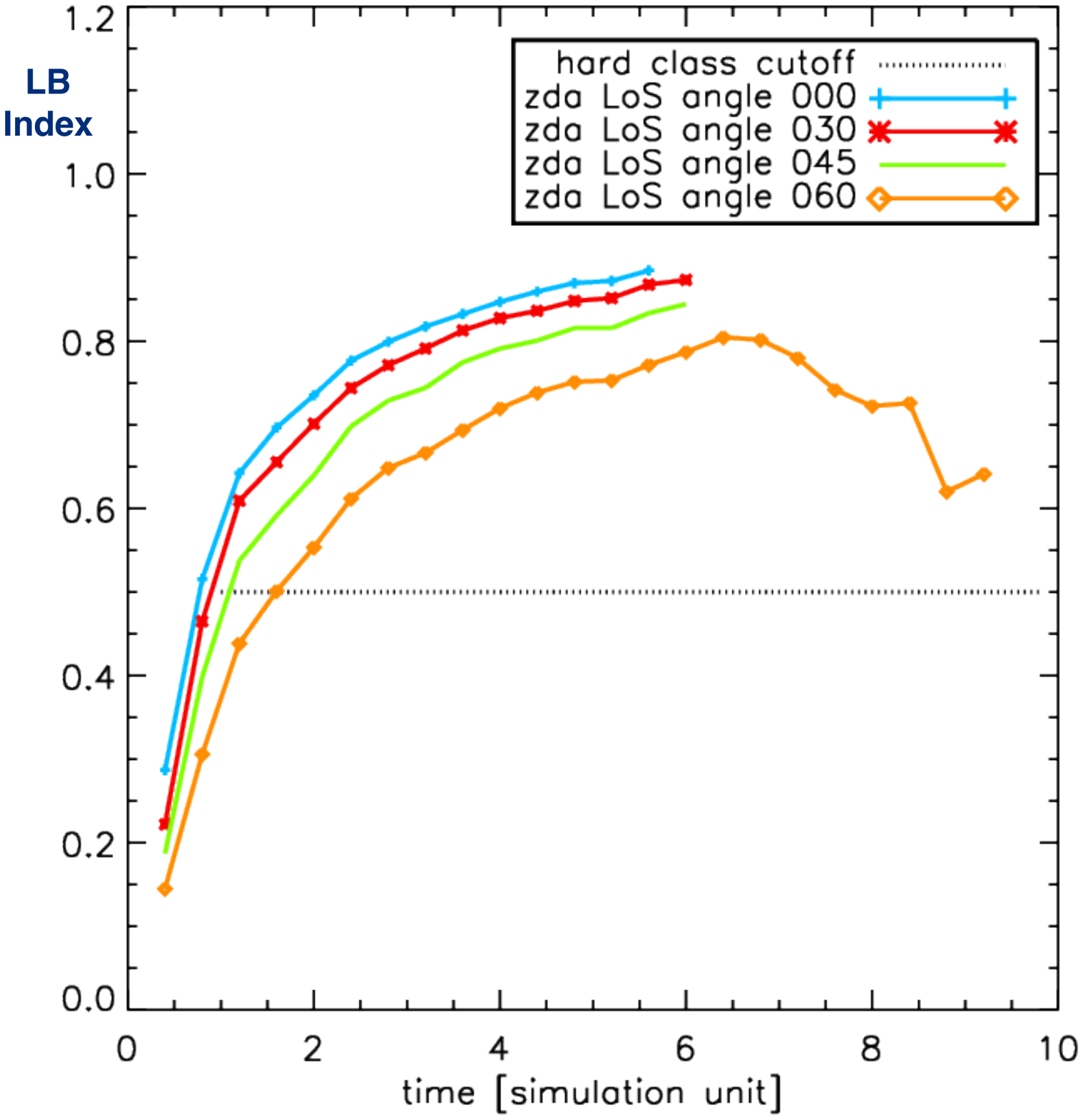} &  \hspace{-3.6cm} 
     \includegraphics[width=0.52\textwidth]{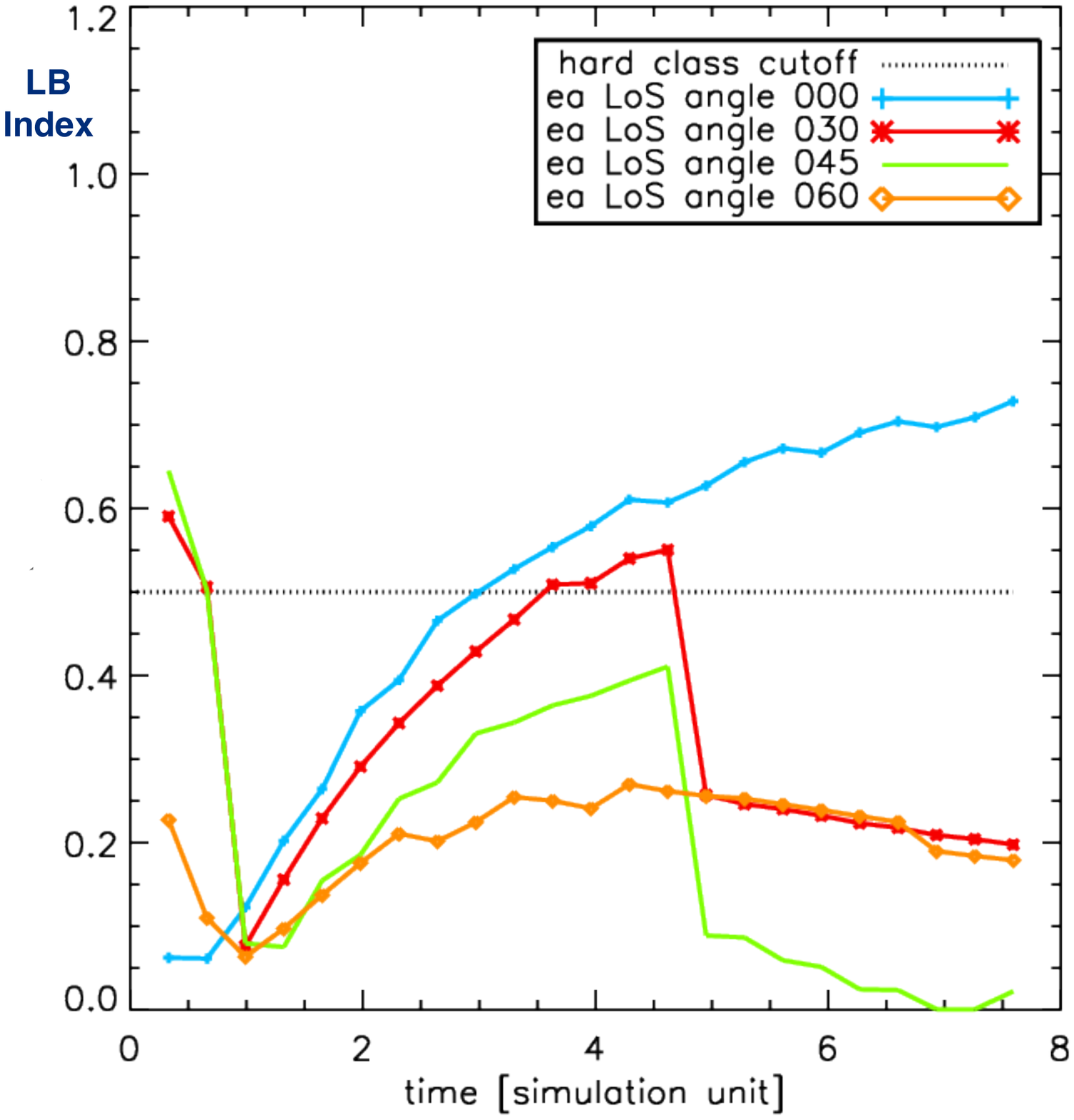} &  \hspace{-3.6cm} 
       \includegraphics[width=0.52\textwidth]{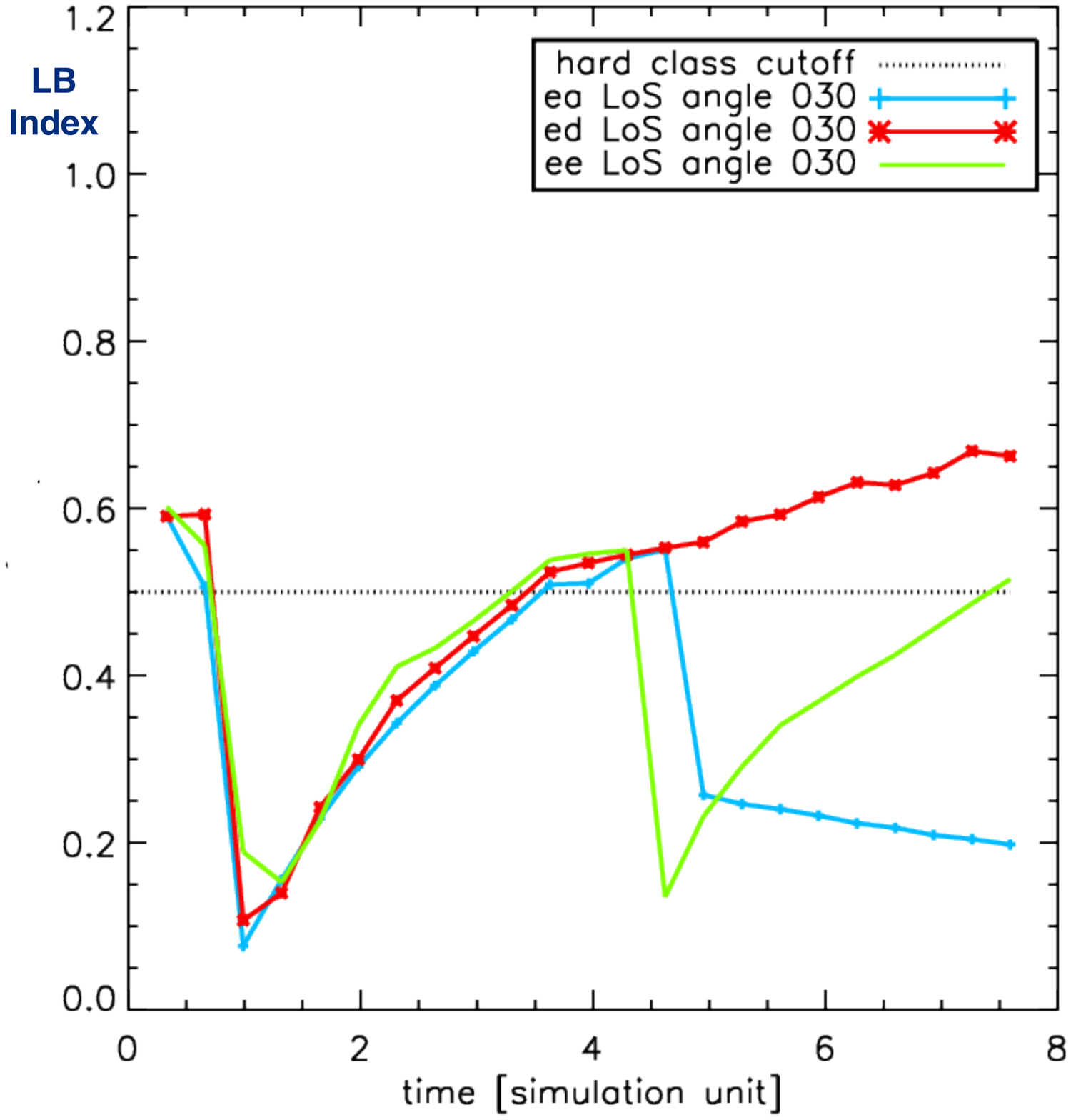} 
 \end{tabular}
 \caption{The LBI Index versus time for the M=6 (left panel) and M=2 (middle panel) simulations of jets essentially straight (1$^\circ$ precession) with a set of viewing angles, and the precessing M=2 jet with precessing angle of
 1$^\circ$ (ea), 10$^\circ$ (ed) and  20$^\circ$ (ee) with a fixed viewing angle of 30$^\circ$ (right panel).}
 \label{frtype-summary}
\end{figure*}

\section{Discussion}

We have presented simulated radio  images  of radio galaxies formed through supersonic hydrodynamic jets. Only the synchrotron  process has been considered without consideration for radiative losses or particle acceleration.
Motivated by the unknown nature of powerful radio galaxies in the distant Universe, we performed straight and precessing jet simulations. In the era in which galaxies form and their central black holes interact and merge, we could expect that the jet direction may be a stronger time variable than at present.  We may also expect relatively high Mach numbers since the detected radio sources are powerful, although this can be debated. Distant radio galaxies can be found by searching for sources with a significantly steep radio spectrum \citep{2008A&ARv..15...67M,2018MNRAS.475.5041S} although the reason for this is quite complex
\citep{2012MNRAS.420.2644K}.
  
We begun this work by investigating the dependence of the radio structure on the radio spectral index by probing  flux-pressure relationships which
may correspond to values of the spectral index of 0.7, 1.0 and 1.3. The latter steep index would be associated with ultra-steep spectra and is used as a tool to discover high redshift radio galaxies \citep{1999ApJ...518L..61V,1997A&A...326..505R,2000A&AS..143..303D,2002A&A...394...59D}.
A constant  ultra-steep spectrum across a lobe is shown here to (1)  considerably diminish the bridge of emission trailing back from the hotspot in an FR\,II and (2) reduce the detectable area of diffuse emission in FR\,Is. However, we conclude that there is a clear  influence on the source designation.

To quantify the differences, we introduced a classification index which is the ratio of the peak emission distance to the total source size (of a single lobe). Here we refer to this as the LB Index and it is shown in Fig\,\ref{frtype-summary} for the slightly perturbed straight jets with Mach numbers of 6 (left panel) and 2 (middle panel). This figure clearly demonstrates  contrasting structures at all viewing angles with the M=2 jets almost exclusively in the FR\,I category (below the 0.5 line) when not in the sky plane.

The cause of the difference between FR types has been well discussed in the literature 
\citep{2008A&ARv..15...67M, 2016A&ARv..24...10T}.  \citet{2019ApJ...871..259G} propose a sequence involving the FR\,0 as an intermediary stage. However, 
\citet{2019MNRAS.488.2701M} find that, within the large LoTSS survey,  radio luminosity does not reliably predict the FR type with the host galaxy properties also important. Jets of low power may still  form hotspots in lower mass hosts.

From this study, the evolutionary sequence from FR\,I to FR\,II  is not suggested  even though the figures may indicate this.  As a jet enters the grid, it creates a reverse shock in the jet and a broad bubble.  However, this is a temporary low-resolution phenomenon, associated with the technical side of initiating the run.
Hence, the initial evolution from low to high LB Index must be seen in this context. On the other hand, we do find evidence for transition from FR\,II to FR\,I in some particular circumstances with both orientation and resolution as important factors.

When a high precession rate is  introduced,  less material is diverted into a single channel of least resistance. Therefore, there is some clear change  of classification  predicted as the head of the jet
  shifts position.  The index was shown to vary quite erratically for precessing jets but generally to remain above about 0.4 for standard runs with a Mach number of 6. However, there are two effects which will act to reduce the LB Index: the viewing angle and the rate of precession.  Interestingly, slow precession has the greatest influence in the reduction of the LB Index with fast precession tending to  behave as if the jet has a wide but constant direction (hence more like an FR\,II).
 
  Besides direction, jets are likely to vary in power.
  Considerable work has also been expended in switching the jet on an off throughout its evolution. Some works in this area include that of \citet{2010ApJ...710..180O,2014MNRAS.439.3969W,1991ApJ...369..308C}. This is a mechanism that could be crucial to early-universe radio galaxies which may experience many mergers of their cores. 

The evolution of the structure of  Giant Radio Galaxies (GRGs), with projected linear sizes exceeding 0.7\,Mpc, may also be strongly
  controlled by intermittancy over periods of order 100\,Myr. 
They typically exhibit FR\,II radio morphologies \citep{1996MNRAS.279..257S}. 
and correspond in their properties to scaled-up versions of the much more common smaller FR\,II sources. However,
there are numerous examples where a hotspot is well recessed from the lobe extremities, cases where 
the jet appears to have been intermittent as well as recently restarted  \citep{2019ApJ...875...88B}  and
the possibility of being switched off for long periods
\citep{2016A&A...585A..29B}. One may thus arrive at a combined giant FR\,II with a newborn FR\,I forming inside. 
These hybrid sources appear similar to some of the synthetic images uncovered here especially where precession has occurred as shown in 
Fig.\,\ref{zdg_precession_rate}. The relevant time scale to generate these images for a GRG
as described by \citet{2016MNRAS.458..558D} is 290\,Myr (for a GRG at 10 simulation time units).  
   
As expected, Mach 2 jets are inherently unstable and so generate FR\,I sources unless the jet is both straight and in the plane of the sky. This is illustrated in the right panel of
  Fig\,\ref{frtype-summary}. The turbulent dissipation of energy within low Mach number jets was postulated by \citet{1985PASAu...6..130B} and is clearly relevant to the radio structure of low-powered radio galaxies.   However, it is also established that FR\,I jets are intrinsically different in composition: -- it is not just a matter of Mach number \citep{2018MNRAS.476.1614C}.    In addition, the jet speed itself is a potential factor as discussed fro the compact FR\,0 sources by  \citet{2019MNRAS.482.2294B},
  
It is conjectured that mechanical feedback from active galactic nuclei is a key  process in the evolution of galaxy clusters \citep[e.g.][]{2012NJPh...14e5023M}. The transfer of energy can suppress  cooling flows in the gas and consequently stop star formation in the central galaxies. The energy transferred far exceeds the work done in creating the space occupied by the lobes because of the global shock front which expands laterally to several times the lobe width, consistent with observations  \citep{2014ApJ...794..164S}. The end result is that most of the bulk energy of the jet is dissipated as thermal energy in to the environment \citep{2016MNRAS.458..558D}.

\section{Conclusions}  
\label{conclusions}

A current issue to be faced is how to automatically  classify radio galaxy shapes in large surveys with wide ranges of redshifts and  sizes whilst taking into account .resolution and interferometric sensitivity to structure  \citep{2019MNRAS.487.1729L}. The simulations analysed here suppose that the jets associated with distant radio galaxies are subject to relatively rapid variations as the energy supply and engine alter direction.
The LB Index may provide a means of diagnosing radio galaxy structure. However, these simulations confirm that   the interpretation of the intrinsic three dimensional structure will remain a matter of conjecture given the dependence on viewing angle, precession parameters and resolution. Moreover, a single radio galaxy simulation can appear in many guises given the wide range of viewing angles through axis rotation as well as axis orientation. Therefore caution needs to be taken when applying analytical models to complete sample of FR\,II type radio galaxies where their use as cosmological standard candles has been developed \citep{2019MNRAS.486.1225T}.

Consider a radio galaxy of the size of Cygnus A which possesses lobes of length $\sim$ 60\,kpc at a redshift 0f 0.056.
We move it to a distance corresponding to z $\sim$ 1.6 where the angular diameter distance is a minimum for standard cosmological models. 
This would reduce the resolution by a factor of 8, not a factor that would alter the morphological classification at high radio frequencies, as confirmed by the results here.

However, we have found that significant morphological changes occur when an image is smoothed by a factor of about ten
as well illustrated in the top panels of Fig.\, \ref{fig_zdas_redshift-emission-graph}. 
Remarkably, further smoothing does not have any additional effect: once the lobe emission has overtaken the hotspot emission, further smoothing will not significantly alter the classification. 
 
We have found that the total radio luminosity (the integrated flux in a waveband)  of a simulated lobe of a  radio galaxy  is a very weak function of 
both the source age and size. This can be clearly explained by first inspecting Panel (d) of 
Fig.\,\ref{fig_zda_zdb_zdc_zdm_emission-class}. Note the narrow luminosity range of this and the other panel (d) plots. This figure corresponds to the straight jet which drives a hotspot ahead of it with a fixed size and flux.
Therefore, if the bright hotspots in other simulations also dominate the integrated flux, then we may only expect mild increases as the diffuse lobe
inflates or even mild decreases as a hotspot pressure is reduced due to rapid precession..
This is an interesting result, consistent with the fact that
the jet kinetic power is mainly transferred into the ambient medium via shocks and compression as shown in Paper 1.

Relativistic flows are generally expected to yield similar flow patterns. The hydrodynamic equations can be rewritten in a similar form and
the Mach number must be appropriately redefined \citep{1981MNRAS.194..771S} . The  effective specific heat ratio is complex unless the gas is ultra-relativistic. In addition, the jet-ambient density ratio could be extremely low, creating wide lobes and cavities. However, the resulting radio maps may well display unique features due to the light travel time across the lobe: the rear of the lobe contributing features from an earlier time.  This has a general tendency to rotate the cube of data away from the line of sight which may increase the limb brightening of an approaching lobe \citep{1983Natur.303..779E} while acting to limb-darken a receding lobe. This does not take into account how the source structure has evolved in the meantime and so makes the relativistic version of this research of importance.

We conclude that very distant radio galaxies could appear systematically more limb-darkened  due to merger-related re-direction and precession as well as due to the sensitivity limitation. 
However, resolution itself may not directly influence the classification since the angular diameter size remains quite constant beyond the redshift of 1.5.

\section{Acknowledgements}  
\label{acks}

JD wishes to thank SEPnet for supplying infrastructure funding. 

\bibliography{jets}

\appendix
\label{appendix}
\begin{table*}
\caption{Breakdown of simulation names and main parameters used in this paper. Boundary legend where the jet travels left to right, \ $\text{left}\dfrac{\text{top}}{\text{bottom}}\text{right}$, $r$: reflextive, $o$: outflow and $os$: outflow with an added staggered grid around the main resolution.}
\label{simulation_name}
\begin{tabular}{cccccc}
\hline
File Name & Boundary &  Density Ratio & Mach & Precession $\frac{\pi}{180}$(rad) & Comments \\ 
\hline
&&&&&\textbf{Resolution 75x75x75} \\
\hline
zaa & $r \dfrac{o}{o} o$  & 0.1 & 6 & {1} &  \\
 \hline
&&&&&\textbf{Resolution 150x150x150} \\
\hline
ba & $o \dfrac{o}{o} o$   & 0.1 & 6 & {1} &  \\
bb & $o \dfrac{o}{o} o$   & 0.01 & 6 & {1} &  \\
bc & $o \dfrac{o}{o} o$   & 0.001 & 6 & {1} & \\
bd & $o \dfrac{o}{o} o$   & 0.1 & 6 & {10} & \\
bg & $o \dfrac{o}{o} o$   & 0.1 & 6 & {20} & \\
bm & $o \dfrac{o}{o} o$   & 0.0001 & 6 & {1} & \\
bn & $o \dfrac{o}{o} o$   & 0.0001 & 6 & {10} & \\
bo & $o \dfrac{o}{o} o$   & 0.0001 & 6 & {20} & \\
zbm & $r \dfrac{o}{o} o$  & 0.0001 & 6 & {1} & \\
zbn & $r \dfrac{o}{o} o$  & 0.0001 & 6 & {10} & \\
zbo & $r \dfrac{o}{o} o$  & 0.0001 & 6 & {20} & \\
zba & $r \dfrac{o}{o} o$  & 0.1 & 6 & {1} & \\
zbd & $r \dfrac{o}{o} o$  & 0.1 & 6 & {10} & \\
zbg & $r \dfrac{o}{o} o$  & 0.1 & 6 & {20} & \\
ea & $o \dfrac{o}{o} o$  & 0.1 & 2 & {1} & \\
ee & $o \dfrac{o}{o} o$  & 0.1 & 2 & {20} & \\
fa & $o \dfrac{o}{o} o$  & 0.1 & 4 & {1} & \\
fe & $o \dfrac{o}{o} o$  & 0.1 & 4 & {20} & \\
ga & $o \dfrac{o}{o} o$  & 0.1 & 8 & {1} & \\
ge & $o \dfrac{o}{o} o$  & 0.1 & 8 & {20} & \\
ha & $o \dfrac{o}{o} o$  & 0.1 & 12 & {1} & \\
he & $o \dfrac{o}{o} o$  & 0.1 & 12 & {20} & \\
ia & $o \dfrac{o}{o} o$  & 0.1 & 24 & {1} & \\
ie & $o \dfrac{o}{o} o$  & 0.1 & 24 & {20} & \\
ja & $o \dfrac{o}{o} o$  & 0.1 & 48 & {1} & \\
je & $o \dfrac{o}{o} o$  & 0.1 & 48 & {20} & \\
 \hline
&&&&&\textbf{Resolution 225x225x225} \\
\hline
zca & $r \dfrac{o}{o} o$  & 0.1 & 6 & {1} & \\ 
\hline
&&&&&\textbf{Resolution 300x300x300} \\
\hline
zda & $r \dfrac{o}{o} o$  & 0.1 & 6 & {1} & \\ 
zdas & $r \dfrac{os}{os} os$  & 0.1 & 6 & {1} & \\ 
zdgs & $r \dfrac{os}{os} os$  & 0.1 & 6 & {20} & \\ 
zdgs\_3 & $r \dfrac{os}{os} os$  & 0.1 & 6 & {20} & 2x precession rate\\ 
zdgs\_4 & $r \dfrac{os}{os} os$  & 0.1 & 6 & {20} & 4x precession rate\\ 
zdos\_2 & $r \dfrac{os}{os} os$  & 0.1 & 6 & {20} & 2x precession rate\\ 
zdos\_3 & $r \dfrac{os}{os} os$  & 0.1 & 6 & {20} & 4x precession rate\\ 
\hline
\end{tabular}
\end{table*}
\label{lastpage}

\end{document}